\pdfoutput=1
\RequirePackage{ifpdf}
\documentclass[cits,hyper]{JINST} 

\usepackage{graphicx} 
\DeclareGraphicsExtensions{.pdf}

\usepackage{atlasphysics} 
\usepackage{subfigure} 
\usepackage{multicol} 
\usepackage{multirow}

\usepackage{lineno}

\makeatletter
\g@addto@macro\bfseries{\boldmath}
\makeatother

\def\chisq{\ensuremath{\raise.4ex\hbox{$\chi$}^2}} 

\title{Operation and performance of the ATLAS semiconductor tracker}

\author{The ATLAS Collaboration}

\abstract{
  The semiconductor tracker is a silicon microstrip detector forming part of
  the inner tracking system of the ATLAS experiment at the LHC. The operation 
  and performance of the semiconductor tracker during the first years of LHC 
  running are described. More than 99\% of the detector modules were operational
  during this period, with an average intrinsic hit efficiency of (99.74$\pm$0.04)\%. 
  The evolution of the noise occupancy is discussed, and measurements of the
  Lorentz angle, $\delta$-ray production and energy loss presented. 
  The alignment of the detector is found to be stable at the few-micron level
  over long periods of time. Radiation damage measurements, which include the evolution
  of detector leakage currents, are found to be consistent with predictions and
  are used in the verification of radiation background simulations. 
}


\begin{document}

\section{Introduction}
The ATLAS detector~\cite{bib:ATLASDetectorPaper} is a multi-purpose apparatus designed to study a wide 
range of physics processes at the Large Hadron Collider (LHC)~\cite{bib:LHCPaper} at CERN. In addition to 
measurements of Standard Model processes such as vector-boson and top-quark production, the properties
of the newly discovered Higgs boson~\cite{bib:ATLASHiggs,bib:CMSHiggs} are being investigated and 
searches are being carried out for as yet undiscovered particles such as those predicted by theories
including supersymmetry. 
All of these studies rely heavily on the excellent performance of the ATLAS inner detector tracking 
system. The semiconductor tracker (SCT) is a precision silicon microstrip detector which forms an 
integral part of this tracking system.

\begin{figure}[hptb]
\begin{center}
\includegraphics[width=\textwidth]{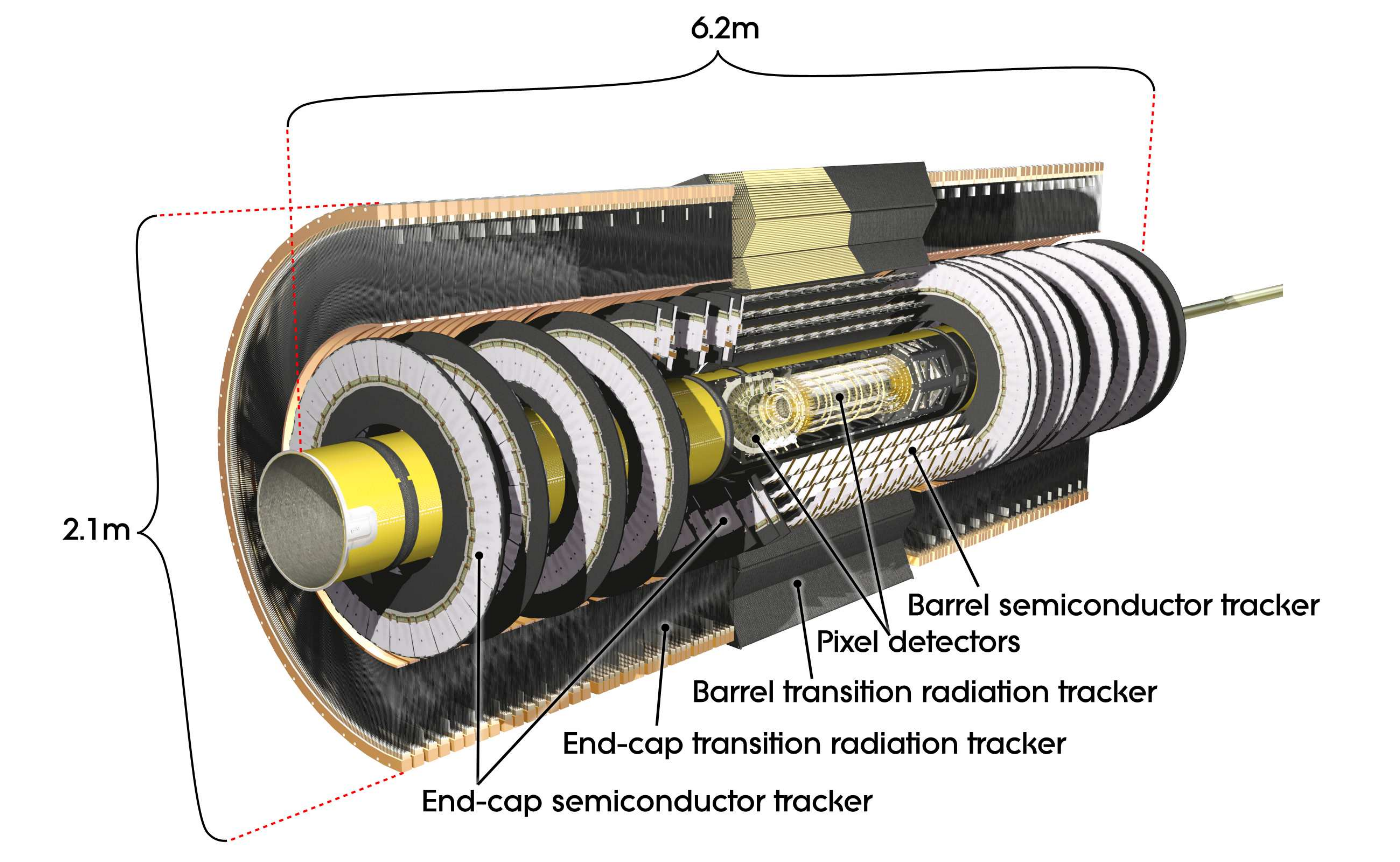}
\end{center}
\caption{A cut-away view of the ATLAS inner detector.}
\label{fig:IDLayout}
\end{figure}

The ATLAS detector is divided into three main components. A high-precision toroid-field muon 
spectrometer surrounds electromagnetic and hadronic calorimeters, which in turn surround the
inner detector. This comprises three complementary subdetectors: a silicon pixel 
detector covering radial distances\footnote{ATLAS uses a right-handed coordinate system with its origin at
the nominal interaction point in the centre of the detector and the $z$-axis along the beam-pipe.
The $x$-axis points from the interaction point to the centre of the LHC ring and the $y$-axis points
upwards. Cylindrical coordinates ($r,\phi$) are used in the transverse plane, $\phi$ being the
azimuthal angle around the beam-pipe. The pseudorapidity $\eta$ is defined in terms of the polar angle 
$\theta$ as $\eta = -\ln\tan(\theta/2)$.} between 50.5\,mm and 150\,mm, the SCT covering radial distances
from 299\,mm to 560\,mm and a transition radiation tracker (TRT) covering radial distances from 563\,mm
to 1066\,mm.  These detectors are surrounded by a superconducting solenoid providing a 2\,T axial 
magnetic field.
The layout of the inner detector, showing the SCT together with the pixel detector and transition 
radiation tracker, is shown in figure~\ref{fig:IDLayout}.
The inner detector measures the trajectories of charged particles within the 
pseudorapidity range $|\eta| < 2.5$. It has been designed to provide a 
transverse momentum resolution, in the plane perpendicular to the beam axis, of 
$\sigma_{\pT}/\pT = 0.05\%\times\pT[\GeV]\oplus 1\%$ and a transverse impact parameter resolution of 
10\,$\mu$m for high-momentum particles in the central pseudorapidity region~\cite{bib:ATLASDetectorPaper}.

After installation in the ATLAS cavern was completed in August 2008, the SCT underwent an extensive period
of commissioning and calibration before the start of LHC proton--proton collisions in late 2009.
The performance of the detector was measured using cosmic-ray data~\cite{bib:IDCosmic}; 
the intrinsic hit efficiency and the noise occupancy were found to be well within the design requirements.
This paper describes the operation and performance of the
SCT during the first years of LHC operation, from autumn 2009 to February 2013, referred to as `Run 1'.  
During this period the ATLAS experiment collected proton--proton collision data at centre-of-mass 
energies of $\sqrt{s}$ = 7~\TeV\ and 8~\TeV\ corresponding to integrated luminosities~\cite{bib:ATLASLumi} of
5.1~\ifb\ and 21.3~\ifb\ respectively, together with small amounts at 
$\sqrt{s}$ = 900~\GeV\ and 2.76~\TeV. In addition, 158~$\mu{\rm b}^{-1}$ of lead--lead collision data 
at a nucleon--nucleon centre-of-mass energy of 2.76~\TeV\ and 30~nb$^{-1}$ of proton--lead
data at a nucleon--nucleon centre-of-mass energy of 5~\TeV\ were recorded. 
The collision data are used to measure the intrinsic hit efficiency 
of the silicon modules and the Lorentz angle. Compared with the previous results using cosmic-ray
data~\cite{bib:IDCosmic}, the efficiency measurements are now extended to the endcap regions of
the detector, and the large number of tracks has allowed the Lorentz angle to be studied in more detail.  
In addition, studies of energy loss and $\delta$-ray production in the silicon have been performed. 

The layout of this paper is as follows. The main components of the SCT are described briefly in 
section 2. The operation of the detector is discussed in section 3. Offline reconstruction and 
simulation are outlined in section 4, and monitoring and data quality assessment discussed in section 5. 
Section 6 presents performance results, including detector occupancy in physics running, noise occupancy,
alignment, efficiency, measurements of the Lorentz angle and energy loss in the silicon and a study of 
$\delta$-ray production in silicon. Finally, section 7 describes the effects of radiation on the 
detector.

\section{The SCT detector}
\label{sec:detector}
The main features of the SCT are described briefly in this section; full details can be found
in ref.~\cite{bib:ATLASDetectorPaper}. 

\begin{figure}[htb]
\begin{center}
\includegraphics[width=\textwidth]{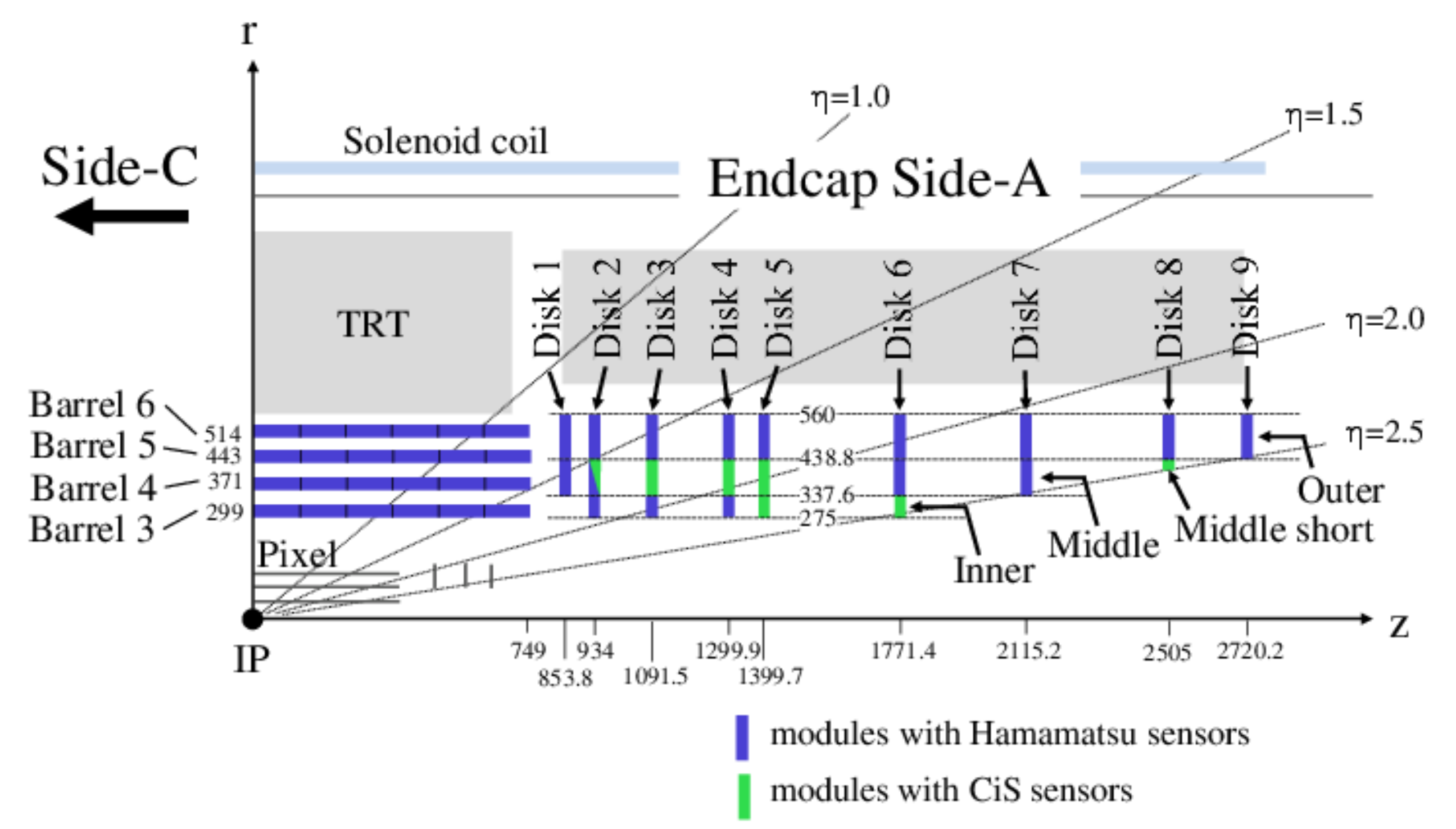}
\end{center}
\caption{A schematic view of one quadrant of the SCT. The numbering scheme for barrel layers and endcap 
         disks is indicated, together with the radial and longitudinal coordinates in millimetres. The 
         disks comprise one, two or three rings of modules, referred to as inner, middle and outer 
         relative to the beam pipe. The figure distinguishes modules made from Hamamatsu (blue) or 
         CiS (green) sensors; the middle ring of disk 2 contains modules of both types. The two sets 
         of endcap disks are distinguished by the labels A (positive $z$) and C (negative $z$).}
\label{detector:Quadrant}
\end{figure}

\subsection{Layout and modules}
The SCT consists of 4088 modules of silicon-strip detectors arranged in four concentric barrels 
(2112 modules) and two endcaps of nine disks each (988 modules per endcap), as shown in 
figure~\ref{detector:Quadrant}. 
Each barrel or disk provides two strip measurements at a stereo angle which are combined to build space-points. The SCT typically provides eight strip measurements (four 
space-points) for particles originating in the beam-interaction region.
The barrel modules~\cite{bib:SCTBarrelModules} are of a uniform design, with strips approximately 
parallel to the magnetic field and beam axis. 
Each module consists of four rectangular silicon-strip sensors~\cite{bib:SCTSensors} with strips with a 
constant pitch of 80~$\mu$m; two sensors on each side are daisy-chained together to give 768 strips of 
approximately 12~cm in length.
A second pair of identical sensors is glued back-to-back with the first pair at a
stereo angle of 40~mrad. The modules are mounted on cylindrical supports
such that the module planes are at an angle to the tangent to the cylinder 
of 11$\degr$ for the inner two barrels and 11.25$\degr$ for the outer two barrels, 
and overlap by a few millimetres to provide a hermetic tiling in azimuth.

Each endcap disk consists of up to three rings of modules~\cite{bib:SCTEndcapModules} with trapezoidal 
sensors. The strip direction is radial with constant azimuth and a mean pitch of 80~$\mu$m.
As in the barrel, sensors are glued back-to-back at a stereo angle of 40~mrad to provide 
space-points. Modules in the outer and middle rings consist of two daisy-chained sensors on each side, 
whereas those in the inner rings have one sensor per side.

All sensors are 285~$\mu$m thick and are constructed of high-resistivity n-type bulk silicon with 
p-type implants. Aluminium readout strips are capacitatively coupled to the implant strips. 
The barrel sensors and 75\% of the endcap sensors were supplied by Hamamatsu 
Photonics,\footnote{Hamamatsu Photonics Co.\ Ltd.,1126-1 Ichino-cho, Hamamastu, Shizuoka 431-3196, Japan.}
while the remaining endcap sensors were supplied by CiS.\footnote{CiS Institut f\"{u}r Mikrosensorik 
gGmbH, Konrad-Zuse-Stra{\ss}e 14, 99099 Erfurt, Germany.} 
Sensors supplied by the two manufacturers meet the same performance
specifications, but differ in design and processing details~\cite{bib:SCTSensors}.  
The majority of the modules are constructed from silicon wafers with crystal lattice orientation
(Miller indices) $<$111$>$. However, a small number of modules in the barrel ($\sim$90) use wafers 
with $<$100$>$ lattice orientation. For most purposes, sensors from the different manufacturers or 
with different crystal orientation are indistinguishable, but differences in e.g.\ noise performance 
have been observed, as discussed in section~\protect\ref{sec:noise}.

Measurements often require a selection on the angle of a track incident on a silicon module. 
The angle between a track and the normal to the sensor in the plane defined by the normal 
to the sensor and the local $x$-axis (i.e.\ the axis in the plane of the sensor perpendicular 
to the strip direction) is termed $\phi_{\rm local}$.
The angle between a track and the normal to the sensor in the plane defined by the normal 
to the sensor and the local $y$-axis (i.e.\ the axis in the plane of the sensor parallel 
to the strip direction) is termed $\theta_{\rm local}$.

\subsection{Readout and data acquisition system}
\label{sec:DAQ}
The SCT readout system was designed to operate with 0.2\%--0.5\% occupancy in the 6.3 million sampled 
strips, for the original expectations for the LHC luminosity of 1$\times$10$^{34}$~cm$^{-2}$s$^{-1}$ and 
pile-up\footnote{The term pile-up refers to multiple {\it pp} interactions per bunch crossing.} of up to 23 
interactions per bunch crossing. 
The strips are read out by radiation-hard front-end ABCD chips~\cite{bib:SCTABCD3TA}
mounted on copper--polyimide flexible circuits termed the readout hybrids.  
Each of the 128 channels of the ABCD has a preamplifier and shaper stage; the output has a shaping time of 
$\sim$20~ns and is then discriminated to provide a binary output. A common discriminator threshold is applied to 
all 128 channels, normally corresponding to a charge of 1~fC, and settable by an 8-bit DAC. To compensate 
for variations in individual channel thresholds, each channel has its own 4-bit DAC (TrimDAC) used to offset the 
comparator threshold and enable uniformity of response across the chip. The step size for each TrimDAC setting 
can be set to four different values, as the spread in uncorrected channel-to-channel variations is 
anticipated to increase with total ionising dose.

The binary output signal for each channel is latched to the 40~MHz LHC clock and stored in a 132-cell pipeline; 
the pipeline records the hit sequence for that channel for each clock cycle over $\sim$3.2~$\mu$s. 
Following a level-1 trigger, data for the preceding, in-time and following bunch crossings are compressed 
and forwarded to the off-detector electronics. Several modes of data compression have been used, as 
specified by the hit pattern for these three time bins:
\begin{itemize}
\item Any hit mode; channels with a signal above threshold in any of the three bunch crossings are read out. 
      This mode is used when hits outside of the central bunch crossing need to be recorded, for example to 
      time in the detector or to record cosmic-ray data.
\item Level mode (X1X); only channels with a signal above threshold in the in-time bunch crossing are read 
      out. This is the default mode used to record data in 2011--2013, when the LHC bunch spacing was 50~ns,
      and was used for all data presented in this paper unless otherwise stated. 
\item Edge mode (01X); only channels with a signal above threshold in the in-time bunch crossing and no hit 
      in the preceding bunch crossing are read out. This mode is designed for 25~ns LHC bunch spacing, 
      to remove hits from interactions occurring in the preceding bunch crossing. 
\end{itemize}

The off-detector readout system~\cite{bib:SCTDAQ} comprises 90 9U readout-driver boards (RODs) and 90 
back-of-crate (BOC) cards, housed in eight 9U VME crates.
A schematic diagram of the data acquisition system is shown in figure~\ref{fig:DAQ}.
Each ROD processes data for up to 48 modules, and the BOC provides the optical interface  between the ROD and the modules; the BOC also transmits the reformatted data for those 48 modules to the ATLAS data acquisition (DAQ) chain via a single fibre known as an `S-link'.
There is one timing, trigger and control (TTC) stream per module from the BOC to the module, and two data streams returned from each module corresponding to the two sides of the module.
The transmission is based on vertical-cavity surface-emitting lasers (VCSELs) operating at a wavelength of 850~nm and uses radiation-hard fibres~\cite{bib:SCTOpticalLinks}. 
The TTC data 
are broadcast at 40\ Mb/s to the modules via 12-way VCSEL arrays, and converted back to electrical signals by silicon {\it p-i-n} diodes on the 
on-detector optical harnesses. Data are optically transmitted back from the modules  at 40\ Mb/s and received by arrays 
of silicon {\it p-i-n} diodes on the BOC. 

\begin{figure}[htb]
\begin{center}
\includegraphics[width=0.9\textwidth]{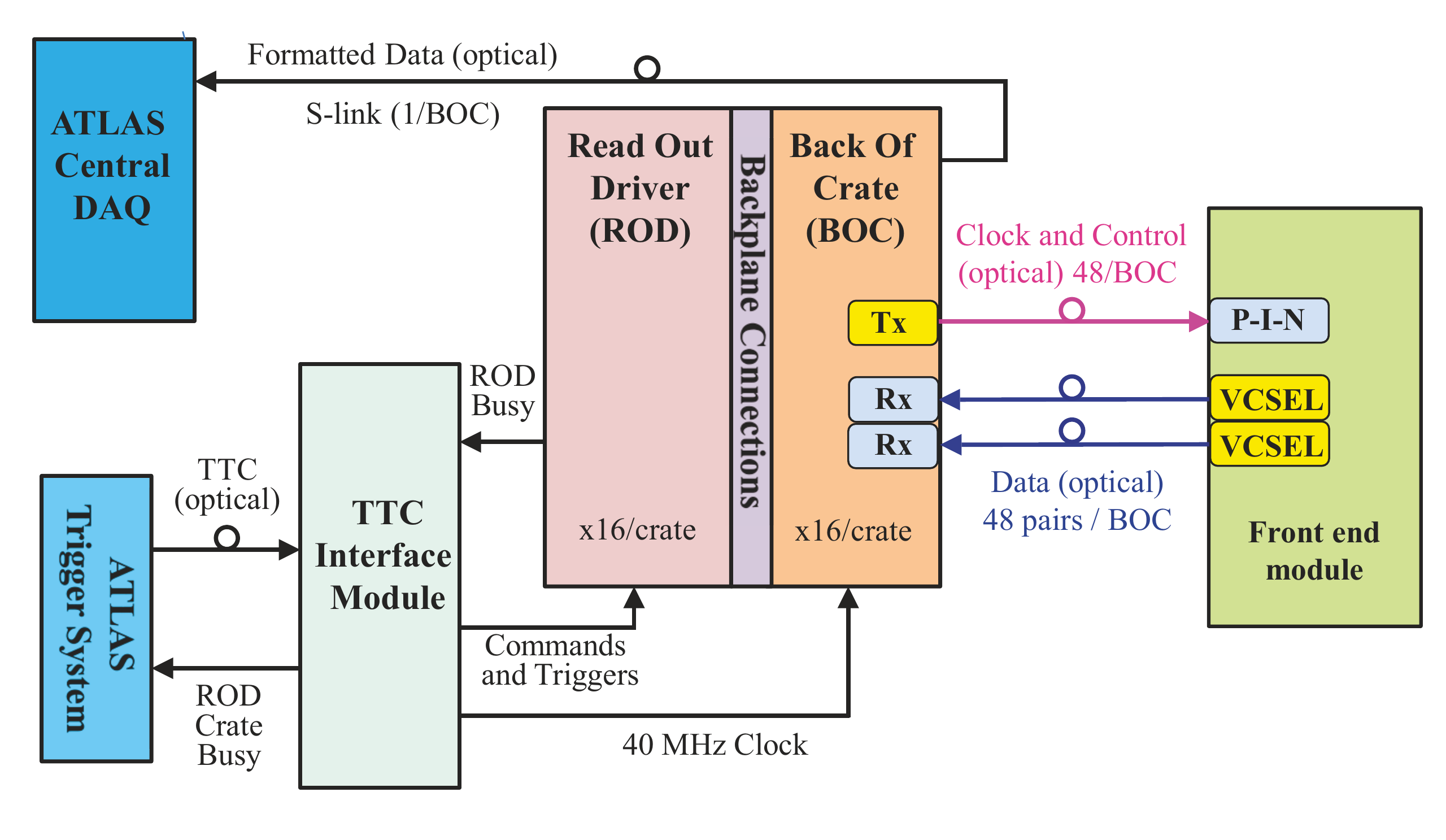}
\end{center}
\caption{A schematic diagram of the SCT data acquisition hardware showing the main connections between
         components.} 
\label{fig:DAQ}
\end{figure}

Redundancy is implemented for both the transmitting (TX) and receiving (RX) optical links, in case of fibre breaks, VCSEL 
failures or diode problems.
Redundancy is built into the TX system by having electrical links from one module to its neighbour. 
If a module loses its TTC signal for any reason, an electrical control line can be set which results in the neighbouring module sending a copy of its TTC data to the module with the failed signal, 
without any impact on operation. 
For the data links, one side of the module can be configured to be read out through the other link. 
Although readout of both sides of the module though one RX link inevitably reduces the readout bandwidth, this
is still within design limits for SCT operation. 
For the barrel modules, readout of both sides through one link also results in the loss of one ABCD chip on the re-routed link 
because the full redundancy scheme was not implemented due to lack of space on the readout hybrid.

Table~\ref{redundancy} shows the number of optical links configured to use the redundancy mechanism  at the end of data-taking in February 2013. 
The RX links configured to read out both module sides were mainly due to connectivity defects during installation of the SCT, and the number remained stable throughout Run~1. The use of the TX redundancy mechanism varied significantly due to VCSEL failures, discussed in section~\protect\ref{sec:operationissues}.

\begin{table}[htdp]
\caption{Numbers of optical links configured to use the redundancy mechanism in February 2013, 
         in the barrel, the endcaps and the whole SCT. The last column shows the corresponding 
         fraction of all links in the detector.}
\begin{center}
\begin{tabular}{lcccc}
\hline
\hline
\multicolumn{5}{c}{Links using redundancy mechanism} \\
\hline
Links & Barrel & Endcaps & SCT & Fraction [\%] \\
\hline
TX & 9 & 5 & 14 & 0.3 \\
RX & 41 & 91 & 132 & 1.6 \\
\hline
\hline
\end{tabular}
\end{center}
\label{redundancy}
\end{table}

\subsection{Detector services and control system}
\label{sec:DCS}
The SCT, together with the pixel detector, is cooled by a bi-phase evaporative 
system~\cite{bib:CoolingPaper} which is designed to deliver C$_3$F$_8$ fluid at $-25$$^\circ$C 
via 204 independent cooling loops within the low-mass cooling structures on the detector. 
The target temperature for the SCT silicon sensors after irradiation is $-7$$^\circ$C, which was 
chosen to moderate the effects of radiation damage. 
An interlock system, fully implemented in hardware, using two 
temperature sensors located at the end of a half cooling-loop for 24 barrel modules or
either 10 or 13 endcap modules, prevents the silicon modules from overheating in the event
of a cooling failure by switching off power to the associated channels within approximately
one second. 
The cooling system is interfaced to and monitored by the detector control system, and is operated independently of the status of SCT operation.

The SCT detector control system (DCS)~\cite{bib:SCTDCS} operates
within the framework of the overall ATLAS
DCS~\cite{bib:ATLASDCSPaper}. Custom embedded local monitor
boards~\cite{bib:ELMB}, designed to operate in the strong magnetic
field and high-radiation environment within ATLAS, provide the
interface between the detector hardware and the readout system. Data
communication through controller area network\footnote{CAN in Automation, 
Kontumazgarten 3, DE-90429, N\"{u}renburg, Germany.} (CAN) buses, 
alarm handling and data display are
handled by a series of PCs running the commercial controls software
PVSS-II.\footnote{ETM professional control GmbH, Marketstrasse 3, A-7000 Eisenstadt, Austria.}

The DCS is responsible for operating the power-supply system: setting the high-voltage (HV)
supplies to the voltage necessary to deplete the sensors and the low-voltage (LV) supplies for the 
read-out electronics and optical-link operation. It monitors voltages and currents, and also
environmental parameters such as temperatures of sensors, support structures and cooling pipes, 
and the relative humidity within the detector volume. 
The DCS must ensure safe and reliable operation of the SCT, by taking appropriate action
in the event of failure or error conditions. 

The power-supply system is composed of 88 crates, each controlled
by a local monitor board and providing power for 48 LV/HV channels. For each
channel, several parameters are monitored and controlled, amounting
to around 2500 variables per crate. A total of 16 CAN buses are needed
to ensure communication between the eight DCS PCs and the power-supply
crates. The environmental monitoring system reads temperatures and
humidities of about 1000 sensors scattered across the SCT volume, and
computes dew points. Temperature sensors located on the outlet of the
cooling loops are read in parallel by the interlock system, which can
send emergency stop signals to the appropriate power-supply crate by means of an
interlock-matrix programmable chip in the unlikely event of an
unplanned cooling stoppage. All environmental sensors are read by
local monitor boards connected to two PCs, each using one CAN bus.

The SCT DCS is integrated into the ATLAS DCS via a finite state machine structure  
(where the powering status of the SCT can be one of a finite number of states) forming
a hierarchical tree. States of the hardware are propagated up the tree and
combined to form a global detector state, while commands are
propagated down from the user interfaces. Alarm conditions can be
raised when the data values from various parameters of the DCS go outside of defined limits.

Mitigation of electromagnetic interference and noise pickup from power lines is critical 
for the electrical performance of the detector. Details of the grounding and shielding of 
the SCT are described in ref.~\cite{bib:SCTGrounding}.

\subsection{Frequency scanning interferometry}
Frequency scanning interferometry (FSI) is a novel technique
developed to monitor the alignment stability of the detector by measuring
distances between fiducial points on the support structure with high 
precision~\cite{Gibson:1305878}. It provides information to
augment track-based alignment by determining the internal
distortions of the SCT structure on short, medium and long timescales.
Lengths are measured using 842 laser interferometers arranged in a geodetic
grid covering the detector; the grid layout is shown in figure~\ref{fig:GLIlayout}.
The lengths of the grid lines are measured in real time and compared to a reference 
length provided by an off-detector reference system in a controlled environment. 
Two lasers are used to scan the frequencies in opposite
directions (increasing, decreasing) to cancel drift errors. The light
from each laser is split into two beams to be sent simultaneously to
the endcap and barrel sections of the system. The
beams are then split close to the detector into the hundreds of
interferometers. 
The same light is also sent to the  reference system for later analysis.
The working principle of the individual
interferometers is shown in figure~\ref{fig:GLIdesign}.
The distance is measured between two components of each interferometer: a
`quill' which contains the light delivery and return fibres, and a distant
retro-reflector.
The wide-angle beam emerging from the quill provides
tolerance to small misalignments which may occur during the
planned ten-year operational lifetime. As a trade off, the interferometer
provides a return signal of around only 1~pW per mW of input
power, for a 1\ m interferometer length. 

\begin{figure}[ht!]
\centering
\subfigure[]{\includegraphics[width=\textwidth]{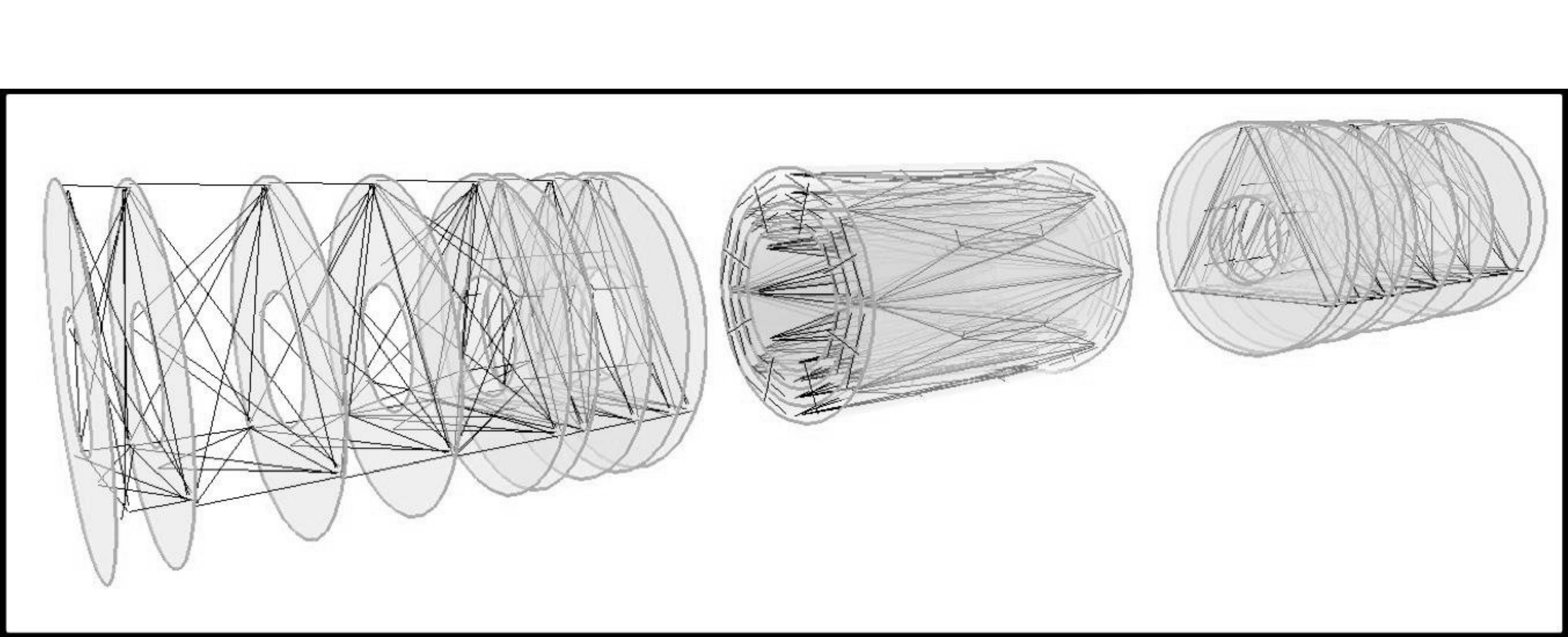} \label{fig:GLIlayout}}
\subfigure[]{\includegraphics[width=0.8\textwidth]{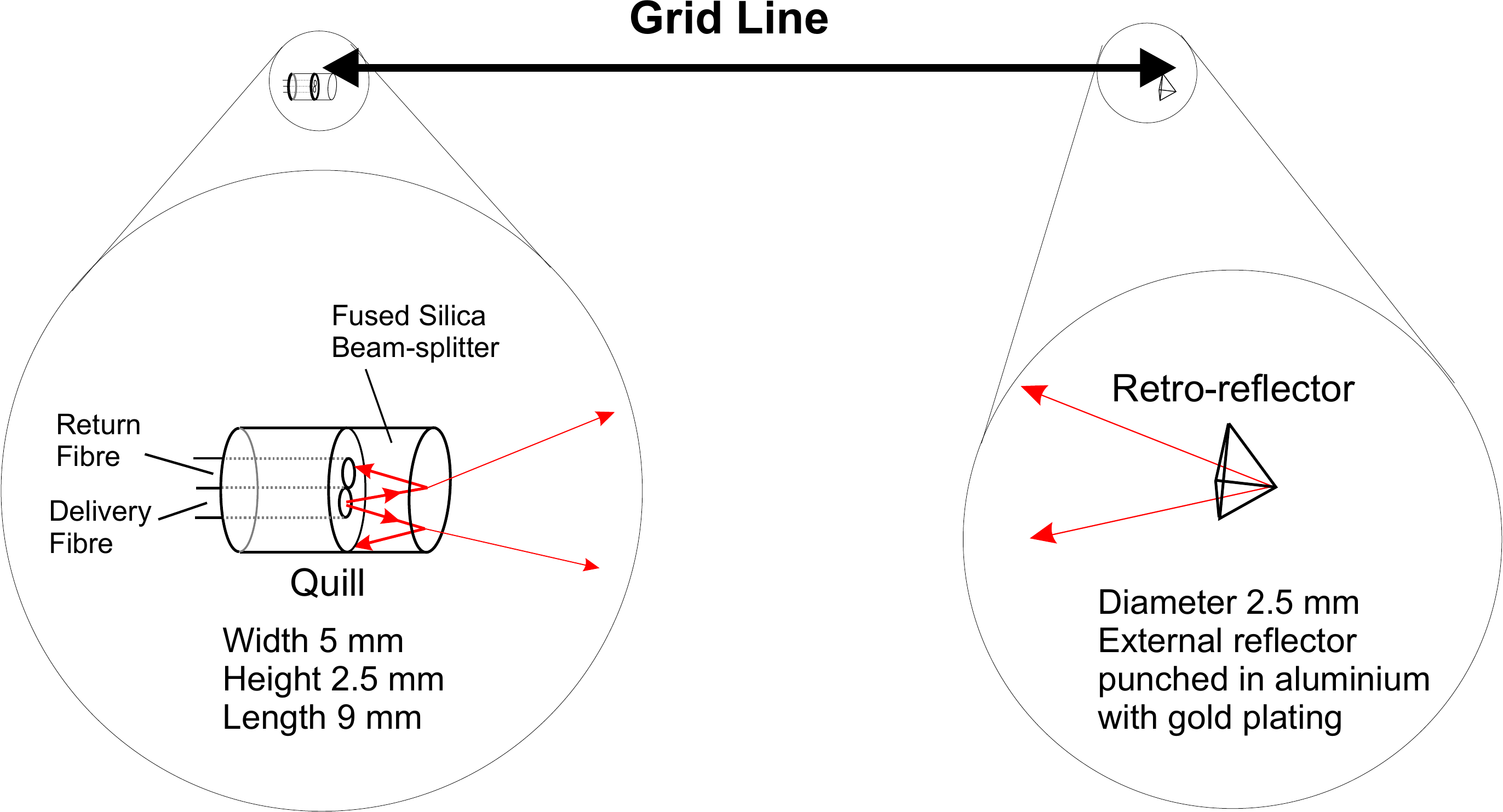} \label{fig:GLIdesign}} 
\subfigure[]{\includegraphics[width=0.7\textwidth]{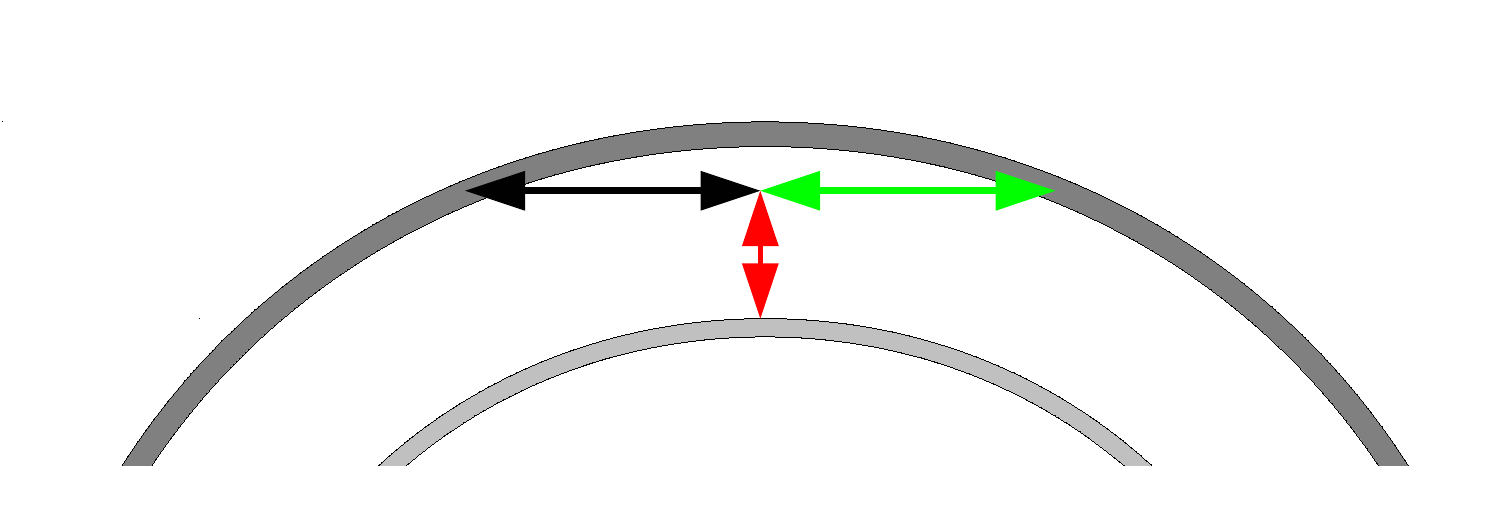} \label{fig:flange}}
\caption[Grid Line Interferometer design]{\label{fig:GLI}{
      (a) The FSI grid layout across the SCT volume.
      (b) The grid-line interferometer design for the FSI system. 
      (c) The arrangement of each group of three grid-line interferometers on the barrel flange. 
       The colours used here for each line in the assembly are replicated in
       section~\protect\ref{sec:fsires}.}}
\end{figure}

The FSI can be operated to measure either the absolute or relative
phase changes of the interference patterns. In the absolute mode, 
absolute distances are measured with micron-level precision over long
periods. In the relative mode, the relative phase change of each interferometer 
is monitored over short periods with a precision on distance approaching 50~nm. 
Both modes can be used with all 842 grid lines. 

The FSI has been in operation since 2009. The nominal power
to the interferometers is 1~mW per interferometer; however, during early
operation, with low trip levels on the leakage current from the
sensors, this power was inducing too much leakage current in the silicon
modules. For this reason, the power output of the two main lasers was
reduced to allow for safe SCT operation. As the leakage
current increased because of radiation damage, this limitation was
relaxed and the trip levels increased. In the current setup, the optical power 
delivered only allows analysis of data from the 144 interferometers that
measure distances between the four circular flanges at each end of the barrel.
These interferometers are grouped into 48 assemblies of three interferometers
each, which monitor displacements between the carbon-fibre support cylinders
in adjacent barrel layers, as illustrated in figure~\ref{fig:flange}.

\subsection{Detector safety}
The ATLAS detector safety system~\cite{bib:DSS} protects the SCT and all other ATLAS detector 
systems by bringing the detector to a safe state in the event of alarms arising from cooling system
malfunction, fire, smoke or the leakage of gases or liquids. Damage to the silicon sensors could also 
arise as a result of substantial charge deposition during abnormal beam conditions.
Beam conditions are monitored by two devices based on radiation-hard polycrystalline chemical 
vapour deposition diamond sensors.

The beam conditions monitor (BCM)~\cite{bib:ATLASDetectorPaper,bib:bcm1,bib:BCMPaper}
consists of two stations, forward and backward, each with four modules located at 
$z = \pm$1.84~m and a radius of 5.5~cm. Each module has two diamond sensors of
1 $\times$ 1~cm${^2}$ surface area and 500~$\mu$m thickness mounted back-to-back. 
The 1~ns signal rise time allows discrimination of particle hits due to collisions
(in-time) from background (out-of-time). Background rates for the two circulating
beams can be measured separately, and are used to assess the conditions before
ramping up the high voltage on the SCT modules. 

The beam loss monitor (BLM)~\cite{bib:BLM} consists of 12 diamond sensors located
at a radius of 6.5~cm at $z \simeq \pm$3.5~m. The radiation-induced currents in the
sensors are averaged over various time periods ranging from 40~$\mu$s to 84~s. The
BLM triggers a fast extraction of the LHC beams if a high loss rate is detected,
i.e.\ if the current averaged over the shortest integration time of 40~$\mu$s exceeds 
a preset threshold simultaneously in two modules on each side of the interaction point.

\section{Operation}
\label{sec:Operations}

The LHC delivered proton--proton collision data at $\sqrt{s}$ = 7~\TeV\ corresponding to  
integrated luminosities of 47~pb$^{-1}$ and 5.6~fb$^{-1}$ in 2010 and 2011 respectively, 
and a further 23.3~fb$^{-1}$ at $\sqrt{s}$ = 8~\TeV\ in 2012. 
There were also three periods of running with heavy ions instead of protons, each approximately one month long. The SCT has been operational throughout all data-taking periods. 
It delivered high-quality tracking data for 99.9\%, 99.6\% and 99.1\%  of the delivered 
proton--proton luminosity in 2010, 2011 and 2012 respectively.

The typical cycle of daily LHC operations involves a period of beam injection and energy ramp, optimisation for collisions, declaration of collisions with stable conditions, a long period of physics data-taking, and finally a dump of the beam.
The SCT remains continuously powered regardless of the LHC status.
In the absence of stable beam conditions at the LHC, the SCT modules are biased at a 
reduced high voltage of 50~V to ensure that the silicon sensors are only partially depleted; in 
the unlikely event of a significant beam loss, this ensures that a maximum of 50\ V is 
applied temporarily across the strip oxides, which is not enough to cause electrical breakdown.
Normal data-taking requires a bias voltage of 150\ V on the silicon in order to maximise hit efficiencies 
for tracking, and the process of switching the SCT from standby at 50\ V to on at 150\ V is 
referred to as the `warm start'. Once the LHC declares stable beam conditions, the SCT is
automatically switched on if the LHC collimators are at their nominal positions for physics, if the background rates measured in BCM, BLM and the ATLAS forward detectors are
low enough, and if the SCT hit occupancy with 50~V is consistent with the expected luminosity.

\subsection{Detector status}
\label{sec:DetConf}

The evaporative cooling system provided effective cooling for the SCT as well as the pixel sub-detector throughout the Run~1 period. The system was usually operated continuously apart from 10--20 days of maintenance annually while the LHC was shut down. Routine maintenance (e.g.\ compressor replacements) could be performed throughout the year without affecting the operation, as only four of the available seven compressors were actually required for operation. In the first year, the system had several problems with compressors, leaks of fluid and malfunctioning valves. However, operation in 2011 and 2012 was significantly more stable, following increased experience and optimisation of maintenance procedures. For example, in 2011 there were only two problems coming from the system itself out of 19 cooling stops. The number of active cooling loops as a function of time during Run~1 is shown in figure~\ref{fig:cooling_loops}. Figure~\ref{fig:cooling_temp} shows the mean temperatures for each
barrel and each endcap ring in the same time period, as measured by sensors mounted on the hybrid of each module. The inner 
three barrels were maintained at a hybrid temperature of approximately $2^{\circ}$C, while the outermost
barrel and the endcap disk temperatures were around $7^{\circ}$C. The mean temperatures of each layer 
were stable within about one degree throughout the three-year period.     

\begin{figure}[htbp]
\begin{center}
\subfigure[]{\includegraphics[width=0.49\textwidth]{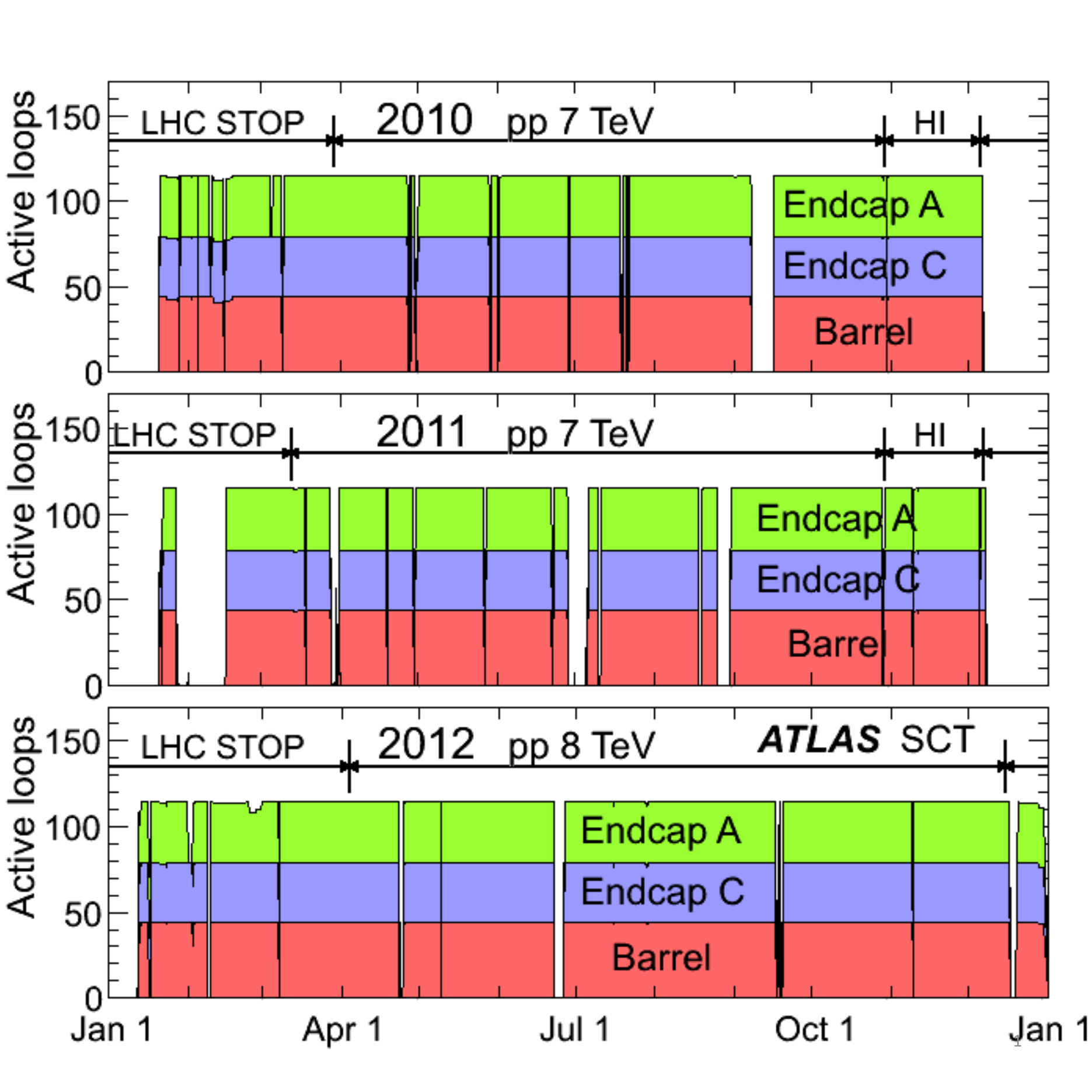} \label{fig:cooling_loops}}
\subfigure[]{\includegraphics[width=0.49\textwidth]{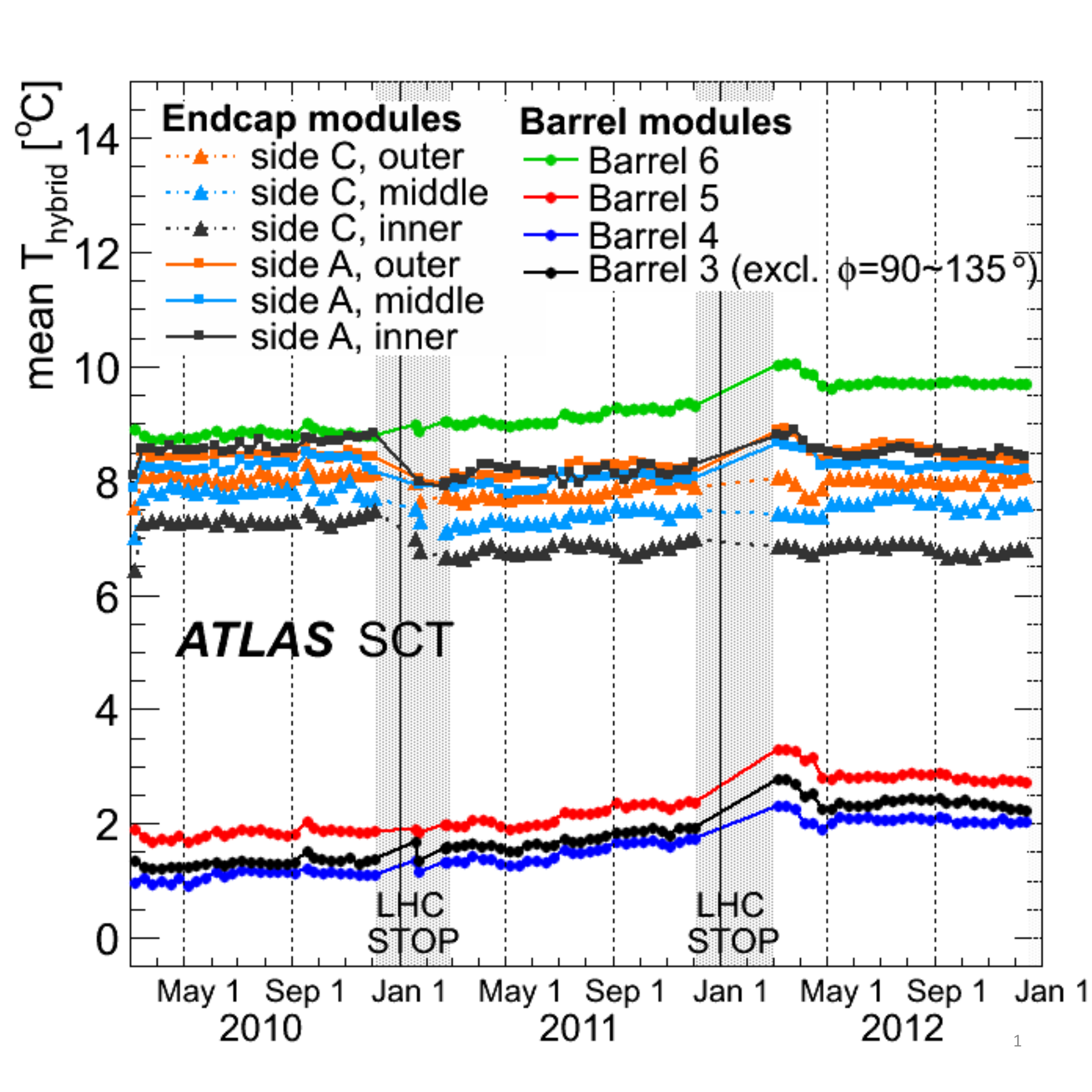} \label{fig:cooling_temp}}
\caption{(a) Number of active cooling loops during 2010 to 2012. Periods of $pp$ collisions, heavy-ion (HI)
             running and LHC shutdown periods are indicated. The periods of a few days with no active
             cooling loops correspond to short technical stops of the LHC.
         (b) Mean hybrid temperature, $T_{\rm hybrid}$, as a function of time (averaged
             over intervals of ten days). Values are shown for each barrel, and 
             for each ring of endcap modules, separately for endcap sides A and C.
             The values for barrel 3 exclude modules served by one cooling loop 
             ($\phi \sim 90^{\circ}-135^{\circ}$), 
             which has a temperature consistently about $4^{\circ}$ higher. The grey bands indicate
             the extended periods without LHC beam around the end of each year.} 
 \end{center}
\end{figure}

The detector operative fraction was consistently high, with at least 99\% of the SCT modules functional and available for tracking throughout 2010 to 2013. 
Table~\ref{tab:disabled} shows the numbers of disabled detector elements at the end of data-taking in February 2013. The numbers are typical and changed minimally during the Run~1 period. The number of disabled strips (mainly due to high noise or unbonded channels) and non-functioning chips is negligible and the largest contribution is due to disabled modules, as detailed in table~\ref{disabledmodules}.
\begin{table}[htd]
\caption{Numbers of disabled SCT detector elements in February 2013, in the barrel, the endcaps and
         the whole SCT. The last column shows the corresponding fraction of all elements in the detector. 
         The numbers of chips (strips) exclude those in disabled modules (modules and chips).}
\begin{center}
\begin{tabular}{lccccc}
\hline
\hline
\multicolumn{5}{c}{Number of disabled elements} \\
\hline
Component & Barrel & Endcaps & SCT   & Fraction [\%] \\
\hline
Modules   & 11     & 19      & 30    & 0.73 \\
Chips     & 38     & 17      & 55    & 0.11 \\
Strips    & 4111   & 7252    & 11363 & 0.18 \\
\hline
\hline
\end{tabular}
\end{center}
\label{tab:disabled}
\end{table}
\begin{table}[htdp]
\caption{Numbers of disabled modules in February 2013 classified according to reason. 
         The first three columns show the numbers of modules affected by each issue for 
         the barrel, endcaps and the whole SCT, while the final column shows the corresponding 
         fraction of all modules in the detector.}
\begin{center}
\begin{tabular}{lccccc}
\hline
\hline
\multicolumn{5}{c}{Number of disabled modules} \\
\hline
          & Barrel & Endcaps & SCT & Fraction [\%] \\
\hline
Cooling   & 0      & 13      & 13  & 0.32 \\
LV        & 6      & 1       & 7   & 0.17 \\
HV        & 1      & 5       & 6   & 0.15 \\
Readout   & 4      & 0       & 4   & 0.10 \\
\hline
Total     & 11     & 19      & 30  & 0.73 \\
\hline
\hline
\end{tabular}
\end{center}
\label{disabledmodules}
\end{table}
Half of the disabled modules are due to one cooling loop permanently disabled as a result of an inaccessible leak in that loop. Fortunately this only affects one quadrant of one of the outermost endcap disks, and has negligible impact on tracking performance. The remaining disabled modules are predominantly due to on-detector connection issues.

\subsection{Calibration}
\label{sec:calibration}
Calibrations were regularly performed between LHC fills. The principle aim is to impose a 1\ fC threshold across all chips to ensure low noise occupancy ($<$5$\times$10$^{-4}$) and yet high hit efficiency ($>$99$\%$) for each channel. Calibrations also provide feedback to the offline event reconstruction, and measurements of electrical parameters such as noise for use in studies of detector performance. There are three categories of calibration tests:
\begin{itemize}
\item Electrical tests to optimise the chip configuration and to measure noise and gain, performed
      every few days.
\item Optical tests to optimise parameters relevant to the optical transmission and reception of data between the back-of-crate cards and the modules, performed daily when possible.
\item Digital tests to exercise and verify the digital functionality of the front-end chips, performed occasionally.

\end{itemize}

The principle method of electrical calibration is a threshold scan. A burst of triggers is issued and the 
occupancy (fraction of triggers which generate a hit) is measured while a chip parameter, usually the 
discriminator setting, is varied in steps.
Most electrical calibrations involve injecting a known amount of charge into the front-end of the chip by 
applying a voltage step across a calibration capacitor.
The response to the calibration charge when scanning the discriminator threshold is parameterised using a
complementary error function. The threshold at which the occupancy is 50\% (vt50) corresponds to the median of 
the injected charge, while the Gaussian spread gives the noise after amplification. The process is repeated 
for several calibration charges (typically 0.5~fC to 8~fC). The channel gain is extracted from the dependence
of vt50 on input charge (slope of the response curve) and the input equivalent noise charge (ENC) is
obtained by dividing the output noise by the gain. 

\subsection{Timing}

The trigger delivered to each SCT module must be delayed by an amount equal to the pipeline length minus all 
external delays (e.g.\ those incurred by the trigger system and cable delays) in order to select the correct 
position in the pipeline.
Prior to collision data-taking, the trigger delay to each of the 4088 modules was adjusted to compensate 
for the different cable and fibre lengths between the optical transmitter on the BOC and the point of trigger 
signal distribution on the module, and for the different times-of-flight for particles from collisions, 
depending on the geometric location of the module. Further adjustments were applied using collision data.

On receipt of a level-1 trigger, the contents of three consecutive pipeline bins are sampled, and the timing 
is considered optimal if the hit pattern arising from a particle from a collision in ATLAS gives 01X (nothing 
in first bin, a hit in the middle bin, and either in the third bin). The procedure for timing in the SCT is to 
take physics data with $pp$ collisions while stepping through the timing delay in (typically) 5~ns steps. 
The optimal delay is derived from the mid-point of the time-delay range for which the fraction of recorded 
hits on reconstructed tracks satisfying 01X is maximal. In general, such a timing scan was performed 
and any necessary timing adjustments applied during the first collision runs in each year. 

A verification of the timing is performed by a check of the hit pattern in the three sampled time
bins during data-taking. For each module, a three-bin histogram is filled according to whether
a hit-on-track is above threshold in each time bin. After timing in, only the time-bin patterns 
010 and 011 are significantly populated. The mean value of the histogram would be 1.0 if all 
hits were 010 and 1.5 if all hits were 011, since the second and third bins would be equally 
populated. The mean value per layer for a high-luminosity $pp$ run in October 2012 is shown in 
figure~\ref{fig:timingplot}; the
error bars represent the r.m.s.\ spread of mean time bin for modules in that layer. 
The plot is typical, and illustrates that the timing of the SCT is uniform. Around 71\% of the 
hits-on-track in this run have a time-bin pattern of 011. This fraction varies by about 1\% from layer 
to layer and 2.5\% for modules within a layer.
\begin{figure}
  \centering
  \includegraphics[width=\columnwidth]{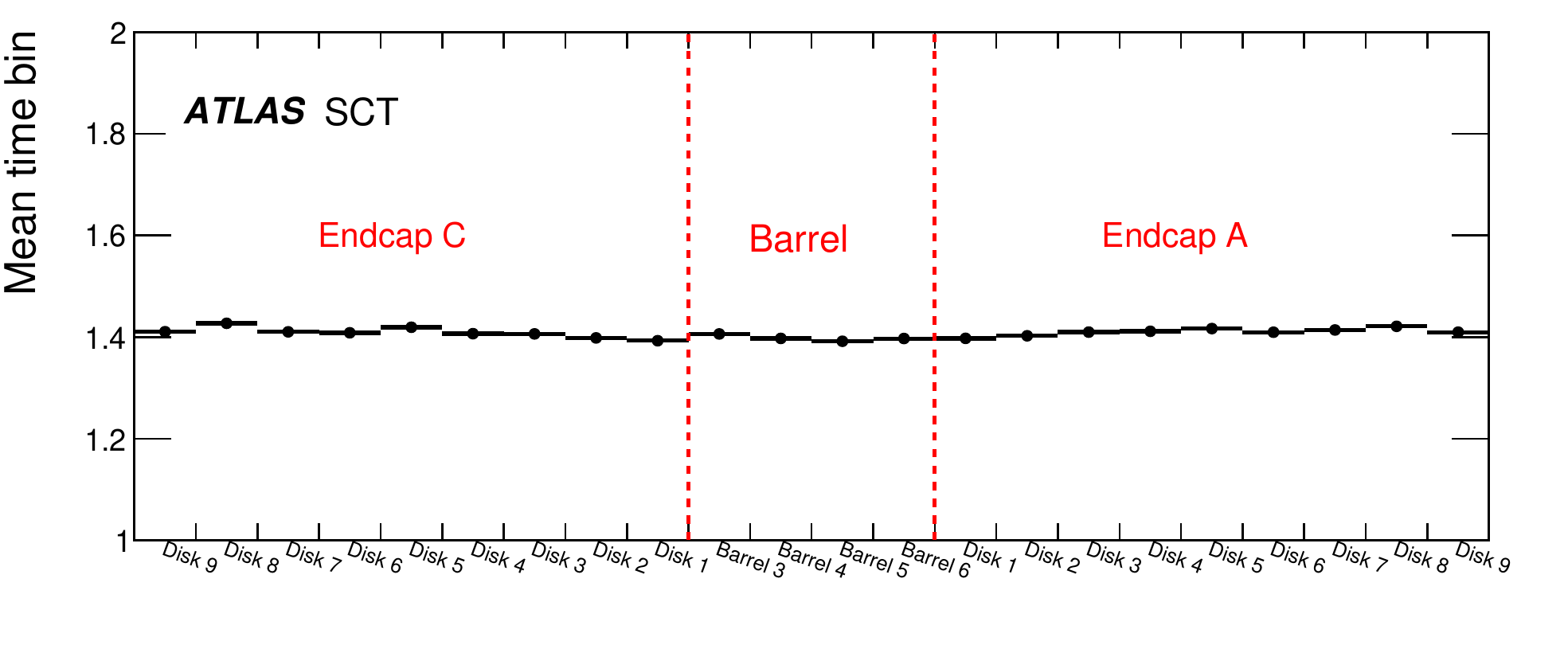}
  \caption{The mean of the three-bin timing distribution across all SCT layers; 010 and 011 hits 
           correspond to 1.0 and 1.5 in the plot, respectively. The error bars represent the 
           r.m.s.\ spread of mean time bin for modules in that layer.}
  \label{fig:timingplot}
\end{figure}

\subsection{Data-taking efficiency}
\label{sec:data-taking_eff}

There are three potential sources of data-taking inefficiency: the time taken to switch on the SCT upon the declaration of stable beam conditions, errors from the chips which flag that data fragments from those chips cannot be used for tracking, and a signal (known as `busy') from the SCT DAQ which inhibits ATLAS data-taking due to a DAQ fault. Of these, the busy signal was the dominant cause of the $\sim$0.9\% loss in data-taking efficiency in 2012; the warm start typically took 60 seconds, which was shorter than the time taken by some other subsystems in ATLAS; the chip errors effectively reduced the detector acceptance but had little impact on overall data-taking efficiency.

 The presence of ROD busy signals was mainly due to the very high occupancy and trigger rates experienced during 2012; the mean number of interactions per bunch crossing during $pp$ collisions routinely exceeded the original design expectations of around 23, often reaching 30 or more with level-1 trigger rates of typically 60--80~kHz. Although the bandwidth of the DAQ was sufficient to cope with these conditions, the high occupancy and rate exposed shortcomings in the on-ROD processing and decoding of the data which led to an increased rate of disabled data links and also ROD busy signals. Specifically, problems within the ROD firmware were exposed by optical noise on the data-link inputs (often associated with failures of the TX optical transmitters and also high current from some endcap modules with CiS sensors), and by specific high-occupancy conditions.
 It is anticipated that these issues will be resolved in later ATLAS runs by upgrades to the ROD firmware.
 The impact on data-taking efficiency was mitigated by introducing the ability to remove
 a ROD which holds busy indefinitely from the ATLAS DAQ, reconfigure the affected modules, and then to re-integrate the ROD without interruption to ATLAS data-taking. 

 The DAQ may flag an error in the data stream if there is an LV fault to the module or if the chips become desynchronised from the ATLAS system due to single-event upsets (soft errors in the electronics). The rate of link errors was minimised by the online monitoring of chip errors in the data and the automatic reconfiguration of the modules with those errors. In addition, in 2011 an automatic global reconfiguration of all SCT module chips every 30 minutes was implemented, as a precaution against subtle deterioration in chip configurations as a result of single-event upsets. Figure~\ref{fig:error_rate} shows the fraction of links giving errors in physics runs during 2012 data-taking; the rate of such errors was generally very low despite the ROD data-processing issues which artificially increased the rate of such errors during periods with high trigger rates and high occupancy levels.  

 \begin{figure}[htb]
 \begin{center}
 \includegraphics[width=\columnwidth]{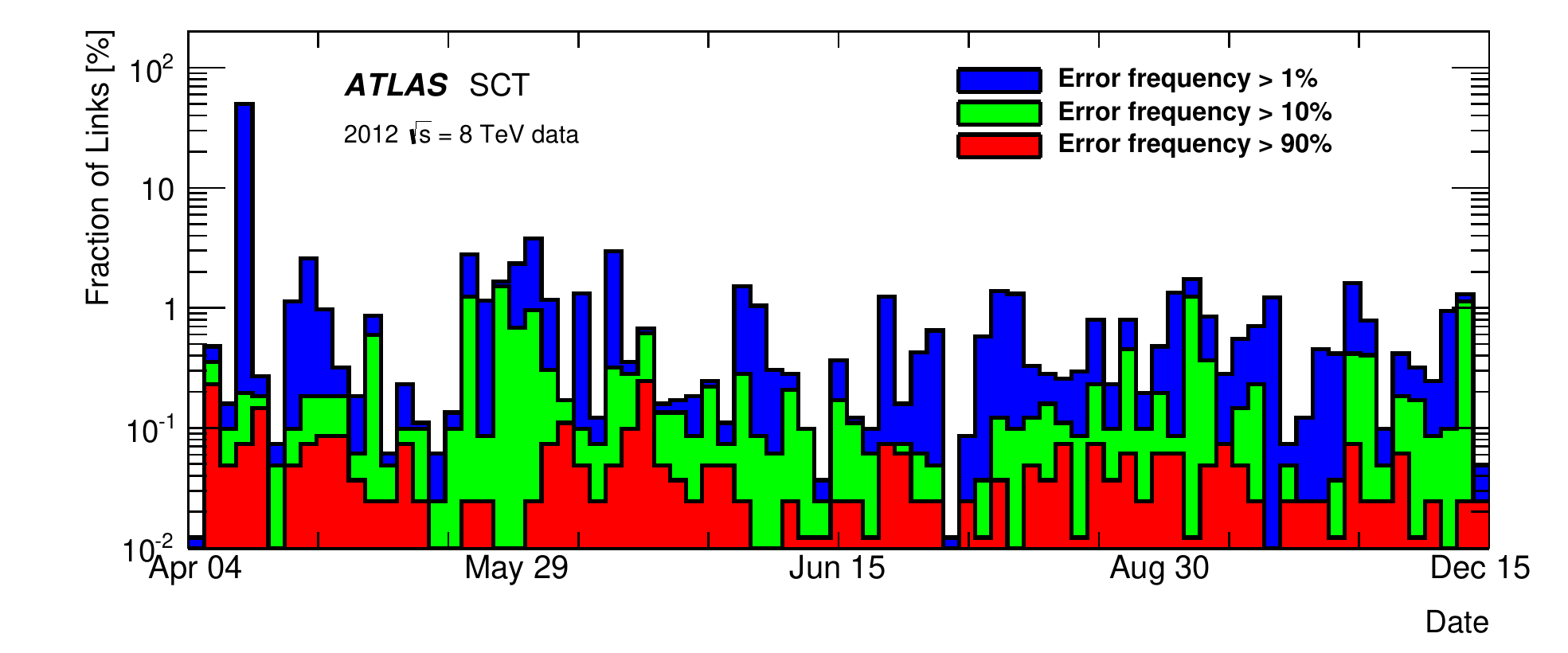}
 \caption{Fraction of data links giving errors during 2012 data-taking.}
 \label{fig:error_rate}
 \end{center}
 \end{figure}

 \subsection{Operations issues}
 \label{sec:operationissues}
 Operations from 2010 to 2013 were reasonably trouble-free with little need for expert intervention, other than performing regular calibrations between physics fills. However, a small number of issues needed vigilance and expert intervention on an occasional basis. Apart from the ROD-busy issues discussed above, the main actions were to monitor and adjust the leakage-current trip limits to counter the increase in leakage currents due to radiation or due to anomalous leakage-current behaviour (see section~\ref{Rad:sec-impact}), and the replacement of failing off-detector optical transmitters.

 The only significant component failure for the SCT has been the VCSEL arrays of the off-detector TX 
 optical transmitters (see section~\ref{sec:DAQ}). The immediate consequence of a TX channel 
 failure was that the module no longer received the clock and command signals, and therefore no longer returned data. VCSEL failures were originally attributed to inadequate electrostatic discharge (ESD) precautions during the assembly of the VCSEL arrays into the TX plug-in, and the entire operational stock of 364 of the 12-channel TXs were replaced in 2009 with units manufactured with improved ESD procedures. 
 Although this resulted in a small improvement in lifetime, further TX channel failures continued at a rate of 10--12 per week.
These failures were finally attributed to degradation arising from exposure of the VCSELs to humidity. 
In 2012, new VCSEL arrays were installed; these VCSELs contained a dielectric layer to act as a moisture barrier.
As expected these TXs  demonstrated improved robustness against humidity, compared to the older arrays, 
when measuring the optical spectra of the VCSEL light in controlled conditions~\cite{bib:weidberg}. 
During 2012, the new VCSELs nonetheless had a small but significant failure rate, suspected to be a result of mechanical stress arising from thermal mismatch between the optical epoxy on the VCSEL surface and the
GaAs of the VCSEL itself. This was despite the introduction of dry air to the ROD racks in 2012, which reduced the relative humidity levels from $\sim$50\% in 2010 to $\sim$30\%. For future ATLAS runs, commercially available VCSELs inside optical sub-assemblies, with proven robustness and reliability, and packaged appropriately on a TX plug-in for the BOC, will be installed. 

Operationally, the impact of the TX failures was minimised by the use of the TX redundancy mechanism, which could be applied during breaks between physics fills. If this was not available (for example, if the module providing the redundancy was also using the redundancy mechanism itself) then the module was disabled until the opportunity arose to replace the TX plug-in on the BOC.
As a consequence, the number of disabled modules fluctuated slightly throughout the years, but rarely increased by more than 5--10 modules.

The RX data links have been significantly more reliable than the TX links, which is attributed to the use of a different VCSEL
technology\footnote{These VCSELs used the older proton-implant technology for current channelling. This technology is more reliable but has become obsolete and has been replaced by oxide-implant VCSELs.} and the fact that the on-detector VCSELs operate at near-zero humidity. There have been nine confirmed failures of on-detector VCSEL channels since the beginning of SCT operations. The RX lifetime will be closely monitored over the coming years; replacements of failed RX VCSELs will not be possible due to the inaccessibility of the detector, so RX redundancy
will remain the only option for RX failures.

\section{Offline reconstruction and simulation}
\subsection{Track reconstruction}
\label{sec:tracking}
Tracks are reconstructed using ATLAS software within the Athena 
framework~\cite{bib:AthenaCore}. Most studies in this paper use tracks reconstructed 
from the whole inner detector, i.e.\ including pixel detector, SCT and TRT hits,
using the reconstruction algorithms described in ref.~\cite{bib:NewTrackingPubNote}.
For some specialised studies, tracks are reconstructed from SCT hits alone.
 
The raw data from each detector are first decoded and checked for error conditions. 
Groups of contiguous SCT strips with a hit are grouped into a cluster. Channels which 
are noisy, as determined from either the online calibration data or offline monitoring 
(see section~\ref{sec:PCL}), or which have other identified problems, are rejected at this 
stage. It is also possible to select or reject strips with specific hit patterns in the
three time bins, for example to study the effect of requiring an X1X hit pattern 
(see section~\ref{sec:DAQ}) on data taken in any-hit mode. 
The one-dimensional clusters from the two sides of a module are combined into 
three-dimensional space-points using knowledge of the stereo angle and the radial 
(longitudinal) positions of the barrel (endcap) modules. 

Pixel clusters are formed from groups of contiguous pixels with hits, in a
fashion similar to clusters of SCT strips. In this case, knowledge of the
position of a single pixel cluster is enough to construct a space-point. 
The three-dimensional space-points in the pixel
detector and the SCT, together with drift circles in the TRT, form the input to
the pattern recognition algorithms.

Track seeds are formed from sets of three space-points in the silicon detectors;
each space-point must originate in a different layer. The seeds are used
to define roads, and further silicon clusters within these roads are added to
form track candidates. Clusters can be attached to multiple track candidates.
An ambiguity-resolving algorithm is used to reject poor track candidates until
hits are assigned to only the most promising one. A track fit is performed to
the clusters associated to each track. Finally, the track is extended into the TRT by 
adding drift circles consistent with the track extension, and a final combined
fit is performed~\cite{bib:GlobalChi2TrackFitter}.

The reconstruction of SCT-only tracks is similar, but only SCT space-points
are used in the initial seeds, and only SCT clusters in the final track fit.

The average number of SCT clusters per track is around eight in the barrel 
region ($|\eta|$ {\raisebox{-3pt}{$\lapprox$}} 1) and around nine in the endcap regions, as shown in 
figure~\ref{fig:SCTHits_eta} for a sample of minimum-bias events recorded at
$\sqrt{s}$ = 8~\TeV. In this figure, data are compared with simulated minimum-bias 
events generated using PYTHIA8~\cite{bib:PYTHIA8} with the A2:MSTW2008LO 
tune~\cite{bib:ATLAS_PYTHIA8_TUNES}. The simulation is reweighted such that the 
vertex $z$ position and track transverse momentum distributions match those observed 
in the data. The tracks are required to have at least six hits in the SCT and at least 
two hits in the pixel detector, transverse ($d_0$) and longitudinal ($z_0$) impact 
parameters with respect to the primary vertex of $|d_{0}| <$ 1.5~mm and 
$|z_{0}\sin\theta| <$ 4~mm respectively,
and a minimum transverse momentum of 100~\MeV. 
The variation with pseudorapidity of the mean number of hits arises primarily
from the detector geometry. The offset of the mean primary-vertex position in $z$ 
with respect to the centre of the detector gives rise to the small asymmetry seen in 
the barrel region. The variation in mean number of hits with pseudorapidity is well 
reproduced by the simulation.  

\begin{figure}[htbp]
\begin{center}
\includegraphics[width=0.5\columnwidth]{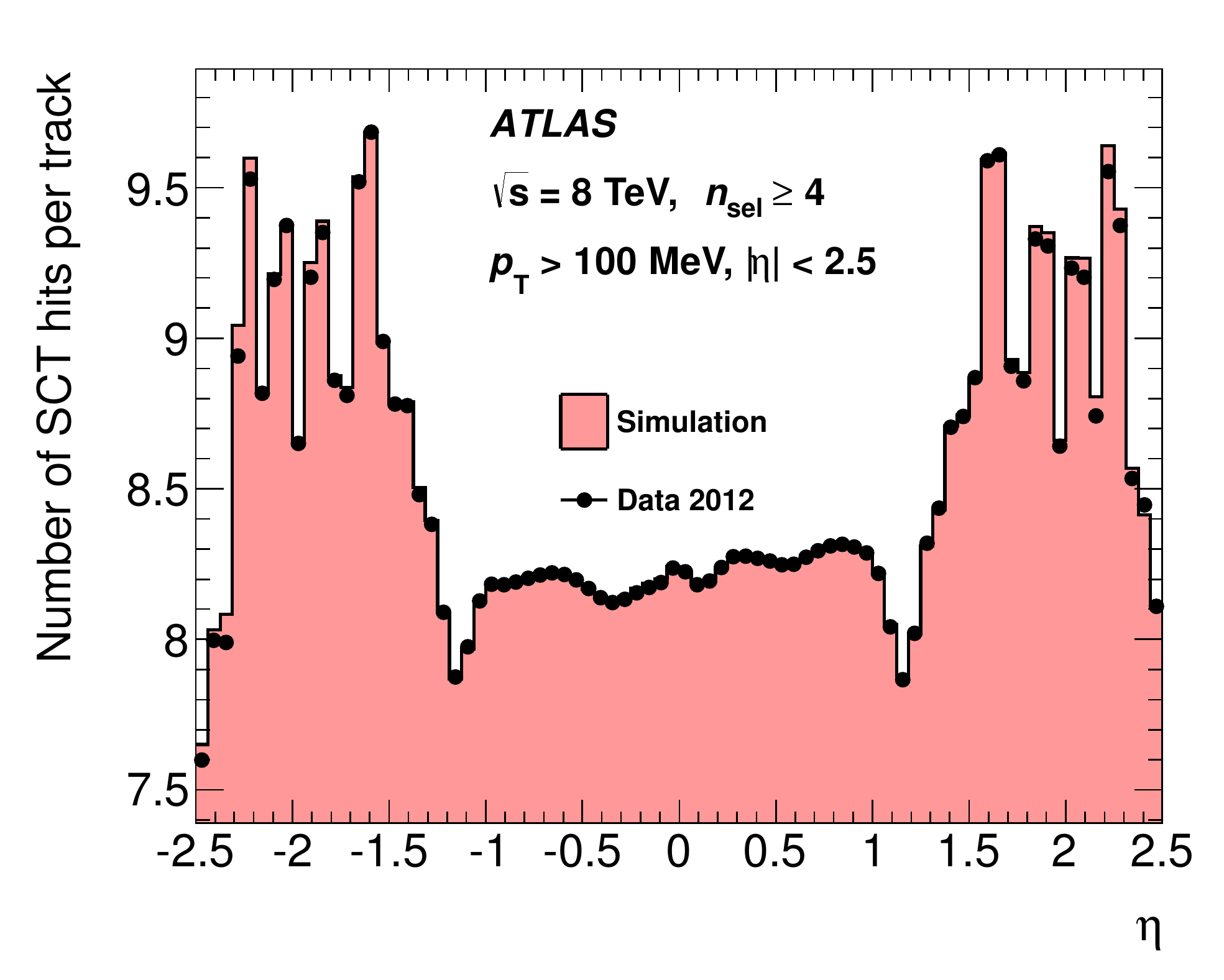}
\caption{Comparison between data (dots) and simulation (histogram) of the 
         average number of SCT clusters (hits) per track as a function of pseudorapidity, 
         $\eta$, measured in minimum-bias events at $\sqrt{s} = $8~\TeV.} 
\label{fig:SCTHits_eta}
\end{center}
\end{figure}

\subsection{Track alignment}\label{trackalign}
Good knowledge of the alignment of the inner detector
is critical in obtaining the
optimal tracking performance. The design requirement is that the resolution of 
track parameters be degraded by no more than 20\% with respect to the 
intrinsic resolution~\cite{bib:ID_TDR}, which means that the SCT modules
must be aligned with a precision of 12~$\mu$m in the direction perpendicular
to the strips. 
The principle method of determining the inner detector alignment 
uses a \chisq{} technique that minimises the residuals to fitted 
tracks from $pp$ collision events. 
Alignment is performed sequentially at different
levels of detector granularity starting with the largest structures
(barrel, endcaps) followed by alignment of individual layers and
finally the positions of individual modules are optimised. The number of 
degrees of freedom at the different levels increases from 24 at the first 
level to 23328 at the module level. This kind of alignment addresses the
`strong modes', where a misalignment would change the \chisq{}
distribution of the residuals to a track. There is also a class of
misalignment where the \chisq{} fit to a single track is not affected
but the value of measured momentum is systematically shifted. These are the
`weak-mode' misalignments which are measured using a variety of techniques,
for example using tracks
from the decay products of resonances like the \jpsi{} and the $Z$ boson. 
Details of the alignment techniques used in the ATLAS experiment can be found
elsewhere~\cite{ATLAS:2010nca,Hsu:2011zza}.

The performance of the alignment procedure for the SCT modules is validated using high-\pT\ tracks 
in $Z\ra\mu\mu$ events in $pp$ collision data at $\sqrt{s}$ = 8~\TeV\ collected in 2012.
Figure~\ref{fig:residual} shows residual distributions in $x$ (i.e.\ perpendicular to the 
strip direction) for one representative barrel and one endcap 
disk, compared with the corresponding distributions from 
simulations with an ideal geometry. The good agreement between the data and
simulation in the measured widths indicates that the single plane position resolution of 
the SCT after alignment is very close to the design value of 17~$\mu$m in $r$--$\phi$.
Similar good agreement is seen in all other layers.
From a study of the residual distances of the second-nearest cluster to a track in each
traversed module, the two-particle resolution is estimated to be $\sim$120~$\mu$m.

\begin{figure}[ht!]
\centering
\begin{tabular}{cc}
\includegraphics[width=0.49\textwidth]{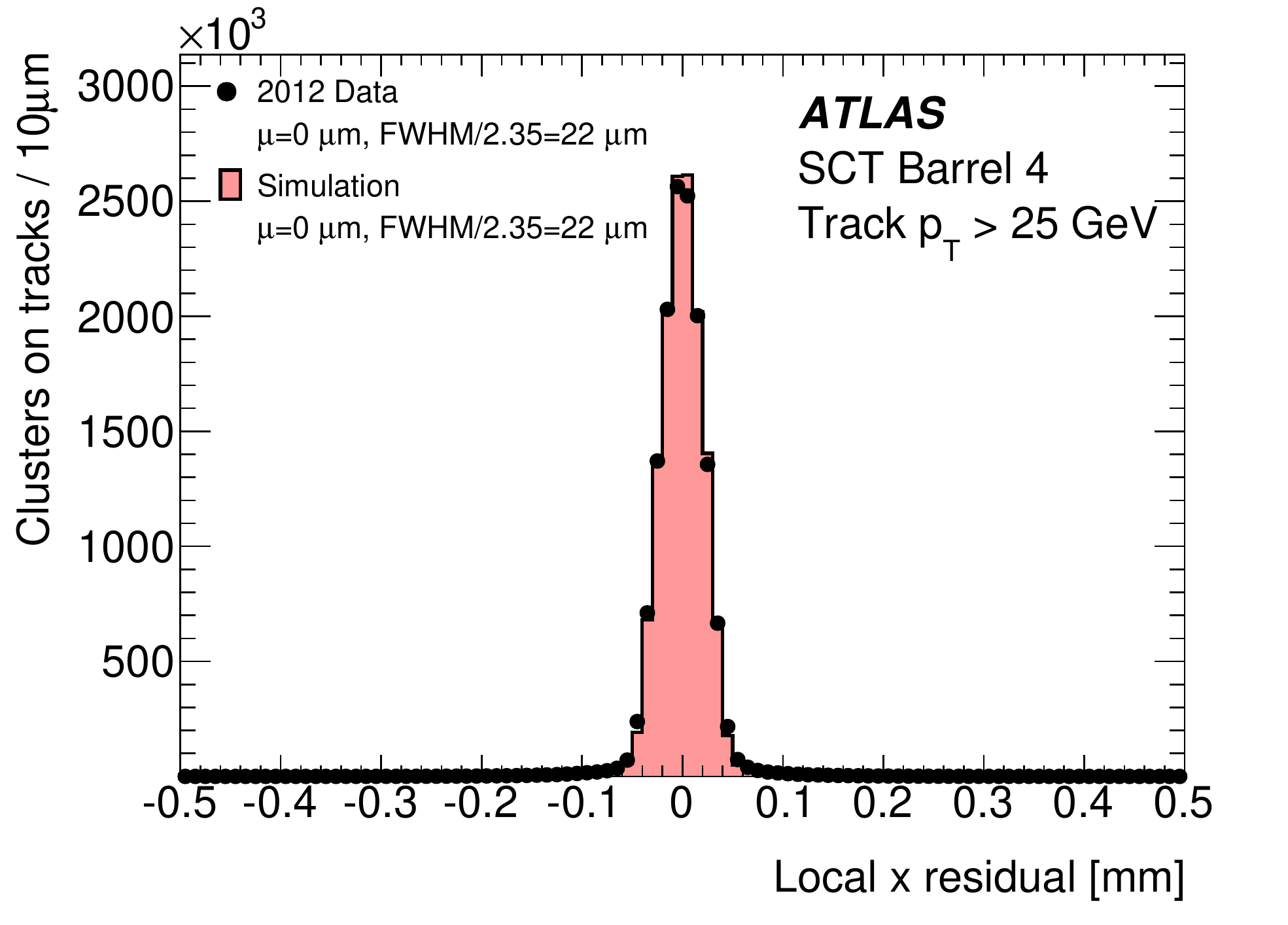} &
\includegraphics[width=0.49\textwidth]{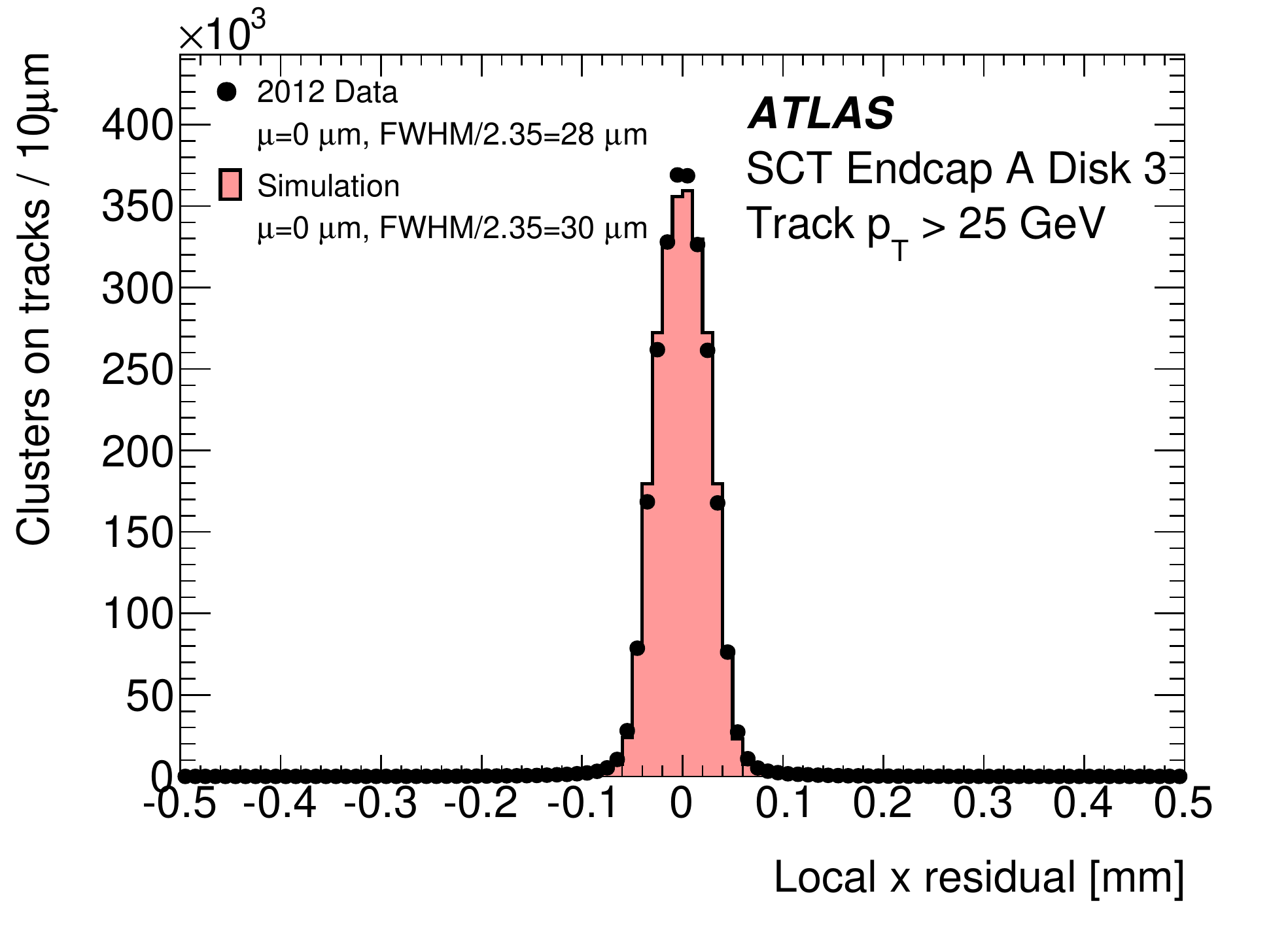} \\
(a) & (b) \\
\end{tabular}
\caption{Distributions of residuals measured in $Z\ra\mu\mu$
         events at $\sqrt{s}$ = 8~\TeV\ for data (points) 
         compared with simulation (histogram), for (a) barrel 4 and 
         (b) disk 3 of endcap A.}
\label{fig:residual}
\end{figure}

\subsection{Simulation}
Most physics measurements performed by the ATLAS collaboration rely on Monte Carlo simulations,
for example to calculate acceptances or efficiencies, to optimise cuts in
searches for new physics phenomena or to understand the performance of the detector.
Accurate simulation of the detector is therefore of great importance. Simulation
of the SCT is carried out using GEANT4~\cite{bib:geant4a} within the ATLAS simulation
framework~\cite{bib:ATLASMCPaper}. A detailed model of the SCT detector geometry is 
incorporated in the simulation program. The silicon wafers are modelled with a
uniform thickness of 285~$\mu$m and planar geometry; the distortions measured 
in the real modules~\cite{bib:SCTBarrelModules} are small and therefore not simulated. 
Modules can be
displaced from their nominal positions to reflect those measured in the track 
alignment procedure in data. All services, support structures and other inert
material are included in the simulation. The total mass of the SCT in simulation
matches the best estimate of that of the real detector, which is known to a 
precision of better than 5\%.

Propagation of charged particles through the detector is performed by GEANT4. An 
important parameter is the range cut, above which secondary particles are 
produced and tracked. In the silicon wafers this cut is set to 50~$\mu$m,
which corresponds to kinetic energies of 80~\keV, 79~\keV\ and 1.7~\keV\ for
electrons, positrons and photons, respectively. During the tracking, the
position and energy deposition of each charged particle traversing a silicon
wafer are stored. These energy deposits are converted into strip hits in
a process known as digitisation, described below. The simulated data are 
reconstructed using the same software as used for real data. Inoperative
modules, chips and strips are simulated to match average data conditions.

\subsubsection{Digitisation model}
\label{sec:default_digi}
The digitisation model starts by converting the energy deposited in each
charged-particle tracking step in the silicon into a charge on the readout
electrodes. Each tracking step, which may be the full width of the wafer, is
split into 5~$\mu$m sub-steps, and the energy deposited shared uniformly among
these sub-steps. The energy is converted to charge using the mean electron--hole
pair-creation energy of 3.63~\eV/pair, and the hole charge is drifted to the 
wafer readout surface in a single step taking into account the Lorentz angle, $\theta_{\rm L}$,
and diffusion. The Lorentz angle is the angle between the drift
direction and the normal to the plane of the sensor which arises when the charge 
carriers are subject to the magnetic field from the solenoid as well as the electric 
field generated by the bias voltage. 
A single value of the Lorentz angle, calculated
as described in section~\ref{sec:LA} assuming a uniform value of the electric
field over the entire depth of the wafer, is used irrespective of the original 
position of the hole cloud. The drift time is calculated as the sum of two components:
one corresponding to drift perpendicular to the detector surface, calculated 
assuming an electric field distribution as in a uniform flat diode, and a second 
corresponding to drift along the
surface, introduced to address deficiencies in the simple model and give better
agreement with test-beam data. 

The second step in the digitisation process is the simulation of the electronics
response. 
For each charge arriving at a readout strip, the amplifier response at three times, 
corresponding to the three detector readout bins, is calculated. The cross-talk 
signal excited in each neighbouring strip, which is a differential form of the
main strip pulse, is also calculated. Electronic noise is added to the strips
with charge, generated from a Gaussian distribution with mean zero and standard 
deviation equal to the equivalent noise charge taken from data. The
noise is generated independently for each time bin. Finally, strips with a signal
above the readout threshold, normally 1~fC, are recorded. Further random strips
from among those without any charge deposited are added to the list of those
read out, to reproduce the noise occupancy observed in data.     

\subsubsection{Induced-charge model}

In order to understand better the performance of the detector, and to check the
predictions of the simple digitisation model, a full, but time-consuming, calculation was used. 
In this `induced-charge model' the drift of both electrons and holes in the silicon is
traced step-by-step from the point of production to the strips or HV plane, and the
charge induced on the strips from this motion calculated using a weighting potential
according to Ramo's theorem~\cite{bib:Ramo,bib:Radeka}.

The electron and hole trajectories are split into steps with corresponding time duration
$\delta t$ = 0.1~ns or 0.25~ns. The magnitude of the drift velocity $v_{\rm d}$ at each step 
is calculated as:
\begin{eqnarray}
v_{\rm d} = \mu_{\rm d} E
\end{eqnarray}  
where $\mu_{\rm d}$ is the mobility and $E$ is the magnitude of the electric field strength. 
The effect of diffusion is included by choosing the actual step length, independently in each
of two perpendicular distributions, from a Gaussian distribution of width $\sigma$ given by: 
\begin{eqnarray}
\sigma = \sqrt{2 D \delta t}; \hspace{1cm}  {\rm with} \hspace{0.5cm} D = k_{\rm B} T \mu_{\rm d}/e
\end{eqnarray}  
where the diffusion coefficient $D$ depends on the temperature $T$; 
$k_{\rm B}$ is Boltzmann's constant and $-e$ the electron charge. 
The electric field at each point is calculated using a two-dimensional finite-element 
model (FEM); the dielectric constant (11.6 $\epsilon_{0}$) and the donor concentration in 
the depleted region are taken into account in the FEM calculation. 
In the presence of a magnetic field, the direction of drift is rotated by the local 
Lorentz angle.

The induced charge on a strip is calculated from the motion of the holes and electrons
using a weighting potential obtained using the same two-dimensional FEM by setting the 
potential of one strip to 1~V with all the other strips and the HV plane at ground.
In the calculation the space charge is set to zero since its presence does not affect the
validity of Ramo's theorem~\cite{bib:Cavalleri}. 
The simulation of the electronics response follows the same procedure as in the default 
digitisation model above.

The induced-charge model predicts an earlier peaking time for the charge collected on a strip
than the default model: up to 10~ns for charge deposited midway between strips. However, after
simulation of the amplifier response and adjusting the timing to maximise the fraction of clusters 
with a 01X time-bin pattern, the output pulse shapes from the two models are similar.     
The mean cluster widths predicted by the induced-charge model are slightly larger than those
predicted by the default digitisation model. The difference is about 0.02 strips (1.6~$\mu$m) for 
tracks with incident angles near to the Lorentz angle. The position of the minimum cluster
width occurs at incident angles about 0.1$^\circ$ larger in the induced-charge model.   
Since the differences between the two models
are small, the default digitisation model described in section~\ref{sec:default_digi}
is used in the ATLAS simulation.  

\subsection{Conditions data}
The offline reconstruction makes use of data describing the detector conditions
in several ways. First, bad channels are rejected from cluster formation.
In the pattern recognition stage, knowledge of dead detector elements is used in
resolving ambiguities. Measured values of module bias voltage and temperature are used to
calculate the Lorentz angle in the track reconstruction. In addition, measured chip 
gains and noise are used in the simulation. The conditions data, stored in the ATLAS 
COOL database~\cite{bib:COOL}, arise from several sources:
\begin{itemize}
\item Detector configuration. These data include which links are in the readout, 
use of the redundancy mechanism, etc., and are updated when the detector configuration is
changed. They cannot be updated during a run.  
\item DCS. The data from the detector control system (see section~\ref{sec:DCS}) 
most pertinent to reconstruction are module bias voltage values and temperatures. 
Data from modules which are not at their nominal bias voltage value are excluded 
from reconstruction at the cluster-formation stage. This condition removes modules 
which suffer an occasional high-voltage trip, resulting in high noise occupancy, for 
the duration of that trip.
\item Online calibration. In reconstruction, use of online calibration data is 
limited to removing individual strips with problems. These are mostly ones which were found to be noisy in the latest preceding calibration runs.
\item Offline monitoring. Noisy strips are identified offline run-by-run, as
described in section~\ref{sec:PCL}. These are also removed during cluster
formation.
\end{itemize}

\section{Monitoring and data quality assessment}
Continuous monitoring of the SCT data is essential to ensure good-quality data for
physics analysis. Data quality monitoring is performed both online and offline. 

\subsection{Online monitoring}
The online monitoring provides immediate feedback on the condition of the SCT,
allowing quick diagnosis of issues that require intervention during a run. These may 
include the recovery of a module or ROD that is not returning data, or occasionally
a more serious problem which requires the early termination of a run and restart.

The fastest feedback is provided by monitoring the raw hit data. Although
limited in scope, this monitoring allows high statistics, minimal trigger bias and 
fast detector feedback. The number of readout errors, strip hits and simple 
space-points (identified as a coincidence between hits on the two sides of a module 
within $\pm$128 strips of each other) are monitored as a function of time and features of a run 
can be studied. During collisions the hit rate increases by several orders of magnitude. 
This monitoring is particularly useful for providing speedy feedback on the condition of the beam, 
and is used extensively during LHC commissioning and the warm-start procedure. 

In addition to raw-data monitoring, a fraction of events is fully reconstructed 
online, and monitoring histograms are produced as described below for the offline 
case. Automatic checks are performed on the 
histograms~\cite{1742-6596-119-2-022033, CuencaAlmenar:2011zz} and 
warnings and alarms issued to the shift crew (`shifters').  

\subsection{Offline monitoring}
Offline monitoring allows the data quality to be checked in more detail, and an assessment
of the suitability of the data for use in physics analyses to be made. A subsample of events is 
reconstructed promptly at the ATLAS Tier0 computer farm. Monitoring plots are produced 
as part of this reconstruction, and used to assess data quality. These plots typically
integrate over a whole run, but may cover shorter periods in specific cases. 
Histograms used to assess the performance of the SCT include:

\begin{itemize}
\item Modules excluded from the readout configuration (see section~\ref{sec:DetConf}).
\item Modules giving readout errors. In some cases the error rates are calculated per 
luminosity block so that problematic periods can be determined.
\item Hit efficiencies (see section~\ref{sec:eff}). The average hit efficiency for 
each barrel layer and endcap disk is monitored, as well as the efficiencies of 
the individual modules.
Localised inefficiencies can be indicative of problematic modules, whereas 
global inefficiencies may be due to timing problems or poor alignment constants.
\item Noise occupancy calculated using two different methods. The first counts hits not 
included in space-points, and is only suitable for very low-multiplicity data. The second
uses the ratio of the number of modules with at least one hit on one side only to the number
with no hit on either side.
\item Time-bin hit patterns for hits on a track (section~\ref{sec:DAQ}), which gives 
an indication of 
how well the detectors are timed in to the bunch crossings.
\item Tracking performance distributions. These include the number of SCT hits per
track, transverse momentum, $\eta$, $\phi$ and impact parameter distributions; track
fit residual and pull distributions.
\end{itemize}
Automatic checks are performed on these histograms using the offline data quality
monitoring framework~\cite{Adelman:2010zza}, and they are also reviewed by a shifter.
They form the basis of the data quality assessment discussed in the next section.

\subsection{Data quality assessment}
The data quality is assessed for every run collected, and the results are stored in
a database~\cite{Golling:2011zy} by setting one or more `defect flags' if a problem is found. 
Each defect flag corresponds to a particular problem, and may be set for a whole run or only a
short period. Several defect flags may be set for the same period. 
These flags are used later to define good data for physics analysis.  
The SCT defect flags also provide operational feedback as to the current performance of the detector.

The fraction of the data recorded which is affected by each SCT issue is given 
in the upper two parts of table~\ref{tab:SctDefects}. The uppermost part of the table shows 
intolerable defects,
which lead to the affected runs or luminosity-block ranges being excluded from physics 
analyses. The most common problem in this category is the exclusion of two or more RODs 
(96 modules, 2.3\% of the detector) or a whole crate (12\% of the detector) from the readout, 
which results in a loss in tracking coverage in a region of the detector. 
The exclusion of RODs or crates affects a limited but significant region of the detector. 
Other intolerable defects affect the whole detector. The global reconfiguration of all
readout chips occasionally causes data errors for a short period; the SCT can become
desynchronised from the rest of ATLAS; occasionally data are recorded with the modules
at standby voltage (50 V).

For more minor issues, or when the excluded modules have less of an impact on tracking coverage, 
a tolerable defect is set. These defects are shown in the central part of table~\ref{tab:SctDefects}.
One excluded ROD is included in this category. A single excluded ROD in the endcap (the most
usual case) has little impact on tracking because it generally serves modules on one disk only.
A ROD in the barrel serves modules in all layers in an azimuthal sector, and thus may affect
tracking performance; this is assessed as part of the global tracking performance checks.
Other tolerable defects include more than 40 modules excluded from the readout or giving
readout errors, which indicate detector problems but do not significantly affect tracking
performance. Although noise occupancy is monitored, no defect is set for excessive noise
occupancy.

In addition to SCT-specific checks, the global tracking performance of the inner
detector is also evaluated, and may indicate problems which arise from the SCT but which
are not flagged with SCT defects, or which are flagged as tolerable by the SCT but are intolerable 
when the whole inner detector is considered. This corresponds to situations where 
the pixel detector or TRT have a simultaneous minor problem in a similar angular region. 
The fraction of data affected by issues determined from
global tracking performance is given in the lower part of table~\ref{tab:SctDefects}. There
is significant overlap between the SCT defects and the tracking performance defects.  

The total fraction of data flagged as bad for physics in 2011 (2012) by SCT-specific checks
was 0.44\% (0.89\%). A further 0.33\% (0.56\%) of data in 2011 (2012) was flagged as bad for 
physics by tracking performance checks but not by the SCT alone. The higher fraction in 2012 
mainly resulted from RODs disabled due to high occupancy and trigger rates, as discussed
in section~\ref{sec:data-taking_eff}. 

\begin{table}[htdp]
\caption{\label{tab:SctDefects}The data quality defects recorded for 2011 and 2012, and percentage of the 
         luminosity affected. The data assigned intolerable defects (marked `Y' in the table) are not 
         included in physics analyses. Empty entries indicate that the defect was not defined for that 
         year. Multiple defect flags may be set for the same data.}
\begin{center}
\begin{tabular}{lccc}
\hline
\hline
                     &  Intolerable & \multicolumn{2}{c }{Luminosity}      \\
Defect Description   &  Defect        & \multicolumn{2}{c }{Affected [\%]} \\
\cline{3-4}
                     &                    & 2011   &    2012 \\
\hline
\multicolumn{4}{ l }{SCT intolerable defects} \\ \hline
Crate(s) excluded from readout & Y & 0.23 & 0.07 \\
Two or more RODs excluded from readout  & Y & 0.15 & 0.45 \\
SCT global desynchronisation & Y & 0.04 & 0.00\\
SCT bias voltage at standby (50\,V) & Y & 0.02 & 0.07 \\
Readout errors after global reconfiguration & Y & -- & 0.30 \\
\hline
\multicolumn{4}{ l }{SCT tolerable defects} \\ \hline
Less than 99\% average efficiency in one of barrel or endcap regions & N & 8.75 & 8.33 \\
SCT not at standard bias voltage & N & 1.64 & 0.05\\
More than 40 modules with errors in short period of time & N & 1.13 & 0.43 \\
Exactly one ROD excluded from readout & N & 0.93 & 2.90 \\
More than 40 modules with errors & N & 0.74 & 1.88 \\
Non-standard timing, e.g.\ timing scans& N & 0.06 & 0.21 \\
More than 40 modules excluded & N & -- & 0.82 \\
Too few events for DQ judgement & N & 0.00 & 0.05\\
\hline
\multicolumn{4}{ l }{Tracking performance defects} \\ \hline
Global tracking performance is impacted by SCT issue & N & 1.09 & 2.09\\
Significant loss of tracking coverage & Y & 0.71 & 0.94\\
SCT impacting $b$-tagging performance& N & -- & 0.35\\
SCT seriously impacting $b$-tagging performance& Y & -- & 0.10\\
\hline
\hline
\end{tabular}
\end{center}
\end{table}

\subsection{Prompt calibration loop}
\label{sec:PCL}
Reconstruction of ATLAS data proceeds in two stages. First a fraction of the data from
each run is reconstructed immediately, to allow detailed data quality and detector performance checks. 
This is followed by 
bulk reconstruction some 24--48 hours after the end of the run. 
This delay allows time for updated detector calibrations to be obtained. For the SCT no offline
calibrations are performed during this prompt calibration loop, but the period is used to obtain conditions data. In particular, strips
that have become noisy since the last online calibration period are found and excluded 
from the subsequent bulk reconstruction. Other conditions data, such as dead strips or 
chips, are obtained for monitoring the SCT performance.

The search for noisy strips uses a special data stream triggered on empty bunch crossings, 
so that no collision hits should be present. Events are written to this stream at a rate of 2--10~Hz, 
giving sufficient data to determine noisy strips run by run for all but the shortest runs.
Processing is performed automatically on the CERN
Tier0 computers and the results are uploaded to the conditions database after a check by the 
shifter.

A noisy strip is defined as one with an average occupancy (during the period with HV on)
of more than 1.5\%. Excluding those already identified as
noisy in the online calibration runs (0.06\% of strips in 2012--2013), the number of noisy 
strips found has increased from $\sim$100 during 2010 to a few thousand during 2012, as shown in 
figure~\ref{fig:noisy_strips}. Significant fluctuations are observed from run to run. These
arise from two sources: single-event upsets causing whole chips to become noisy and the
abnormal leakage-current increases observed in some endcap modules (see 
section~\ref{sec:radiation}). 
Both of these effects depend on the instantaneous luminosity, and thus vary from run to run.
The maximum number of noisy strips, observed in a few runs, corresponds to $<0.25\%$ of the 
total number of strips in the detector. 

\begin{figure}[htbp]
  \centering
   \includegraphics[width=1.0\textwidth]{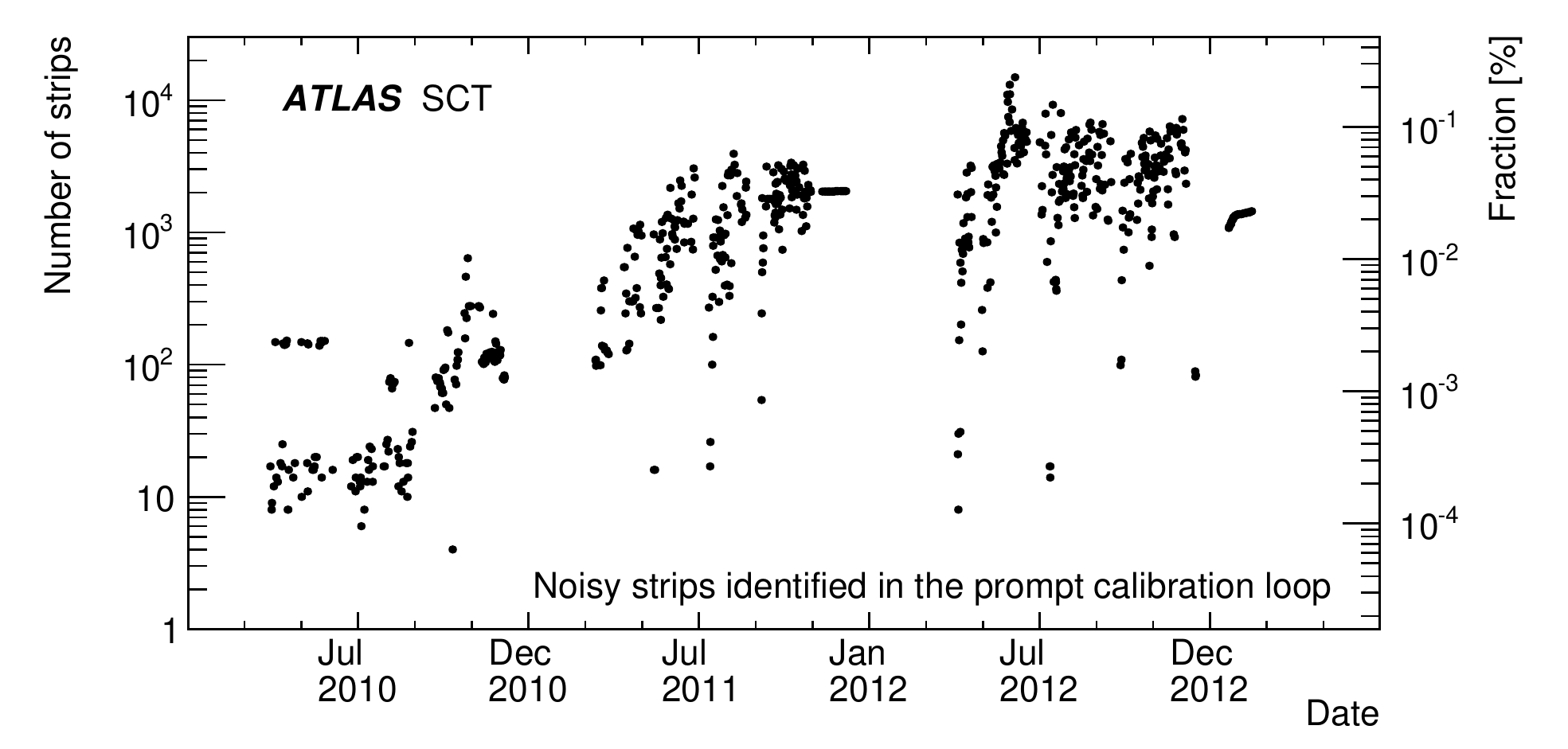}
  \caption{Number of noisy strips (also shown as the fraction of all strips on the right-hand axis) found 
           in the prompt calibration loop for physics runs with
           greater than one hour of stable beams during 2010--2013. Noisy strips identified in 
           online calibration data are excluded from the total.} 
  \label{fig:noisy_strips}
\end{figure}

\section{Performance}
\subsection{Detector occupancy}
\label{sec:occupancy}

The design of the SCT was optimised to have low detector occupancy to reduce 
confusion in pattern recognition that arises in high-multiplicity final states 
resulting from multiple proton--proton interactions. For the initial design luminosity of 
\highL\ at a beam-crossing rate of 40~MHz the mean number of interactions per crossing was 
expected to be 23. In these circumstances, the mean strip occupancy was
expected to be less than 1\%. In 2012 with an LHC bunch-spacing of 50~ns the design goals 
for pile-up were exceeded with no significant loss of tracking efficiency.

In figure~\ref{fig:barrel_occupancy} the occupancy, defined as the number of strips above threshold
divided by the total number of strips, is shown for the four barrels and for the different module 
types in one representative endcap disk (averaging over both endcaps). The SCT was read out in
level mode, i.e.\ demanding a hit in the in-time bunch crossing, and noisy strips identified in
online calibration runs or the prompt calibration loop were removed. 
The data were recorded using a minimum-bias trigger in special high pile-up runs, where average numbers 
of interactions per bunch crossing greater than the 35 routinely seen in physics runs could be achieved,
and events were required to have at least one reconstructed primary vertex.
Values for less than 40 interactions per bunch crossing are from collisions at $\sqrt{s}$ = 7~\TeV, 
those for higher values from collisions at $\sqrt{s}$ = 8~\TeV. The occupancy at 8~\TeV\ is expected 
to be a factor of about 1.03 larger than at 7~\TeV\ for the same number of interactions per bunch
crossing. Allowing for this, good linearity is observed up to 70 interactions per 
crossing, where the average occupancy is less than 2\% in the innermost barrel and $\sim$1.5\% in the 
middle modules of the central disks (the middle modules have the largest pseudorapidity coverage and
hence largest occupancy in the endcaps).  

\begin{figure}[htbp]
  \centering
  \begin{tabular}{cc}
  \includegraphics[width=0.49\columnwidth]{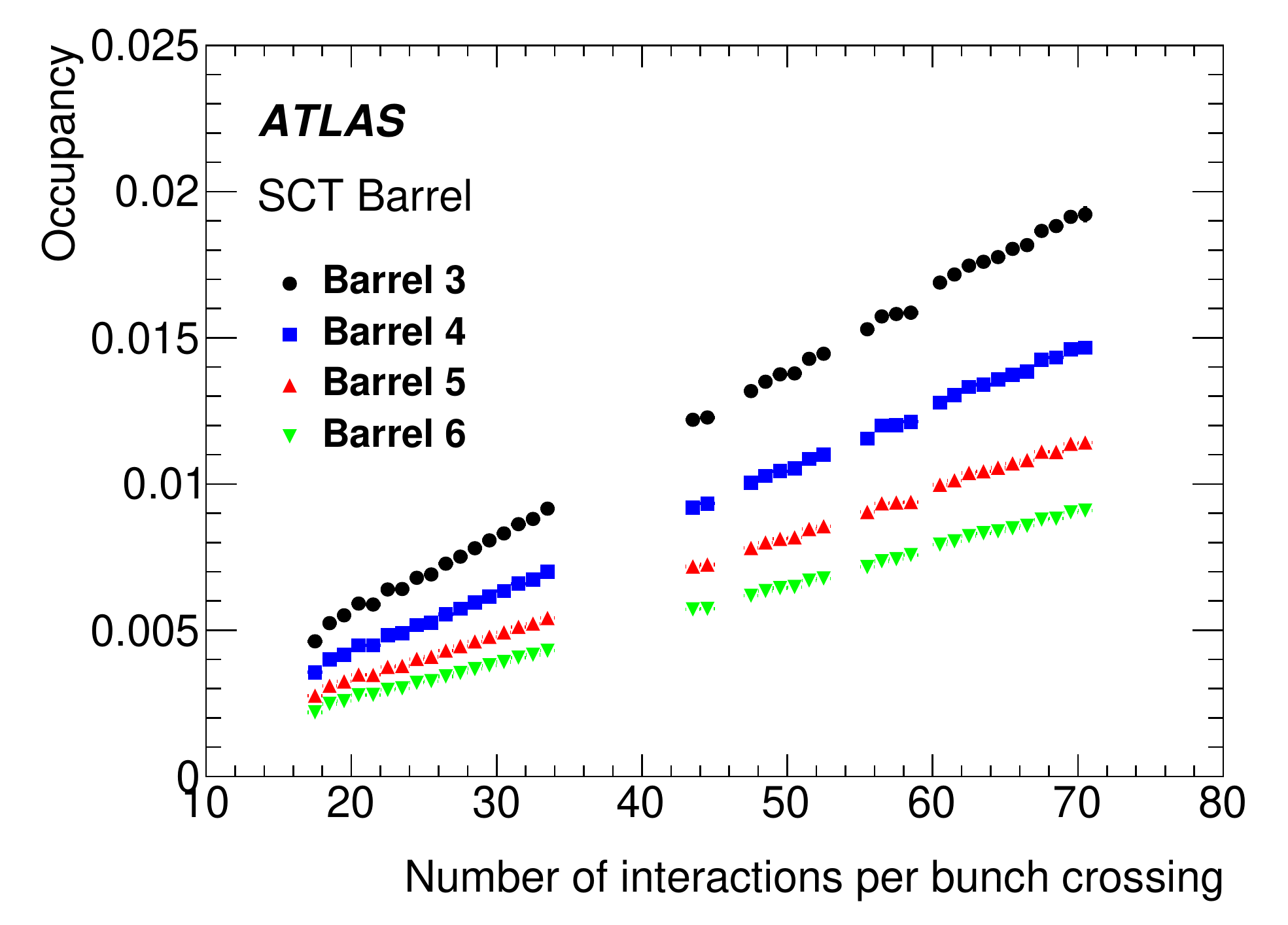} &
  \includegraphics[width=0.49\columnwidth]{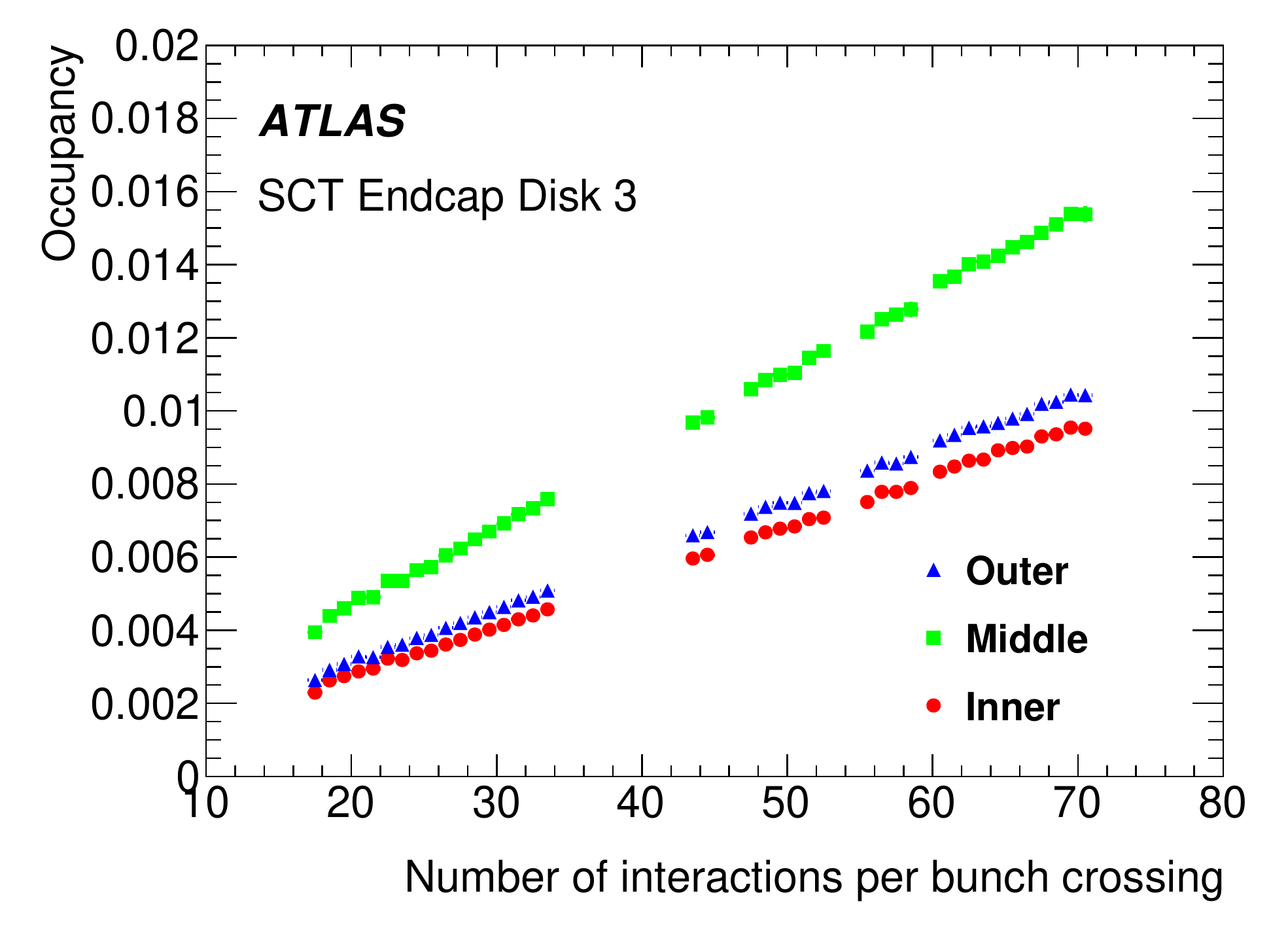} \\
  (a) & (b) \\
  \end{tabular}
  \caption{Mean occupancy of (a) each barrel and (b) inner, middle and outer modules of
           endcap disk 3 as a function of the number of interactions per bunch crossing in minimum-bias
           $pp$ data. Values for less than 40 interactions per bunch crossing are from collisions at
           $\sqrt{s}$ = 7~\TeV, those for higher values from collisions at $\sqrt{s}$ = 8~\TeV, which 
           have an occupancy larger by a factor of about 1.03 for the same number of interactions
           per bunch crossing.}
  \label{fig:barrel_occupancy}
\end{figure}

The readout links were designed to accommodate up to 2\% occupancy at a level-1 trigger rate of
100~kHz without imposing dead-time.
If the trigger rate and/or occupancy is significantly increased beyond these limits, the first 
two limiting bottlenecks within the SCT DAQ are the bandwidth of the data links between the modules and the 
BOCs, and the bandwidth of the data fibres (S-links) that connect the RODs and BOCs to the ATLAS DAQ chain. 
The data volumes in these links were studied in $pp$ collision data at $\sqrt{s}$ = 8~\TeV\ to
identify any potential problems. The studies used the data stream that is reconstructed immediately
for data quality assessment. The detector occupancy in this stream, which comprises a mixture 
of triggers, is a few percent higher than in minimum-bias events. The measured data size for each of the 
optical links was used to calculate the maximum sustainable rate of level-1 triggers as a 
function of the mean number of inelastic $pp$ interactions per bunch crossing, $\mu$. The maximum 
sustainable rate assumes 90\% of the data-link bandwidth is used. The results for each of the 8176 links from
front-end chip to ROD, excluding the small number reading out both module sides, are shown in 
figure~\ref{fig:MaxL1A_8}(a).
Figure~\ref{fig:MaxL1A_8}(b) shows a similar distribution for the transfer of data on
the 90 S-links.  
It can be seen from these plots that the slowest data links are still adequate up to a
pile-up, $\mu$, of 120, while the S-links impose a limit of $\mu < 62$ at the typical 2012 
trigger rate of 70~kHz.

\begin{figure}[htbp]
  \centering
  \begin{tabular}{cc}
  \includegraphics[width=0.49\columnwidth]{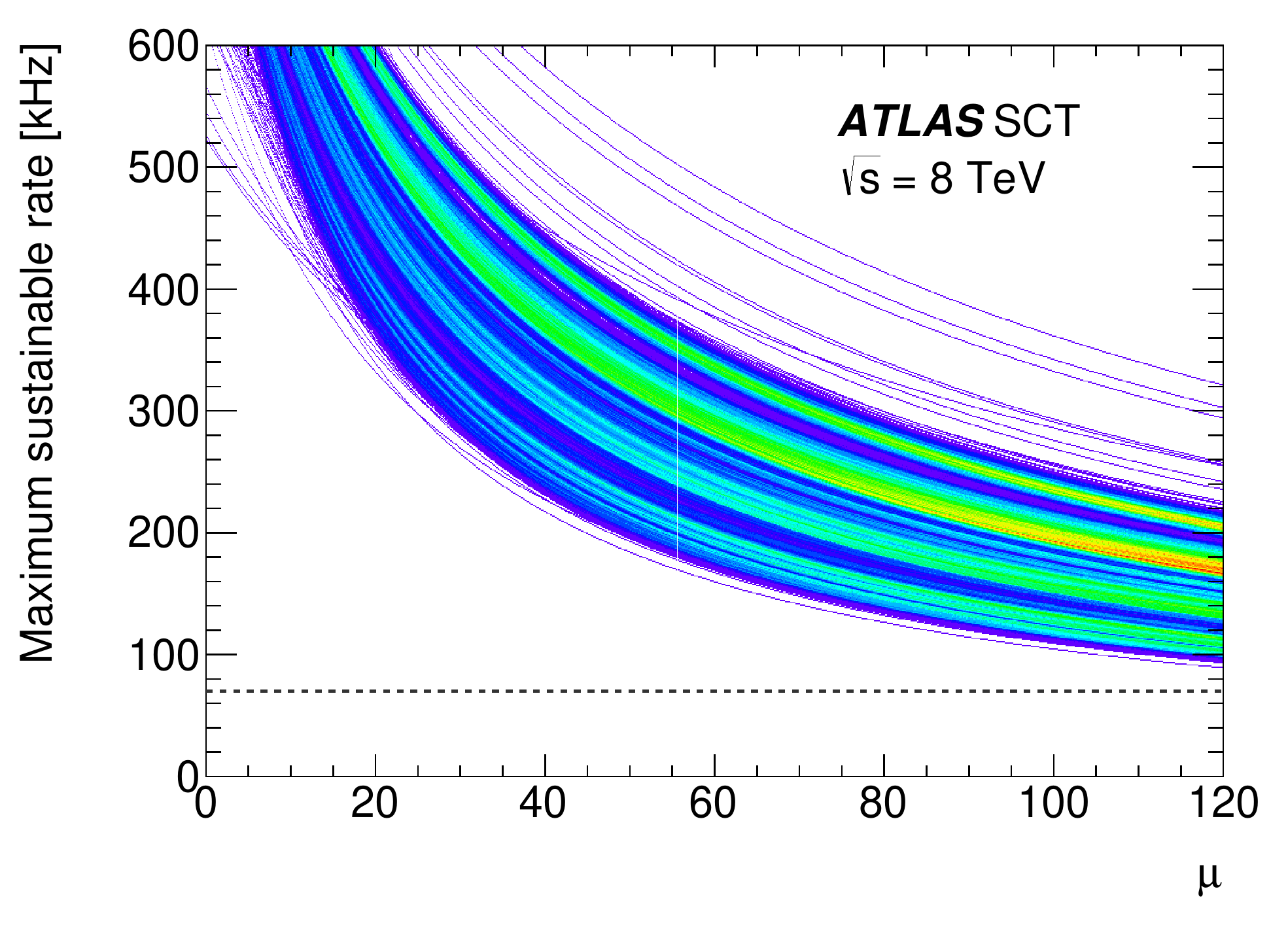} &
  \includegraphics[width=0.49\columnwidth]{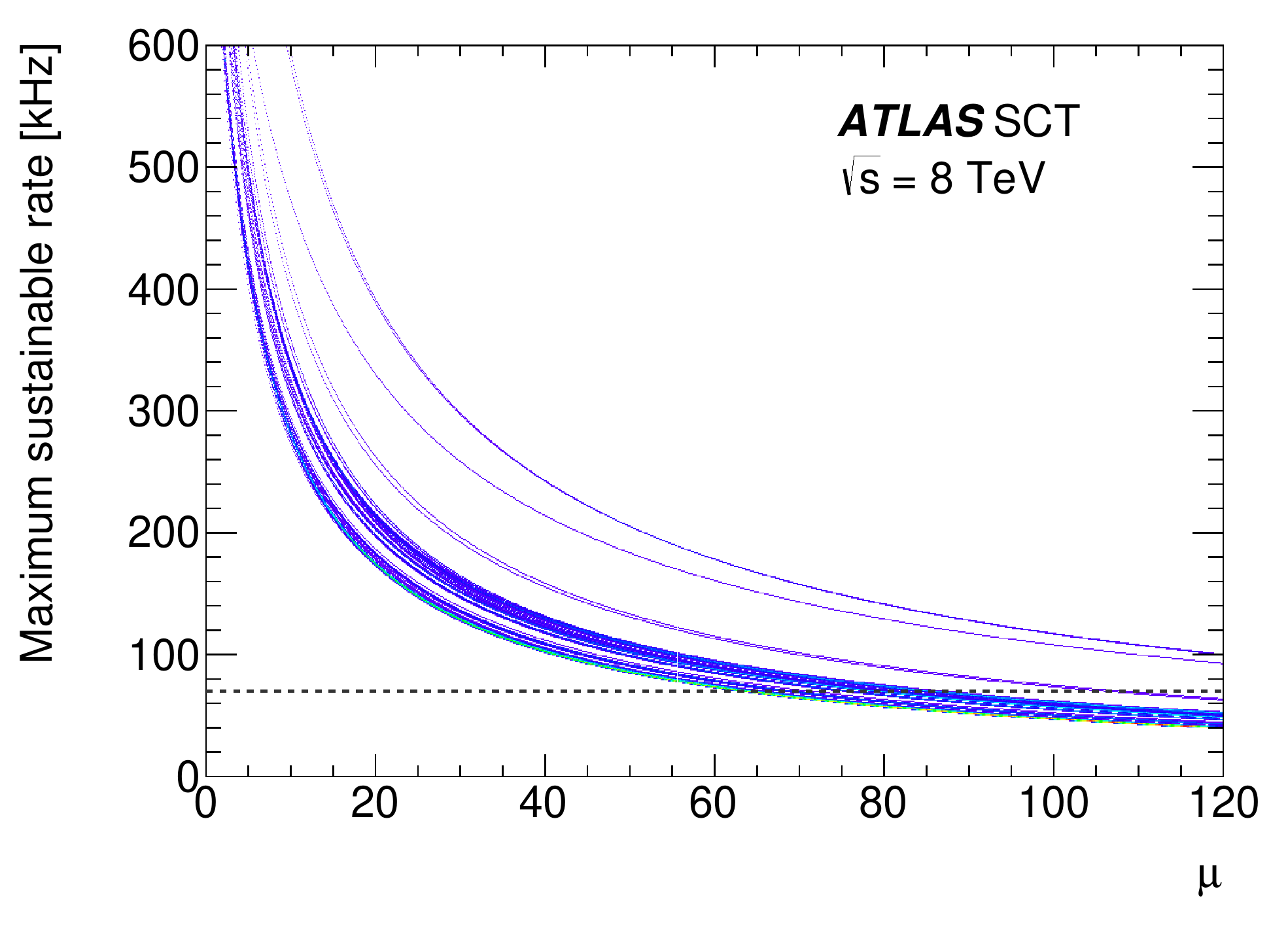} \\
  (a) & (b) \\
  \end{tabular}
  \caption{Maximum sustainable level-1 trigger rate for the transfer of data from (a) front-end
           chip to ROD and (b) ROD to ATLAS DAQ chain, as a function of the mean number of interactions 
           per bunch crossing, $\mu$. Values are calculated from measured event sizes in $pp$ interactions 
           at $\sqrt{s}$ = 8~\TeV\ assuming 90\% of the available data-link bandwidth is used. 
           Each link is shown as a separate curve. Increasing numbers of superimposed curves are
           indicated by changes of colour from blue through green and yellow to red. The dashed lines
           indicate the typical 2012 trigger rate of 70~kHz.}
  \label{fig:MaxL1A_8}
\end{figure}

Significant increases in trigger rate and pile-up are anticipated with the increase in instantaneous 
luminosities foreseen at the LHC beyond 2015 at $\sqrt{s}$ = 14~\TeV. The increase in multiplicity and
trigger rate will impose a reduced limit on $\mu$ of $\sim$87 and $\sim$33 at 100~kHz for the data links 
and S-links respectively, if there are no changes made to the existing DAQ. The limit from the data links 
arises from components mounted on the detector, and therefore cannot be improved.
The limit from the S-links will be improved by increasing the number of RODs, BOCs and S-links from 90 to 
128; the load per S-link will therefore be reduced and balanced in a more equal way. In addition, 
edge-mode readout (01X, see section~\ref{sec:DAQ})) will be deployed to minimise the effect 
of hits from one bunch crossing appearing in another, and an improved data-compression algorithm will be 
deployed in the ROD.
As a result of these changes, the slowest S-link will be able to sustain a level-1 trigger rate of 100~kHz 
at $\mu$ = 87, which matches the hard limit from the front-end data links. This corresponds to a
luminosity of 3.2$\times 10^{34}$~cm$^{-2}$s$^{-1}$ with 25~ns bunch spacing and 2808 bunches per beam.

\subsection{Noise}
\label{sec:noise}

One of the crucial conditions for maintaining high tracking 
efficiency is to keep the read-out thresholds as low as possible.
This is only possible when the input noise level is kept low. 
There are two ways of measuring the input equivalent noise charge (ENC,
in units of number of electrons) in dedicated calibration runs:

\begin{enumerate}

\item {\bf Response-curve noise:} 
In three-point gain-calibration runs, a threshold scan is performed with a known charge
injected into each channel, as described in section~\protect\ref{sec:calibration}.
The Gaussian spread of the output noise value at the injection charge of 2~fC (chosen because 
this value is well clear of non-linearities which can appear below 1~fC) is obtained by fitting 
the threshold curve. The ENC is calculated by dividing the output noise 
by the calibration gain of the front-end amplifier. 

\item {\bf Noise Occupancy (NO):} 
In the noise-occupancy calibration runs, a threshold scan with no charge injection is performed 
to obtain the occupancy as a function of threshold charge.
The ENC is estimated from a linear fit to the dependence on threshold charge of
the logarithm of noise occupancy.

\end{enumerate}

\noindent The former value reflects the width of the 
Gaussian-like noise distribution around $\pm1\sigma$  
while the latter is a measure of the noise tail at around $+4\sigma$,
being sensitive to external sources such as common-mode noise.
Whenever the corresponding calibration runs are performed, the mean
values of the response-curve ENC or NO at 1~fC for each chip 
are saved for later analyses. 
These two noise estimates correlate fairly well 
chip-by-chip, but ENC values deduced from noise occupancies are 
sytematically smaller by about 5\%~\cite{bib:NOvsENC}. 
Below, unless specified, `noise' 
refers to the ENC value from the response-curve width, and is
measured for modules with bias voltage at the normal operating
value of 150 V.  

Figure~\ref{fig:noise_2010_2012} shows the distributions of chip-averaged
response-curve ENC values and noise occupancies at 1~fC 
as of October 2010 and December 2012 for different module types. 
For the barrels, only modules with \textless111\textgreater\ 
sensors are plotted; barrel~6, which has a higher temperature and thus
higher noise, is shown separately from the other three barrels. For the endcaps, 
chips reading out Hamamatsu and CiS sensors are shown separately. 
Clusters around 1000 electrons in figure~\ref{fig:noise_2010_2012}(a)
reflect the shorter strip lengths of endcap inner and middle-short sensors.
Noise occupancies of endcap inner and middle-short modules
are not shown due to large errors on the low occupancy values.
Almost all chips satisfy the SCT requirement of less than 
$5\times10^{-4}$ in noise occupancy for both periods.
  
\begin{figure}[htbp]
  \begin{center}
  \begin{tabular}{cc}
  \includegraphics[width=0.477\columnwidth]{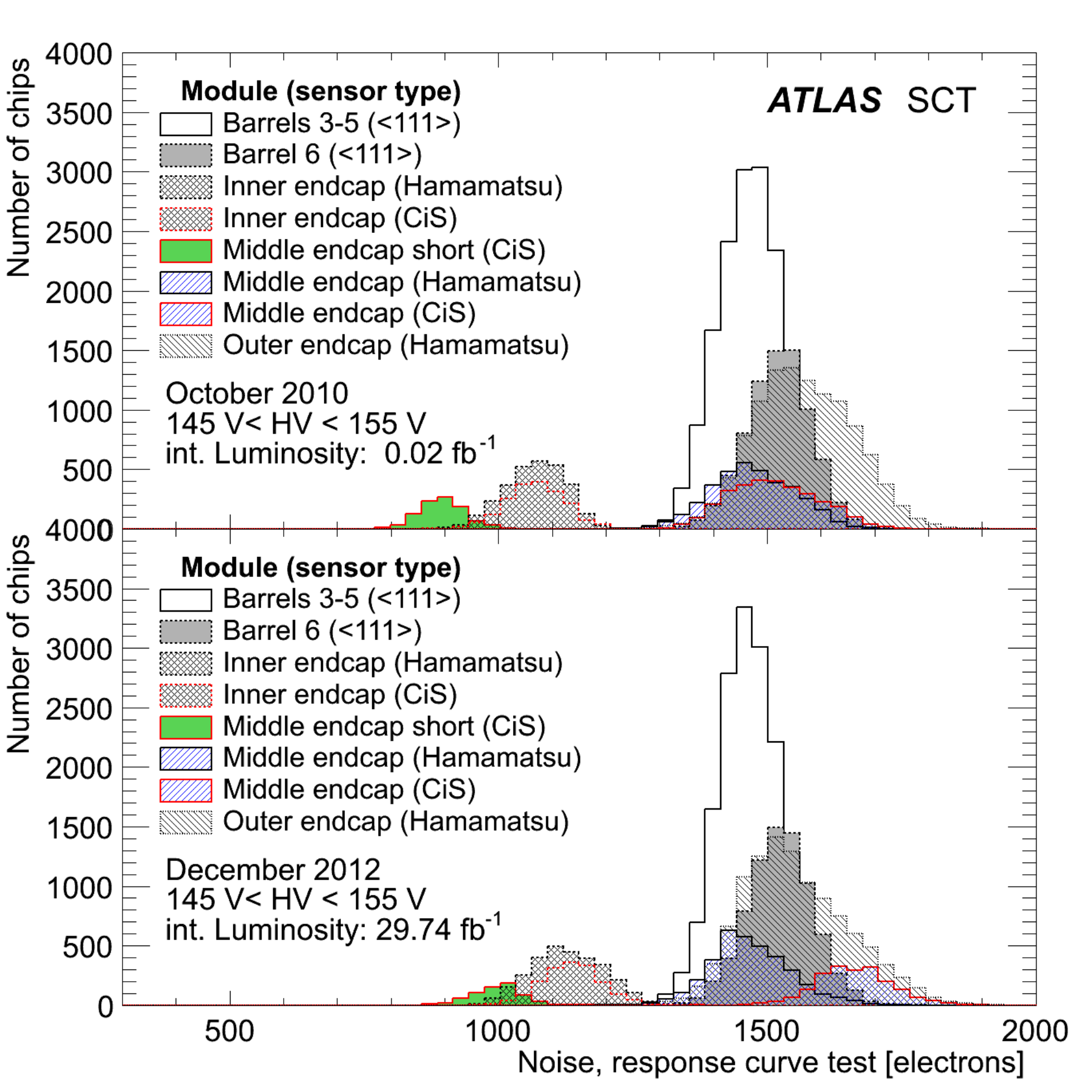} &
  \includegraphics[width=0.49\columnwidth]{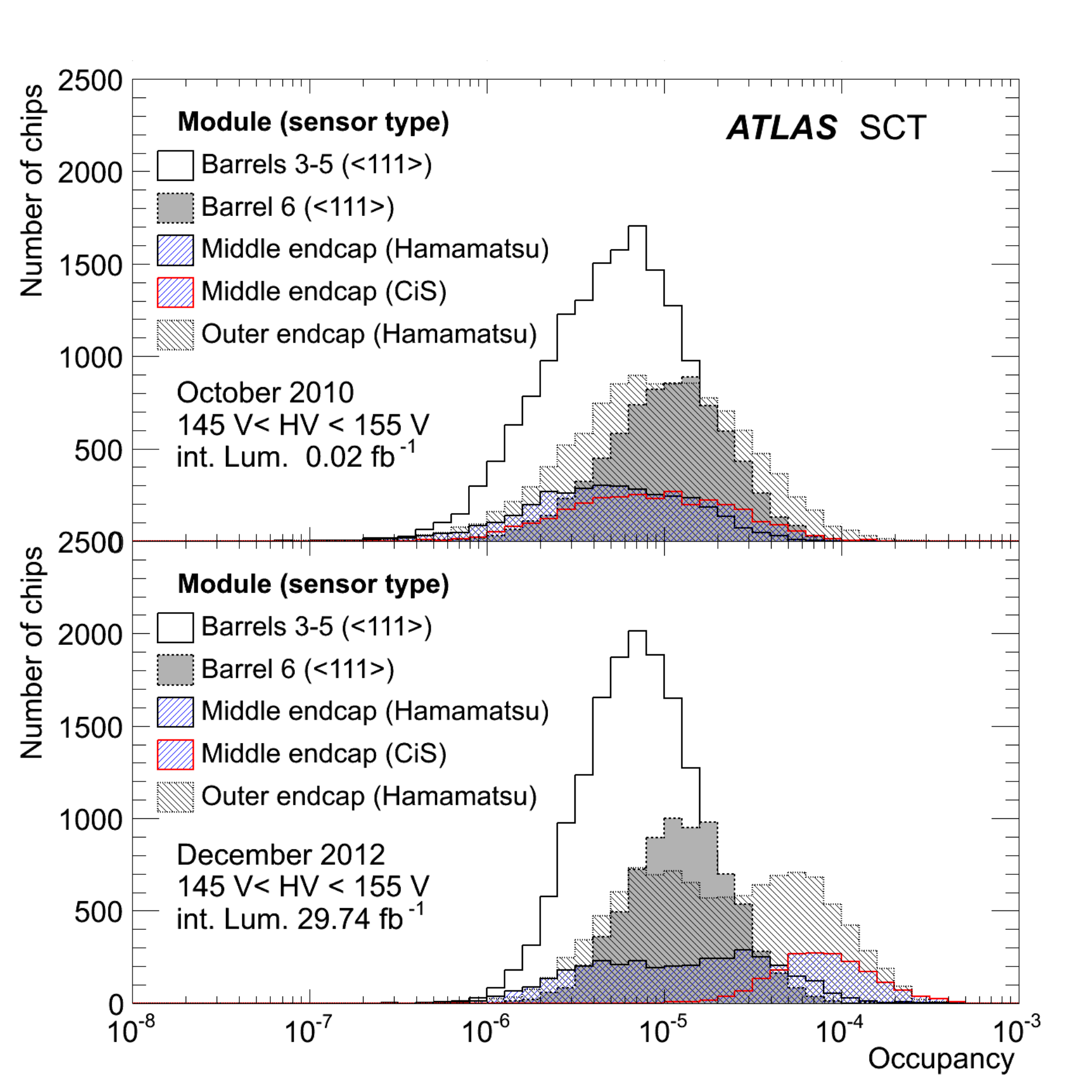}\\
(a) & (b) \\
  \end{tabular}
  \caption{Distributions of chip-averaged (a) ENC from response-curve 
tests and (b) noise occupancy at 1~fC from calibration 
runs as of October 2010 (top) and December 2012 (bottom) 
for different types of module.}
  \label{fig:noise_2010_2012}
  \end{center}
\end{figure}

Table~\ref{tab:noise} shows 
mean values of the response-curve ENCs as well as logarithmic means of
the noise occupancy at 1~fC for all ten types of module. 
The last column of the table shows the ratios of mean ENC values in December 2012
to those in October 2010, corrected (by at most 0.3\%) to compensate for the 
effects of small differences in module temperatures between the two periods.
After receiving 30~\ifb\ of delivered luminosity, 
the noise remained almost unchanged for barrel and endcap outer modules,
while it increased by about 10\% in the endcap middle modules
with CiS sensors and 6\% in endcap inner modules with either 
Hamamatsu or CiS sensors.
The median charge deposited by a minimum-ionising charged particle at normal incidence,
measured in threshold scans in beam tests, is 3.5$\pm$0.1~fC at a bias voltage of 300~V 
(0.15~fC lower at 150~V) for unirradiated modules~\cite{bib:SCTBeamTests}. Thus the
signal-to-noise ratio is about 13 for sensors of length 12~cm and about 20 for 
endcap inner sensors. Even with the observed increase of $\sim$10\% for some
endcap sensors, the signal-to-noise ratio is safely higher than the value of nine needed
for efficient operation.   
   
\begin{table}[htbp]
  \caption{Mean values of ENC from response-curve tests and noise 
occupancies (NO) at 1~fC, obtained in calibration runs in October 2010 and
December 2012. Values are shown for each module type, separated 
according to sensor manufacturer or crystal orientation. The mean temperatures 
measured by thermistors mounted on the hybrid circuits of each module were around
2.5$\degr$C for barrels 3--5, 10$\degr$C for barrel 6 and 8$\degr$C for the endcaps,
and varied by less than one degree between October 2010 and December 2012. 
The last column shows ratios of ENC values measured in 2012 to those measured in 2010 corrected 
for the small temperature differences. Uncertainties in the ratios are less than 1\%.}
\begin{center}
\begin{tabular}{lccccccc}\hline \hline
 \multirow{2}{*}{Module} & \multirow{2}{*}{Type} &  \multirow{2}{*}{Manufacturer} & \multicolumn{2}{c}{October 2010}&\multicolumn{2}{c}{December 2012}
&\multirow{2}{*}{{\large $\frac{<{\rm ENC}>_{2012}}{<{\rm ENC}>_{2010}}$}}\\ 
    & & & {\footnotesize $<{\rm ENC}>$} & {\footnotesize $<{\rm NO}>$}  & {\footnotesize $<{\rm ENC}>$}& {\footnotesize $<{\rm NO}>$} & \\ \hline
 \multirow{2}{*}{Barrels 3--5} & 111 & Hamamatsu& 1470 & $7.5\cdot10^{-6}$ & 1468 & $9.0\cdot10^{-6}$& 0.997\\ 
    & 100 & Hamamatsu& 1362 & $1.6\cdot10^{-6}$ & 1394& $3.1\cdot10^{-6}$ & 1.021\\ \hline
 \multirow{2}{*}{Barrel 6}& 111 & Hamamatsu& 1521 & $1.3\cdot10^{-5}$ & 1524 & $1.5\cdot10^{-5}$ & 1.001 \\
    & 100 & Hamamatsu& 1396 & $2.4\cdot10^{-6}$ & 1436 & $4.8\cdot10^{-6}$ & 1.028\\ \hline
 \multirow{2}{*}{EC inner} & 111 & Hamamatsu& 1069 & --- & 1129 & --- & 1.058 \\ 
    & 111 & CiS & 1070 & --- & 1136 & --- & 1.065 \\ \hline
 \multirow{2}{*}{EC middle}& 111 & Hamamatsu & 1472  & $8.0\cdot10^{-6}$ &1469 & $2.1\cdot10^{-5}$ & 1.000\\ 
    & 111 & CiS & 1513 & $1.4\cdot10^{-5}$ &1662 & $9.5\cdot10^{-5}$ & 1.102\\ 
 middle short & 111 & CiS &  895 &  --- & 1004 & ---  & 1.124 \\ \hline
 EC outer  & 111 & Hamamatsu& 1568 & $1.6\cdot10^{-5}$ &1553 & $3.9\cdot10^{-5}$ & 0.993\\ \hline \hline
\end{tabular}
\label{tab:noise}
\end{center}
\end{table}

The evolution of the response-curve noise as well as the front-end
gains obtained in calibration runs from 2010 to 2012 
is shown in figure~\ref{Noise:noise_gain}; 
modules are divided into ten different groups
as in table~\ref{tab:noise}. 
All the gains are rather stable except the gradual and 
universal change of a few percent in mid 2011 and early 2012,
the reason for which is not clear.

As can be seen in figure~\ref{Noise:noise_gain},
the noise decreased by about 7\% in late 2010 when the
fluence level at the modules was around 
$10^{10}$~cm$^{-2}$, a dose level of a few Gy.
The noise drop varied strongly with the strip location
as shown by the chip-number dependence in 
figure~\ref{Noise:chip_dependence} for barrel~5 as 
an example.
It occurred systematically in all modules except those with
\textless100\textgreater\ sensors. In addition,
the decrease occurred first in barrel 3 and later in barrel 6 indicating 
fluence-dependent effects.
Similar noise decreases were observed in the endcap modules but they strongly depended on 
the module side.
Disconnected strip channels showed no decrease.   
In 2011 and 2012, no such noise decrease was observed, rather
the noise returned to the original value and the chip-number dependence disappeared
as shown by the 2011 points in figure~\ref{Noise:chip_dependence}.
A beam test at the CERN-PS with a low rate reproduced qualitatively
such a trend at a similar dose level.
This unexpected phenomenon is probably caused by charge accumulated 
on the sensor surface leading to effective capacitance changes. 

\begin{figure}[htbp]
\begin{center}
\includegraphics[width=\textwidth]{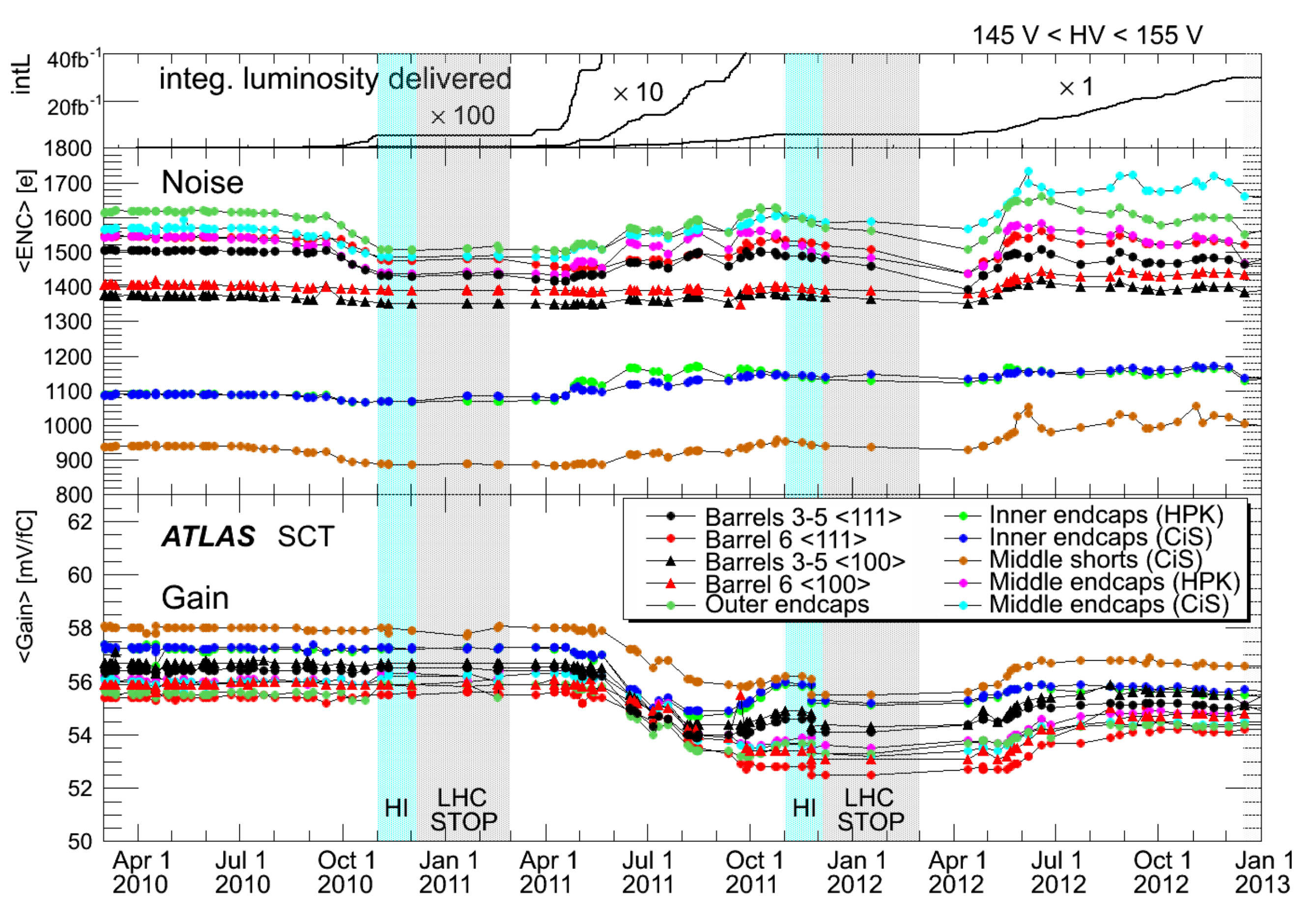}
\end{center}
\caption{Integrated luminosity (top), noise (middle)
and calibration gains of front-end amplifiers (bottom) 
as a function of time. Modules are divided into ten different groups
according to their module type, crystal orientation
\textless111\textgreater\ vs  \textless100\textgreater\ 
and sensor manufacturer (Hamamatsu is labelled HPK). The blue shading and label `HI' indicate
periods of heavy-ion running, while extended periods with no beam in 
the LHC during which the SCT was off are shaded grey.}
\label{Noise:noise_gain}
\end{figure}

\begin{figure}[htbp]
\begin{center}
\includegraphics[width=0.7\textwidth]{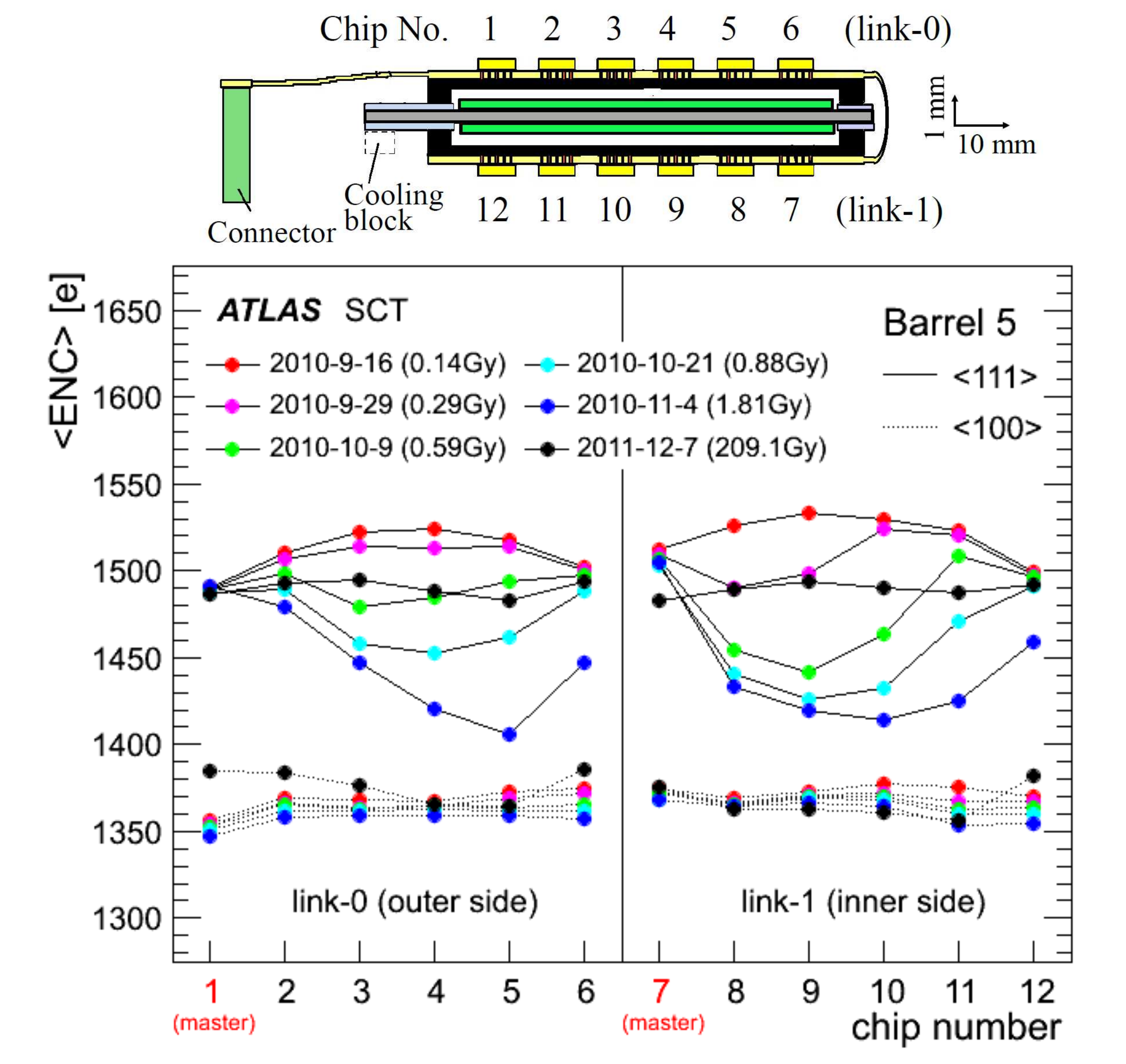}
\end{center}
\caption{Chip-averaged noise as a function of chip number for
several runs during 2010 (coloured points) and at the end of 
year 2011 (black points) for barrel 5. Chips with 
\textless100\textgreater\ sensors are plotted separately. 
Chip numbers and locations are indicated in the top diagram.}
\label{Noise:chip_dependence}
\end{figure}

\subsection{Alignment stability}\label{sec:fsires}
While the alignment is determined using tracks (as described in section~\ref{trackalign}), the FSI system is used to confirm the observation of movement in the SCT.

Figure~\ref{fig:timedep} shows the variation of track-based
alignment translations for the largest structures during the $pp$ collision 
data-taking period from May to June 2011. The corrections shown are for
translations in the global $x$ direction. Movements of the
detector are measured after hardware incidents such as toroid-magnet ramping
as indicated in the first bin of figure~\ref{fig:timedep}.
In between these periods little ($<$1$\mu$m) movement is observed indicating 
that the detector is generally very stable. 

\begin{figure}[ht!]
\centering
\includegraphics[width=\textwidth]{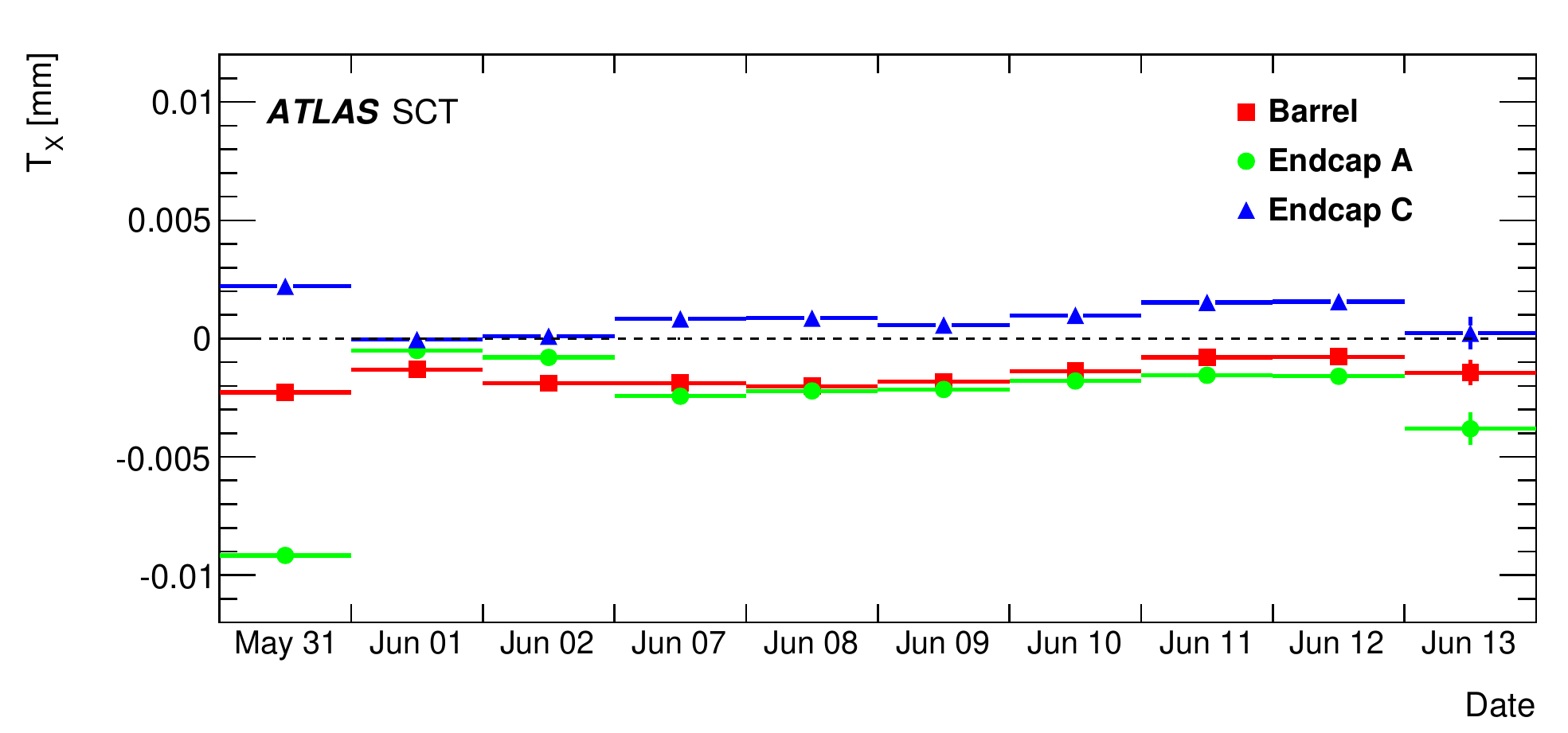}
\caption{Subsystem-level track-based alignment corrections in the global $x$ coordinate,
         $T_{x}$, 
         performed on a run-by-run basis starting from a common set of alignments. The 
         subset shown covers a two-week period during May and June 2011. The first data points here show 
         the effect on the alignment constants of the toroid-magnet ramping.}
\label{fig:timedep}
\end{figure}

The most important result from FSI measurements is that the SCT
detector is found to be stable at the $\mu$m level over extended
periods of time, in agreement with the track-based alignment results
shown in figure~\ref{fig:timedep}. 
Figure~\ref{fig:lumi} shows the SCT barrel-flange
movement for a subset of grid lines during a 72-hour period
in May 2011. The lower plot shows the instantaneous luminosity 
during this period. 
In the upper plot, the top three curves show the
measurements from outer barrel-flange lines (shifted by +3~$\mu$m), the
three curves in the middle show the measurements from the middle flange
lines and the bottom three curves (shifted by --3~$\mu$m) show the
measurements from the inner flange lines. The colours correspond to
those in figure~\ref{fig:flange}. The changes in the lengths of
the flange lines (movements) are most likely correlated to changes in power load 
(and hence temperature) in the front-end amplifiers at high trigger rate,
and indicate a radial expansion and relative rotation of the barrels. 
The scale of the deviation is at the $\mu$m level and when
the heating is removed the detector returns to the same
pre-$pp$-collision position. The very gradual trend observed in some lines
in figure~\ref{fig:lumi} is due to the drift of the lasers over time, and 
does not reflect a physical effect on the grid-line length.

\begin{figure}[ht!]
\centering
\includegraphics[width=0.9\textwidth]{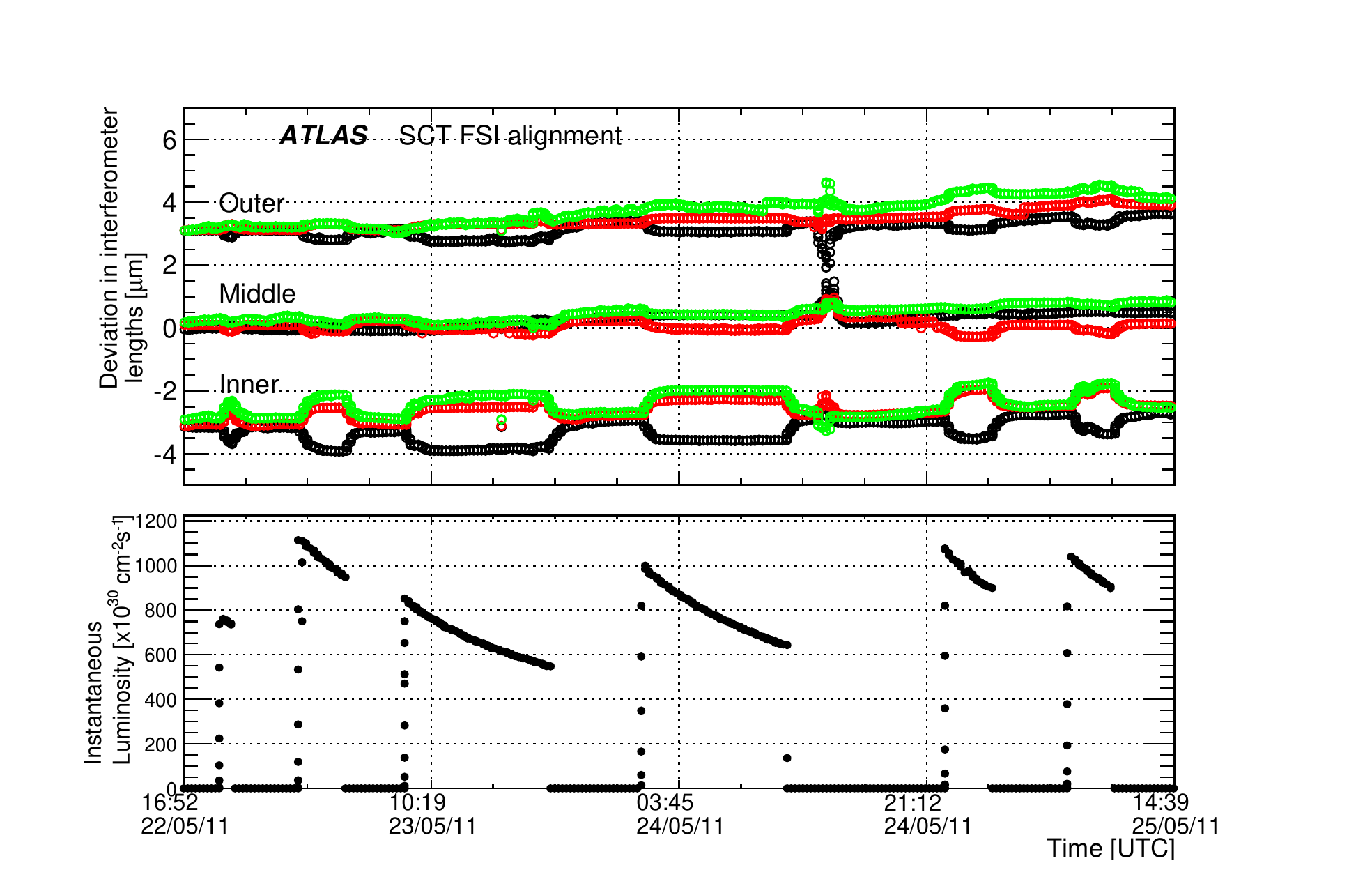}
\caption[Luminosity correlations]{Length deviation of barrel-flange interferometers during 
        three days of LHC collision data-taking in May 2011. In the upper plot, the lower
        three curves (shifted by --3~$\mu$m) show the measurements from the inner layer, the
        three curves in the middle show the measurements from middle layer, the upper three 
        curves (shifted by +3~$\mu$m) show the measurements from the outer layer. The colours
        correspond to those in figure~\protect\ref{fig:flange}. 
        The lower plot shows the instantaneous LHC luminosity.}
\label{fig:lumi}
\end{figure}

Figure~\ref{fig:solenoid} shows the effect observed in the FSI during
ramping of the solenoid magnetic field in December 2009. The upper plots show 
the current in the magnet and the temperature of the gas in the SCT;
the lower plot shows the displacement of the barrel-flange grid lines. 
Displacements up to 3~$\mu$m are observed while the field is
ramping but the detector is observed to return to its initial position
at the end of the ramping process.
In figure~\ref{fig:toroid}, the effect of the toroid field is shown for
three barrel-flange lines. A displacement of 3~$\mu$m is correlated in time 
with a toroidal current variation of 20~kA. This corresponds to the
movement observed in the track-based alignment at the end of May 2011
shown in figure~\ref{fig:timedep}. Another movement of about 0.5~$\mu$m can be 
correlated with the short luminosity pulse observed on the same day at around 20:00~h.

\begin{figure}[ht!]
\centering
\subfigure[]{\label{fig:solenoid}\includegraphics[width=0.48\textwidth]{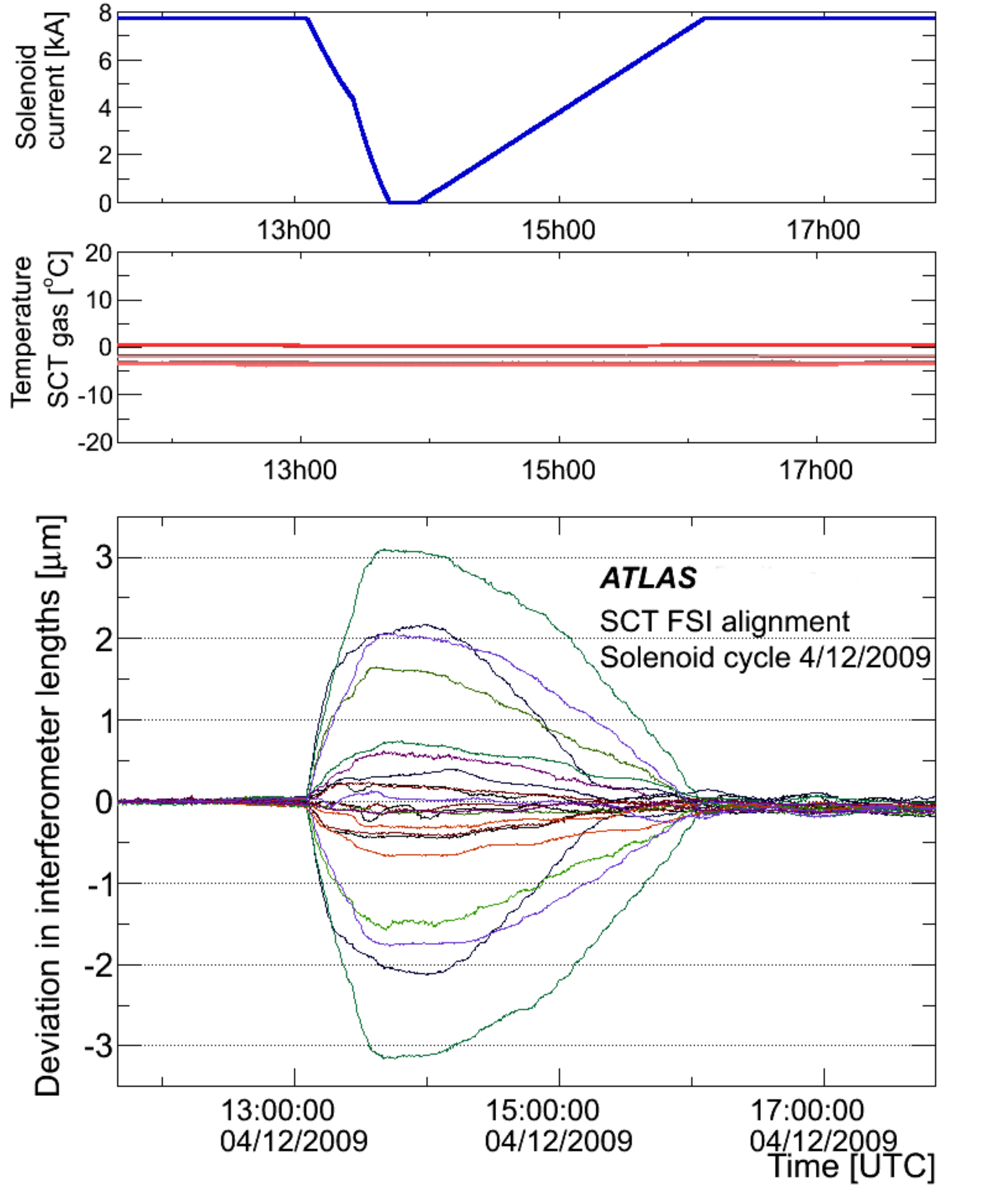}}
\subfigure[]{\label{fig:toroid}\includegraphics[trim = 0cm 1cm 0cm 0cm, clip, width=0.51\textwidth]{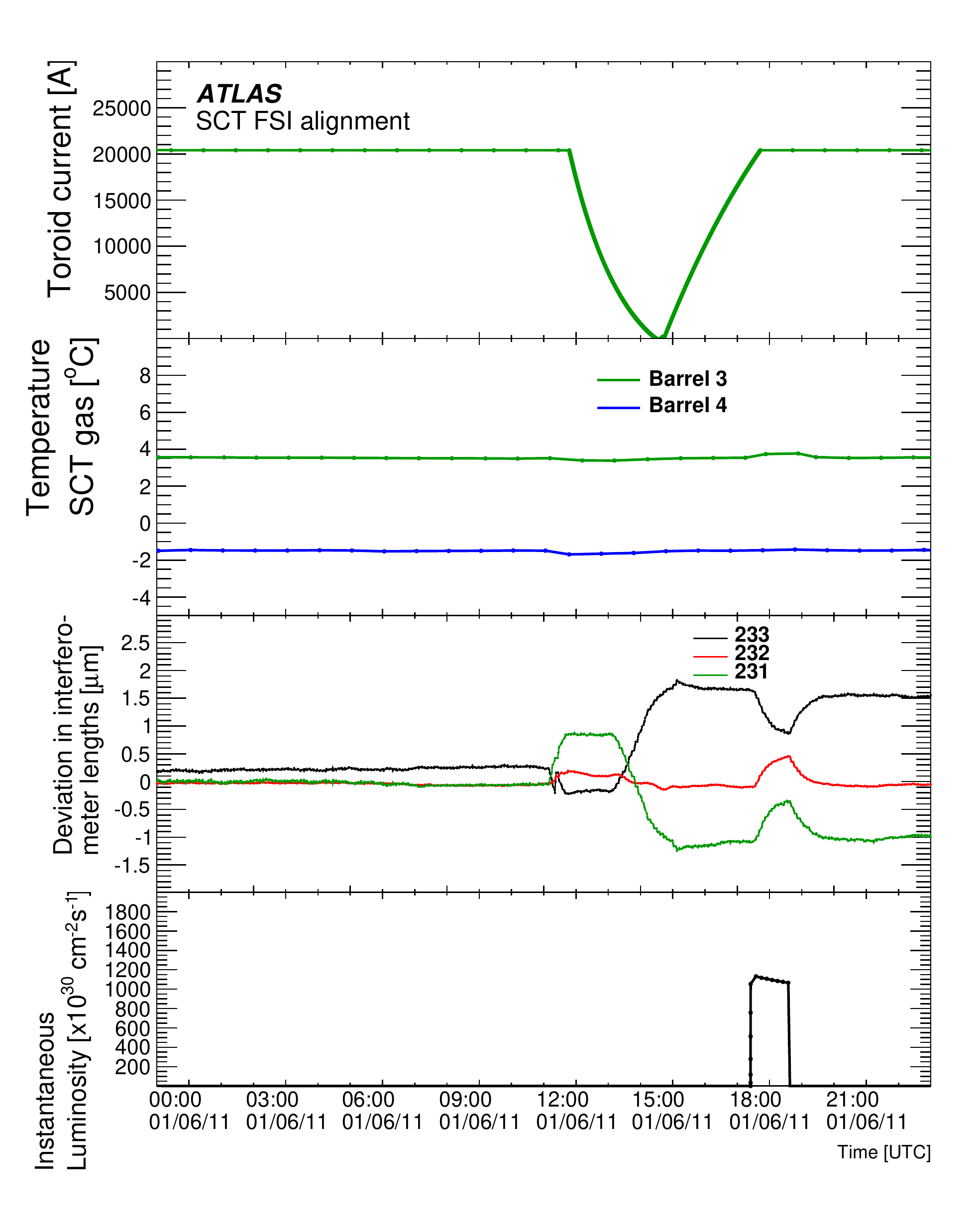}}
\caption[Magnet Field induced movements] {Length deviations of SCT barrel-flange interferometers
        (a) during a cycle of the ATLAS solenoid and (b) during a toroid-magnet ramping. In (a) the upper plot
        shows the solenoid current, the middle plot the temperature of the gas inside the SCT volume
        and the lower plot the measured deviations. In (b) the plots show, from top to bottom, the 
        toroid current, the gas temperature, the measured deviations and the instantaneous luminosity.}
\label{fig:magnetmovement}
\end{figure}

\subsection{Intrinsic hit efficiency}
\label{sec:eff}

The intrinsic detector efficiency measures the probability of a hit being
registered when a charged particle traverses the sensitive part of an
operational detector element. Disabled sensors and chips are excluded
from the measurement. Both a high intrinsic efficiency and a 
low non-operational fraction are essential to ensure good-quality tracking.

The intrinsic efficiencies of the modules are measured 
by extrapolating well-reconstructed tracks through the detector and
counting the numbers of hits (clusters) on the track and `holes' where
a hit would be expected but is not found. The track extrapolation uses 
the full track fit described in section~\ref{sec:tracking} to compute
the intersections of the track with all modules along its trajectory.
If a module side does not have a cluster associated 
to the track and the intersection point is more than 3$\sigma$ from the 
edge of the sensitive area, the absence is called a hole. 
The efficiency, $\varepsilon$, is defined 
as the ratio of the number of clusters found to the number expected:
\begin{eqnarray}
  \varepsilon = \frac{N_{\rm clusters}}{N_{\rm clusters}+N_{\rm holes}}
  \label{equ:eff}
\end{eqnarray}
where $N_{\rm clusters}$ is the number of clusters found and $N_{\rm holes}$
is the number of holes.

Combined inner detector tracks used to measure the efficiency must have at 
least six hits in the SCT excluding the hit or hole under consideration, and
transverse momentum $\pT > 1$~\GeV. 
Events containing more than 250 reconstructed tracks are excluded from the efficiency
measurement, to reduce biases from reconstruction algorithms in high-occupancy events. 
To reduce any bias due to the track fitting and pattern recognition criteria,
which may be affected by residual misalignments, clusters not already assigned to 
a track but within 200~$\mu$m of an intersection are included in $N_{\rm clusters}$ in 
equation~(\ref{equ:eff}) and removed from $N_{\rm holes}$. From a study of the residual
distance of the second-nearest cluster to a track it is estimated that only 4.4\% of
such clusters are due to random associations or noise, and most result from hit-to-track assignment criteria applied during track reconstruction.
The inclusion of these clusters increases the efficiency by 0.014\%.  
The distance for inclusion of unassigned clusters is varied in the assessment of
systematic uncertainties.

Non-functioning complete module sides and chips (see section~\ref{sec:Operations}) are not 
included in the calculation of the intrinsic efficiency; these amount to $\sim$1\% of the detector. 
The measured inefficiency contains a contribution from isolated dead strips, including those which are
disabled in reconstruction due to noise, for which no correction is applied.

The measured efficiency for each barrel and each endcap disk in 2012 is shown in 
figure~\ref{fig:efficiency}. These measurements use data recorded in a proton--proton run
with low pile-up, with less than one interaction per bunch-crossing on average. 
The overall efficiency is $(99.74\pm0.04)\%$, where the error is systematic. 
The statistical error is negligible. 
The systematic uncertainty was estimated by varying the distance within which 
unassigned hits are included from zero to 500~$\mu$m ($^{+0.019}_{-0.014}$\%) and 
by considering the change when only module sides for which a hit was registered 
in the other side are used in the measurement (+0.036\%). The overall systematic uncertainty 
is taken as the (symmetrised) sum in quadrature of these two contributions. As a cross-check, the 
efficiency was determined using different track selection criteria: varying the mimimum transverse 
momentum requirement between 0.5~\GeV\ and 2~\GeV, reducing the requirement on the minimum number of 
other SCT hits from six to five, and placing requirements on the number of hits in the pixel or TRT 
detectors. These variations give values within the uncertainty estimated by requiring a hit on the 
other module side. The efficiencies measured in the barrel and endcap regions separately are
(99.86$\pm$0.03)\% and (99.59$\pm$0.05)\% respectively.   
The efficiency measured in 2012 $pp$ collisions in the barrel region is very similar to the value of
(99.78$\pm$0.01(stat.)$\pm$0.01(syst.))\% measured in cosmic-ray data in 2008~\cite{bib:IDCosmic},
indicating little change in detector performance over this period. 

\begin{figure}[htbp]
  \centering
  \includegraphics[width=\columnwidth]{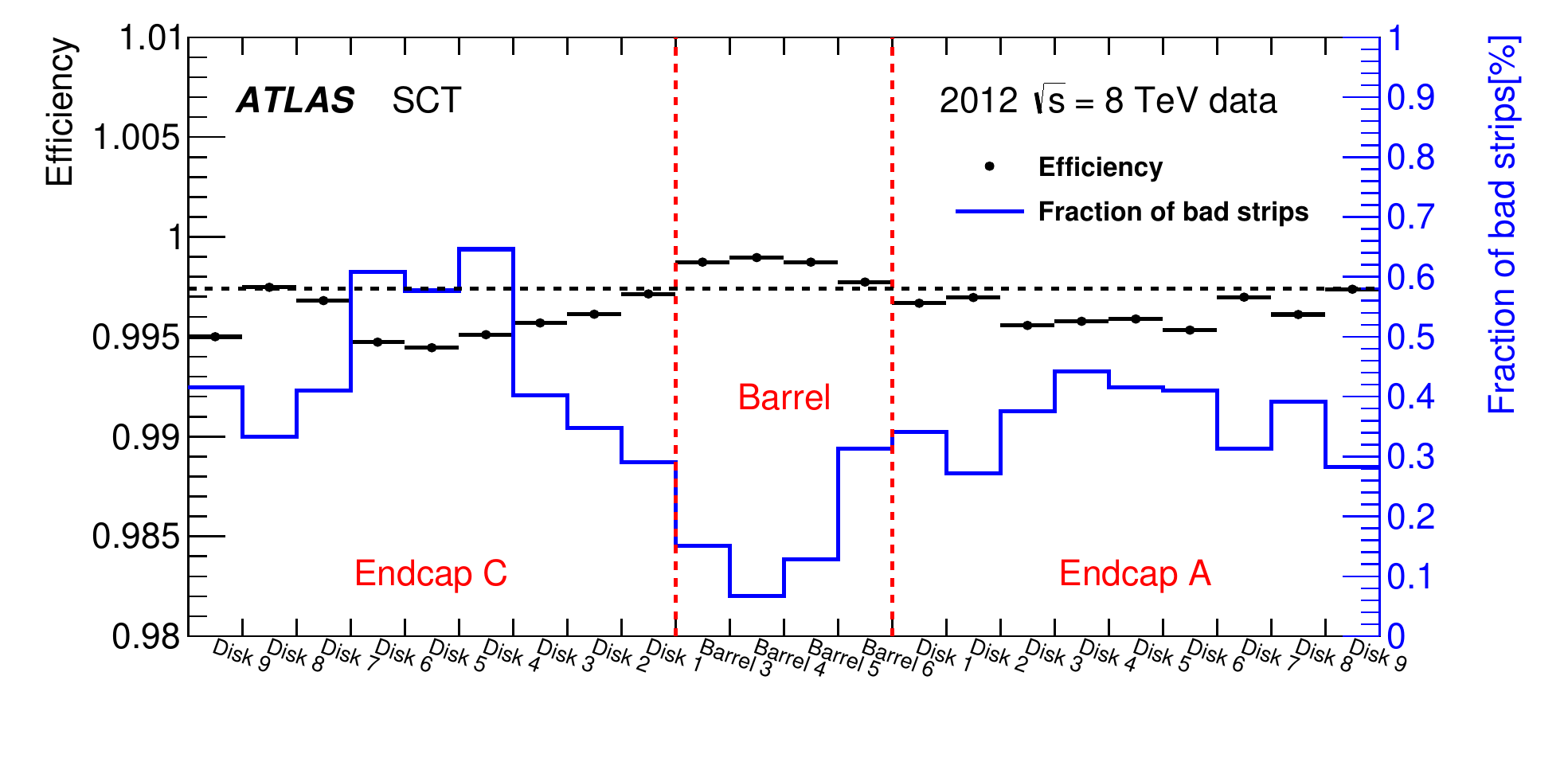}
  \caption{Mean intrinsic hit efficiency for each layer of the SCT measured in proton--proton data 
           at $\sqrt{s}$ = 8~\TeV\ in 2012. The dashed line indicates the overall mean value.
           The blue line and right-hand axis show the fraction of strips in the layer which 
           are disabled (dead or noisy).} 
  \label{fig:efficiency}
\end{figure}

The variation in efficiency from layer to layer seen in figure~\ref{fig:efficiency} primarily 
results from differences in the proportion of isolated disabled strips in each layer. The fraction 
of such strips in each layer is also shown in figure~\ref{fig:efficiency}, and a clear anti-correlation 
with efficiency is seen. While there is some dependence on the distribution of disabled strips (i.e.\
single, pairs or larger groups), on average the inefficiency is linearly dependent on the total 
fraction. This dependence is shown in figure~\ref{fig:eff_correlation}, where it can be seen that the
inefficiency is dominated by the contribution from disabled strips. A 1\% fraction of disabled strips 
gives rise to an efficiency loss of about 0.52\% in the barrel and 0.75\% in the endcap. The lower 
dependence in the barrel arises from the larger mean cluster width. 

\begin{figure}[htbp]
  \centering
  \includegraphics[width=0.7\columnwidth]{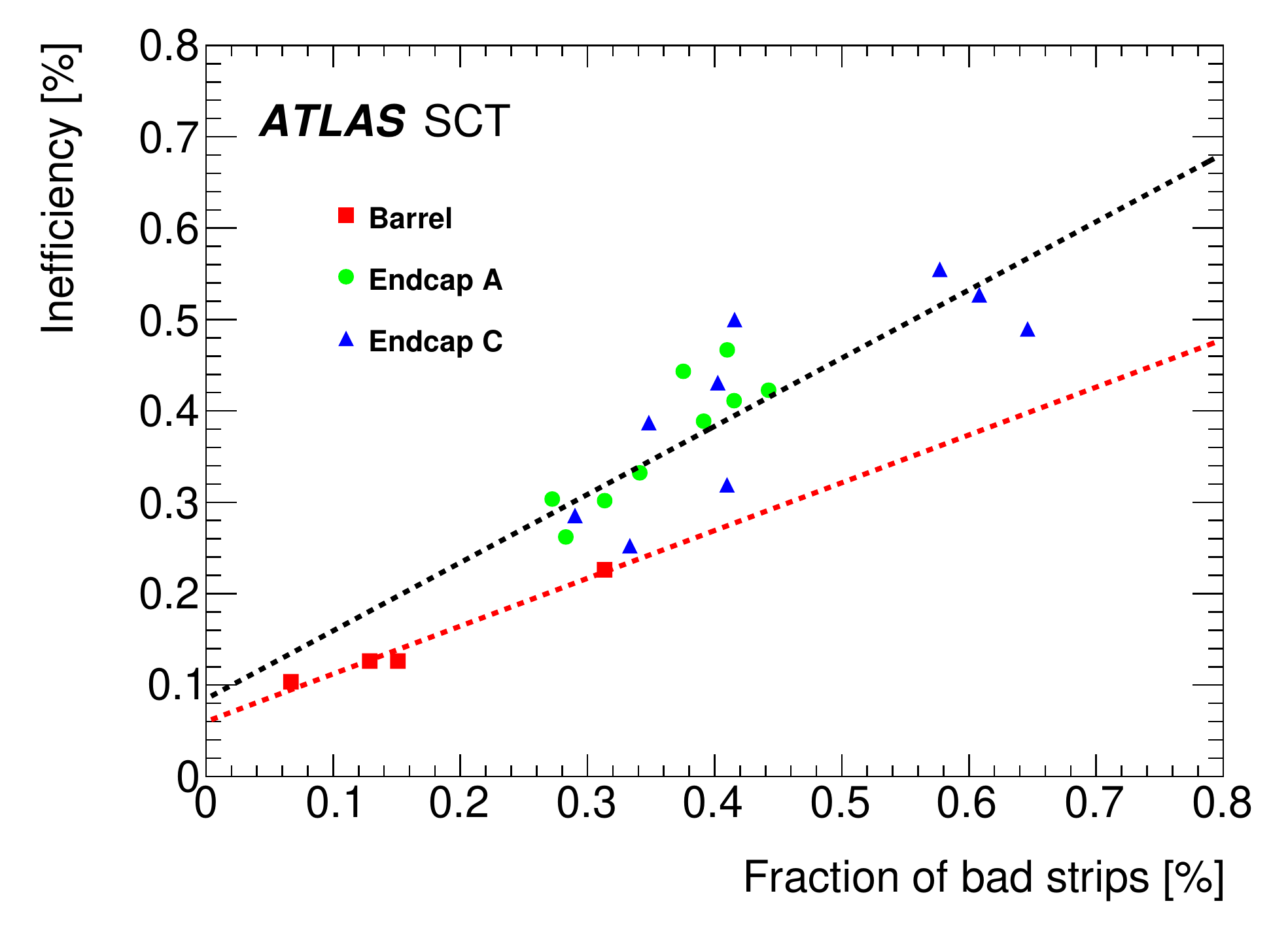}
  \caption{Mean intrinsic hit inefficiency for each layer of the SCT measured in proton--proton 
           collision data at $\sqrt{s}$ = 8~\TeV\ in 2012 as a function of the fraction of 
           individual strips in the layer which are dead or disabled due to noise.
           The statistical errors on the inefficiency values are smaller than the size of
           the data points. The dashed lines show the results of linear fits for the barrel
           and endcaps separately.}  
  \label{fig:eff_correlation}
\end{figure}

\subsection{Lorentz angle}
\label{sec:LA}
Charge carriers in the silicon detectors are subject to the electric 
field, $\mathbf{E}$, generated by the bias voltage and oriented normal to the 
module plane, and to the magnetic field from the solenoid, $\mathbf{B}$. 
In the SCT endcap modules these fields are nearly parallel and the charge carriers drift
directly towards the electrodes. In the barrel modules these fields are perpendicular 
and the charge carriers drift along the Lorentz angle, $\theta_{\rm L}$, with respect to 
the normal to the sensor plane. 
The value of Lorentz angle is given by:
\begin{equation}
\tan\theta_{\rm L} = \mu_{\rm H} B = \gamma_{\rm H} \mu_{\rm d} B 
\end{equation}
where $\mu_{\rm H}$ is the Hall mobility, the product of the charge-carrier mobility in silicon 
$\mu_{\rm d}$ and the Hall factor $\gamma_{\rm H}$, which is of order unity.  
The charge-carrier mobility depends on the bias voltage, the thickness of the depleted region and the 
temperature~\cite{bib:Jacoboni}. 
For fully depleted modules, the average shift in collected charge is approximately 10~$\mu$m, 
which is not negligible with respect to the detector resolution and alignment precision. 
Measurements of the Lorentz angle for the SCT sensors have previously been made in test-beams~\cite{bib:SCTBeamTests} 
and using cosmic-ray data collected prior to LHC beam operation~\cite{bib:IDCosmic}. The latter measurements were
shown to be compatible with the model predictions for the charge-carrier mobility of ref.~\cite{bib:Jacoboni}, 
within the uncertainties on these predictions. The very high number
of tracks in the collision data now allows the Lorentz angle to be studied in more detail
than was possible using cosmic-ray data.      
In particular, differences in the charge-carrier mobility mean that the expected Lorentz angle is 
different for sensors with different crystal lattice orientations. In the studies presented
here, sensors with $<$111$>$ and $<$100$>$ crystal orientations are considered separately.

The Lorentz angle is measured from the dependence of the cluster size on the 
incident angle of the particle. When the incident angle equals the Lorentz 
angle, all charge carriers generated by the particle drift along 
the particle direction and, apart from charge diffusion, are collected
at the same point on the sensor surface, giving a minimum cluster size.
The tilt of the barrel modules with respect to the radial direction and correlations 
between particle transverse momentum and the incident angle for particles originating 
near the beam axis give rise to a range of possible incident angles for positive and 
negative particles. 
Reconstructed tracks are required to have
$\pT > 400$~\MeV, limiting the possible range of incident angles to approximately
$-30^\circ < \phi_{\rm local} < -5^\circ$ for positive particles and 
$-15^\circ < \phi_{\rm local} < 10^\circ$ for negative particles. 
Thus, only the tracks of
negatively charged particles are used to measure the Lorentz angle in collision data.     

\begin{figure}[htbp]
 \centering
 \subfigure[]{\includegraphics[width=0.49\columnwidth]{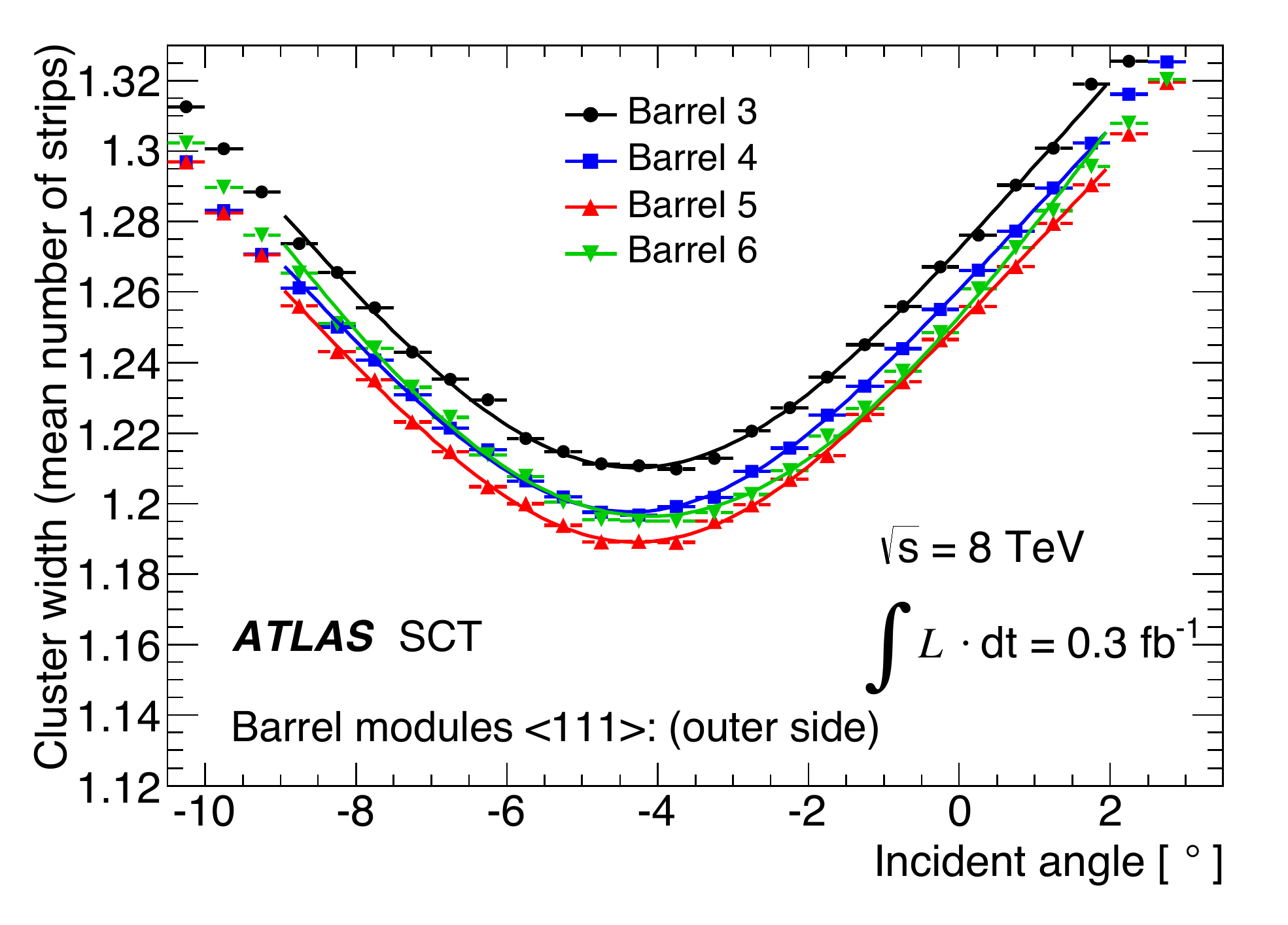}}
 \subfigure[]{\includegraphics[width=0.49\columnwidth]{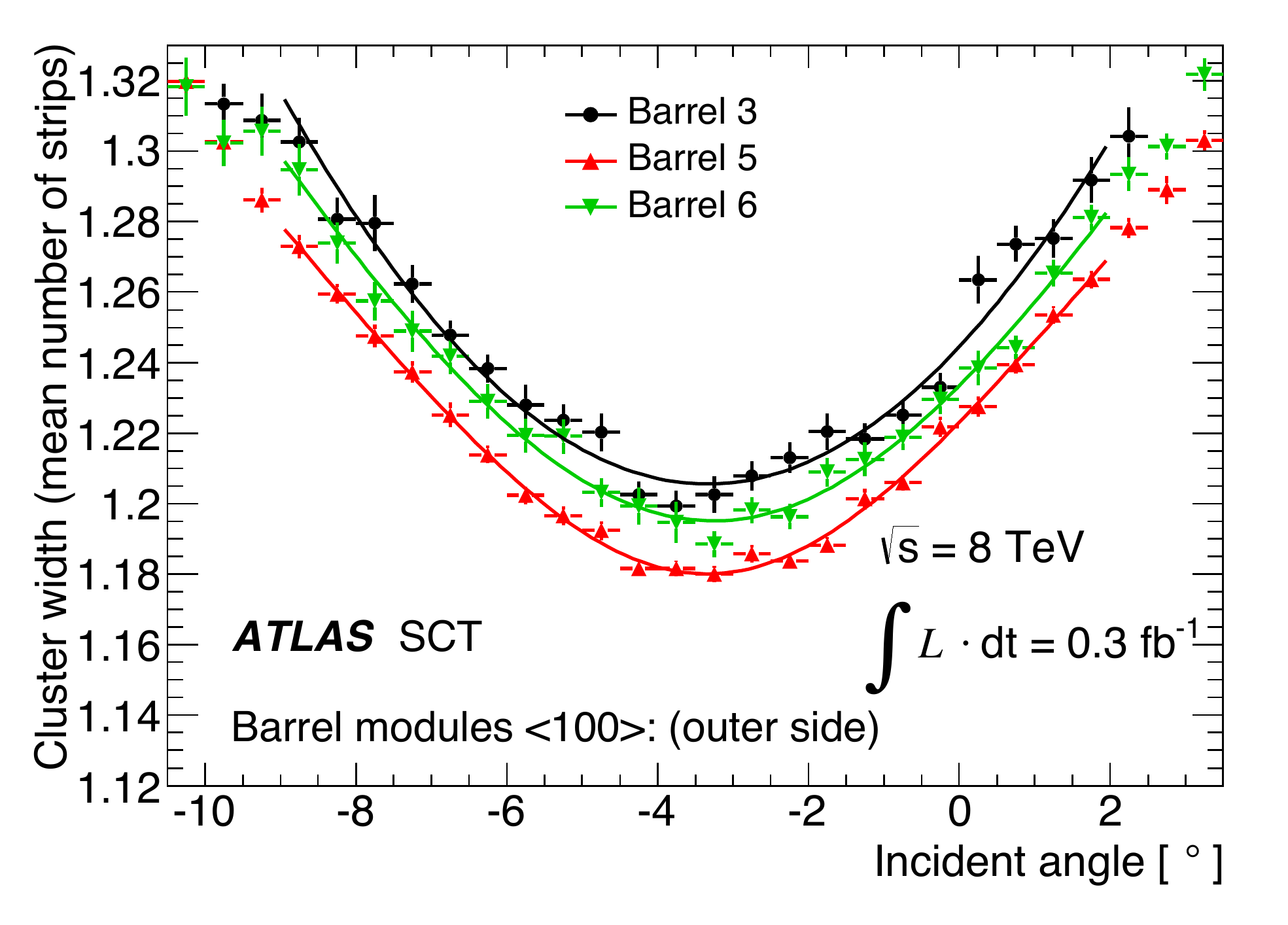}}
 \caption{Cluster-size dependence on the particle incident angle for each SCT barrel
         for (a) $<$111$>$ sensors and (b) $<$100$>$ sensors. 
         The displacement of the minimum from zero is a measurement of the Lorentz angle $\theta_{\rm L}$.}
 \label{fig:LAincidentAngle}
\end{figure}

The dependence of the cluster size on the incident angle 
$\phi_{\rm local}$ is shown in figure~\ref{fig:LAincidentAngle} for data from each
barrel. The figures on the left and on the right show the results for $<$111$>$ and $<$100$>$ 
sensors, respectively.
Data are fitted using a convolution of the function:
\begin{equation}
f(\phi_{\rm local}) = a \left| \tan\phi_{\rm local} - \tan\theta_{\rm L} \right| + b
\label{eq:LorentzGeo}
\end{equation}
with a Gaussian distribution. The fitted parameters are the Lorentz angle, $\theta_{\rm L}$, the shape 
parameters, $a$ and $b$, and the width of the Gaussian. 
The electric field in the silicon, and thus the local Lorentz angle, varies with distance from the 
electrodes. The measured values are averages over particle paths within a sensor.
Fits are performed separately for the inner and outer sides of each barrel layer for several datasets
recorded during 2011 and 2012. Results from the two sides of a layer are compatible and are 
combined. The variation in Lorentz angle expected from the spread of module temperatures in a
barrel of about 1$^\circ$C is negligible.  
Systematic uncertainties arising from the choice of fit
function are assessed by using an alternative asymmetric function, in which the slope parameter 
$a$ is allowed to take different values above and below the minimum. The mean difference in measured 
Lorentz angle in each layer is taken as a systematic uncertainty correlated among datasets, while 
the r.m.s.\ spread of the differences is used as an estimate of the systematic uncertainty 
uncorrelated among datasets.
In addition, a fit was performed on cosmic-ray data with no magnetic field; the measured value 
of the Lorentz angle is compatible with zero, and the precision of this check (0.12$\degr$) 
is included as an additional systematic uncertainty, correlated among all measurements.    

Figure~\ref{fig:LAhistory} shows measured values of the Lorentz angle for the innermost barrel 
for runs during stable-beam periods in 2011 and 2012. The variation in measured Lorentz angle over 
this period is no more than 0.1$\degr$. The other barrel layers show similar stability. 

\begin{figure}[htbp]
 \centering
 \subfigure[]{\includegraphics[width=0.49\columnwidth]{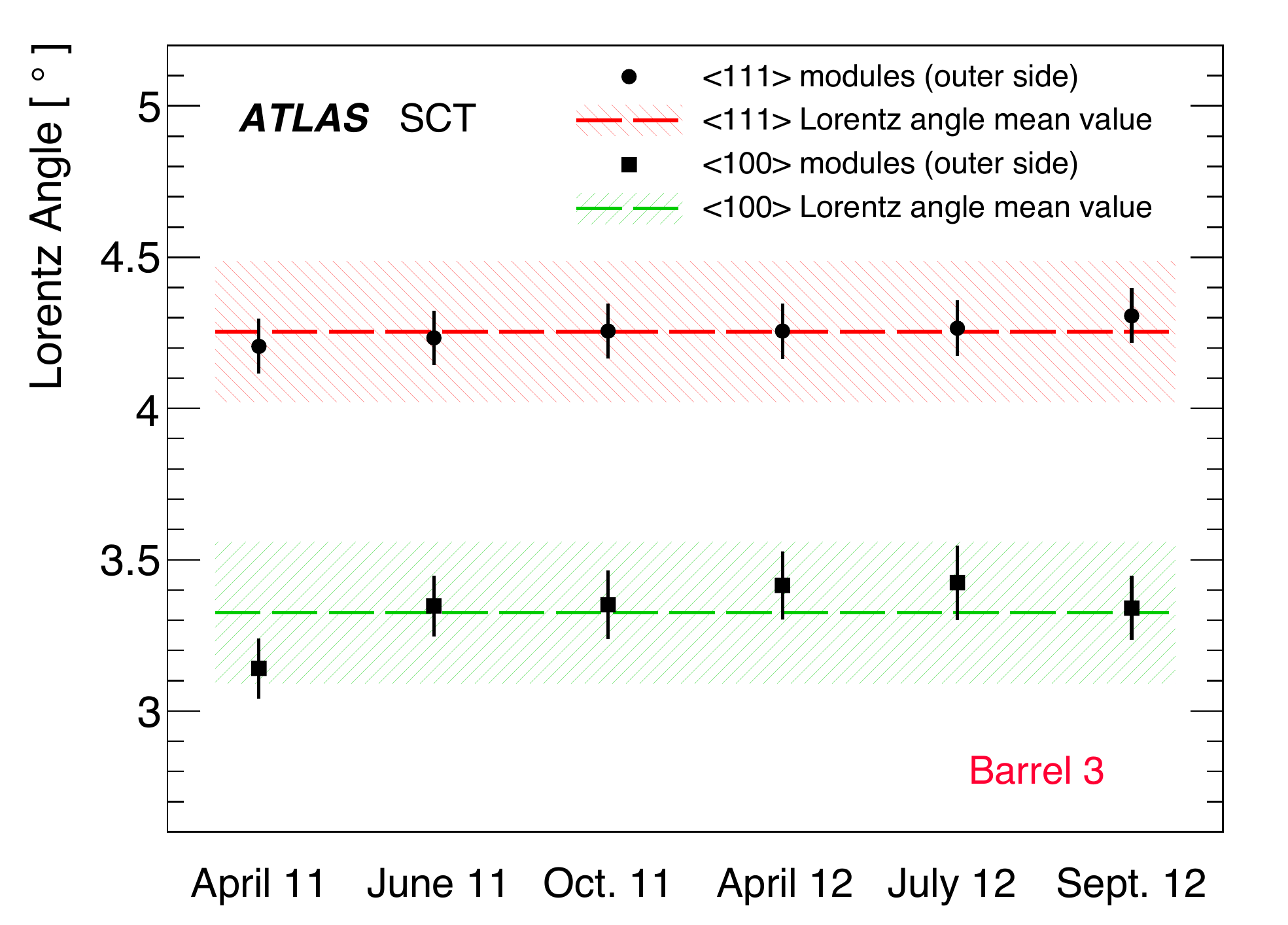}}
 \subfigure[]{\includegraphics[width=0.49\columnwidth]{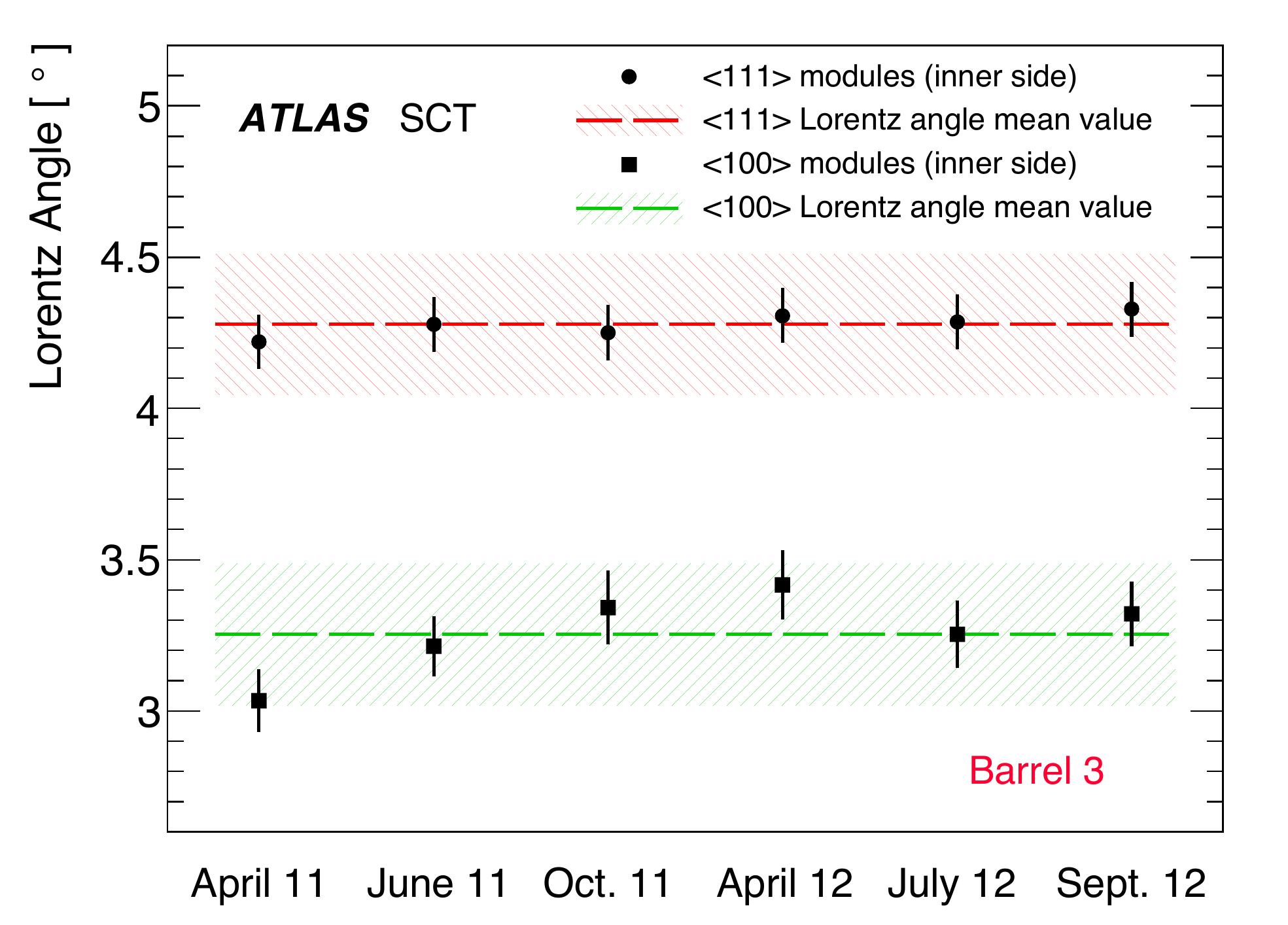}}
 \caption{Average fitted values of the Lorentz angle for $<$111$>$ and $<$100$>$ sensors in the 
          innermost barrel for stable-beam running periods in 2011 and 2012; (a) shows results 
          for the outer modules sides, while (b) shows the inner module sides. The error bars show the 
          combined statistical and uncorrelated systematic uncertainties. 
          In each case, the dashed line shows the average over the full 2011 and 2012 datasets, 
          with the correlated systematic uncertainty indicated by the shaded band.}
 \label{fig:LAhistory}
\end{figure}

The measured values of the Lorentz angle in the 2~T magnetic field, averaged over all datasets,
are shown in tables~\ref{tab:la111} and~\ref{tab:la100} and in figure~\ref{fig:LAdataMC}. 
They are compared with the expectation using the charge-carrier mobility from the 
models in refs.~\cite{bib:Jacoboni} and~\cite{bib:Becker}. The main uncertainty in
the model predictions arises from knowledge of the drift velocity; the uncertainty in this
is assumed to be 5\%, based on measurements using the time-of-flight technique. Non-uniformity
of the electric field and uncertainties in temperature and magnetic field also contribute. 
The measurements are compatible with the model predictions within at most twice the estimated 
uncertainties on these predictions. 
The measured values of the Lorentz angle for the $<$100$>$ sensors are approximately $1^{\circ}$ lower 
than those obtained for $<$111$>$ sensors. This is contrary to the expectation that a higher expected 
charge-carrier mobility in the  $<$100$>$ sensors should result in a higher value of the Lorentz 
angle~\cite{bib:Becker}.

\begin{table}[htbp]
\newcommand\T{\rule{0pt}{2.6ex}}
\newcommand\B{\rule[-1.2ex]{0pt}{0pt}}
\caption{Measured values of the Lorentz angle in $<$111$>$ modules, in a 2~T magnetic field at the 
         average operational temperature in the period from 2011 to 2012, compared with the model 
         expectations of Jacoboni et al.~\cite{bib:Jacoboni} and Becker et al.~\cite{bib:Becker}. 
         The uncertainties on the measurements include both statistical and systematic contributions; 
         those on the model predictions arise from uncertainties in the charge-carrier mobility.
         }
\begin{center}
\begin{tabular}{lcccc}
\hline\hline 
\multicolumn{5}{c}{$<$111$>$}\\
  \hline
  Layer \B \T &$T$ [\degr C]
                               &Measured $\theta_{\rm L} [^\circ]$
                               &Jacoboni et al. Model $\theta_{\rm L} [^\circ]$
                               &Becker et al. Model $\theta_{\rm L} [^\circ]$  \\
  \hline
  Barrel 3 &--2.1             &4.28$\pm$0.23 & 3.88$\pm$0.29  &3.50$\pm$0.34 \\ 
  Barrel 4 &--1.7             &4.26$\pm$0.22 & 3.87$\pm$0.29  &3.50$\pm$0.33 \\
  Barrel 5 &--1.3             &4.30$\pm$0.15 & 3.86$\pm$0.29  &3.49$\pm$0.33 \\ 
  Barrel 6 &\phantom{--}5.6   &4.07$\pm$0.15 & 3.69$\pm$0.26  &3.35$\pm$0.30 \\
  \hline\hline 
  \end{tabular}
\label{tab:la111}
\end{center}
\end{table}

\begin{table}[htbp]
\newcommand\T{\rule{0pt}{2.6ex}}
\newcommand\B{\rule[-1.2ex]{0pt}{0pt}}
\caption{Measured values of the Lorentz angle in $<$100$>$ modules, in a 2~T magnetic field at the 
         average operational temperature in the period from 2011 to 2012, compared with the model 
         expectations of Becker et al.~\protect\cite{bib:Becker}. 
         The uncertainties on the measurements include both statistical and systematic contributions;
         those on the model predictions arise from uncertainties in the charge-carrier mobility.
         Barrel 4 has no $<$100$>$ modules.}
\begin{center}
\begin{tabular}{lccc}
  \hline\hline 
                               \multicolumn{4}{c}{$<$100$>$}\\
 \hline
  Layer \B \T &$T$ [\degr C]
                               &Measured $\theta_{\rm L} [^\circ]$
                               &Becker et al. Model $\theta_{\rm L} [^\circ]$  \\
 \hline
  Barrel 3 &--2.1              &  3.25$\pm$0.23 & 4.01$\pm$0.33 \\
  Barrel 5 &--1.3              &  3.28$\pm$0.15 & 4.00$\pm$0.32 \\
  Barrel 6 &\phantom{--}5.6    &  3.24$\pm$0.15 & 3.82$\pm$0.28 \\
  \hline\hline
  \end{tabular}
\label{tab:la100}
\end{center}
\end{table}

\begin{figure}[htbp]
 \centering
 \subfigure[]{\includegraphics[width=0.49\columnwidth]{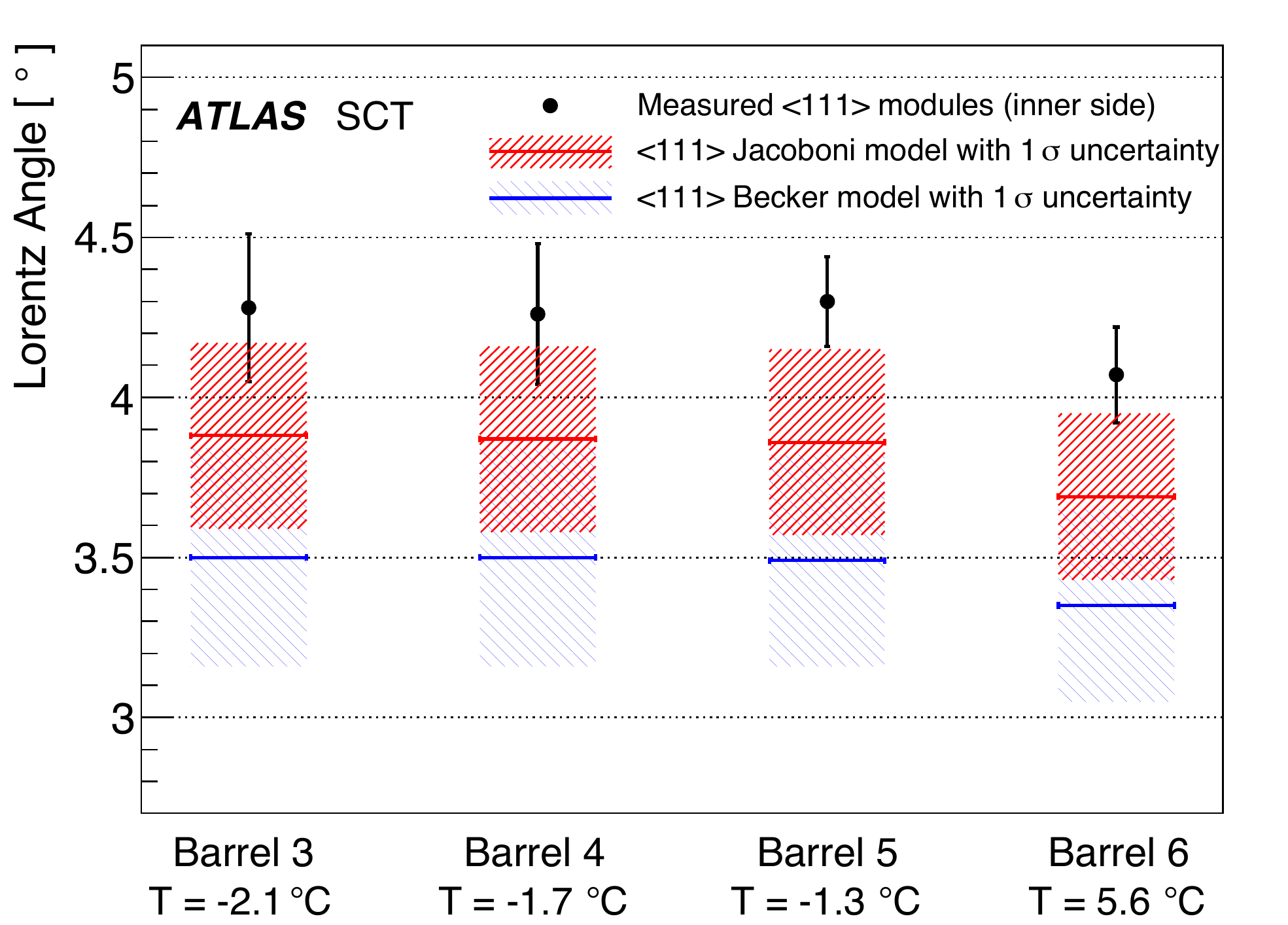}}
 \subfigure[]{\includegraphics[width=0.49\columnwidth]{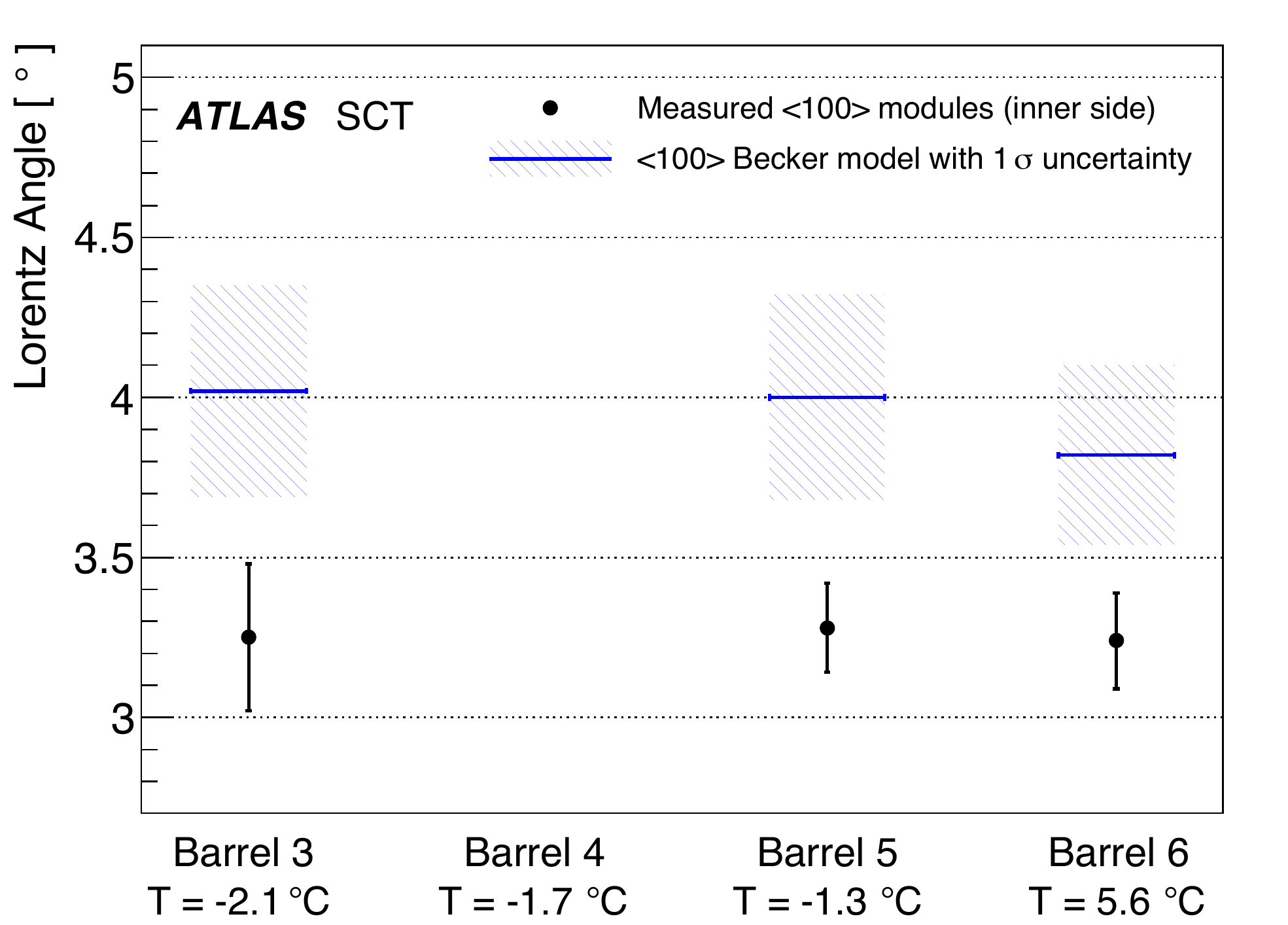}}
 \caption{Measured values of the Lorentz angle $\theta_{\rm L}$ in each layer of the SCT barrel 
          (black points) for (a) $<$111$>$ sensors and (b) $<$100$>$ sensors. The error bars include 
          the statistical and systematic uncertainties. 
          The red and blue lines show the expectations of the models of 
          refs.~\protect\cite{bib:Jacoboni} and~\protect\cite{bib:Becker} respectively,
          with the $1 \sigma$ uncertainties indicated by the hashed areas.
          The mean temperature of each layer is shown with the barrel number;
          a higher temperature is maintained in the outermost layer, barrel 6, resulting in a lower 
          expected Lorentz angle.}
 \label{fig:LAdataMC}
\end{figure}

\subsection{Energy loss and particle identification}
\label{sec:dedx}
Although the SCT is not designed to perform measurements of energy loss, ${\rm d}E/{\rm d}x$, and 
particle identification, some discriminating power is available from the number of time bins above 
threshold and the number of strips in a cluster. A more heavily ionising particle which deposits more 
charge will produce a larger pulse height, which is more likely to be above threshold in the time bin 
corresponding to the bunch crossing after the trigger. Moreover, a shorter path length is required
to deposit enough charge to be above threshold, so such a particle will tend to produce wider clusters.
Both of these expectations are exploited by calculating a quantity related to the energy loss of
a particle, ${\rm d}E/{\rm d}x_{\rm SCT}$, given by:
\begin{eqnarray}
{\rm d}E/{\rm d}x_{\rm SCT} = \frac{1}{N}\sum_{i}{w_{i}\cos\alpha_{i}}
\end{eqnarray}
where the sum runs over all the strips in all clusters assigned to the track, $N$ is the number of
clusters assigned to the track and $w_i$ is a weight depending on the number of time bins 
which are above threshold for that strip; by default $w_i$ is set equal to the number of time bins
above threshold, except for illegal time-bin patterns, which are given a weight of zero. 
The factor of $\cos\alpha_{i}$, where $\alpha_{i}$ is the angle between the track and the normal to the 
silicon sensor, corrects for the path length of the particle within the silicon.

Figure~\ref{fig:dedx_2d} shows ${\rm d}E/{\rm d}x_{\rm SCT}$ as a function of momentum multiplied by
charge for barrel-region tracks from minimum bias data collected in 2010. To increase the number of 
protons in the sample, the fraction of particles originating from secondary interactions in detector material 
is enhanced by selecting tracks with large impact parameters with respect to the primary interaction
vertex. The protons can clearly be seen in the figure, forming a distinct band of highly ionising 
particles in the low-momentum positive-charge region. This band is not seen for negative particles:
the number of anti-protons in the sample enhanced in the products of secondary interactions is expected to be 
much smaller.

\begin{figure}[htbp]
  \centering
  \subfigure[]{\includegraphics[width=0.49\columnwidth]{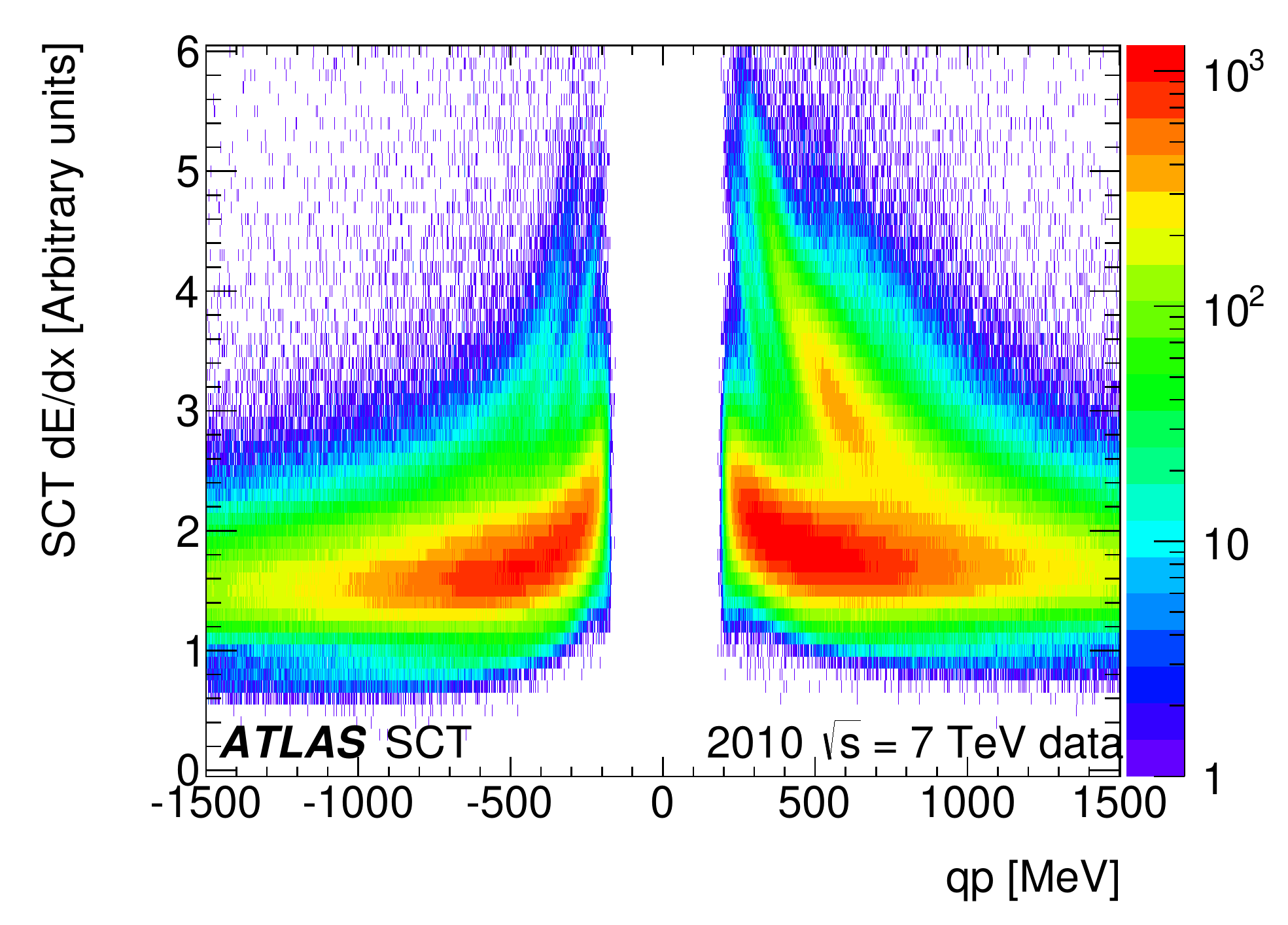} \label{fig:dedx_2d}}
  \subfigure[]{\includegraphics[width=0.49\columnwidth]{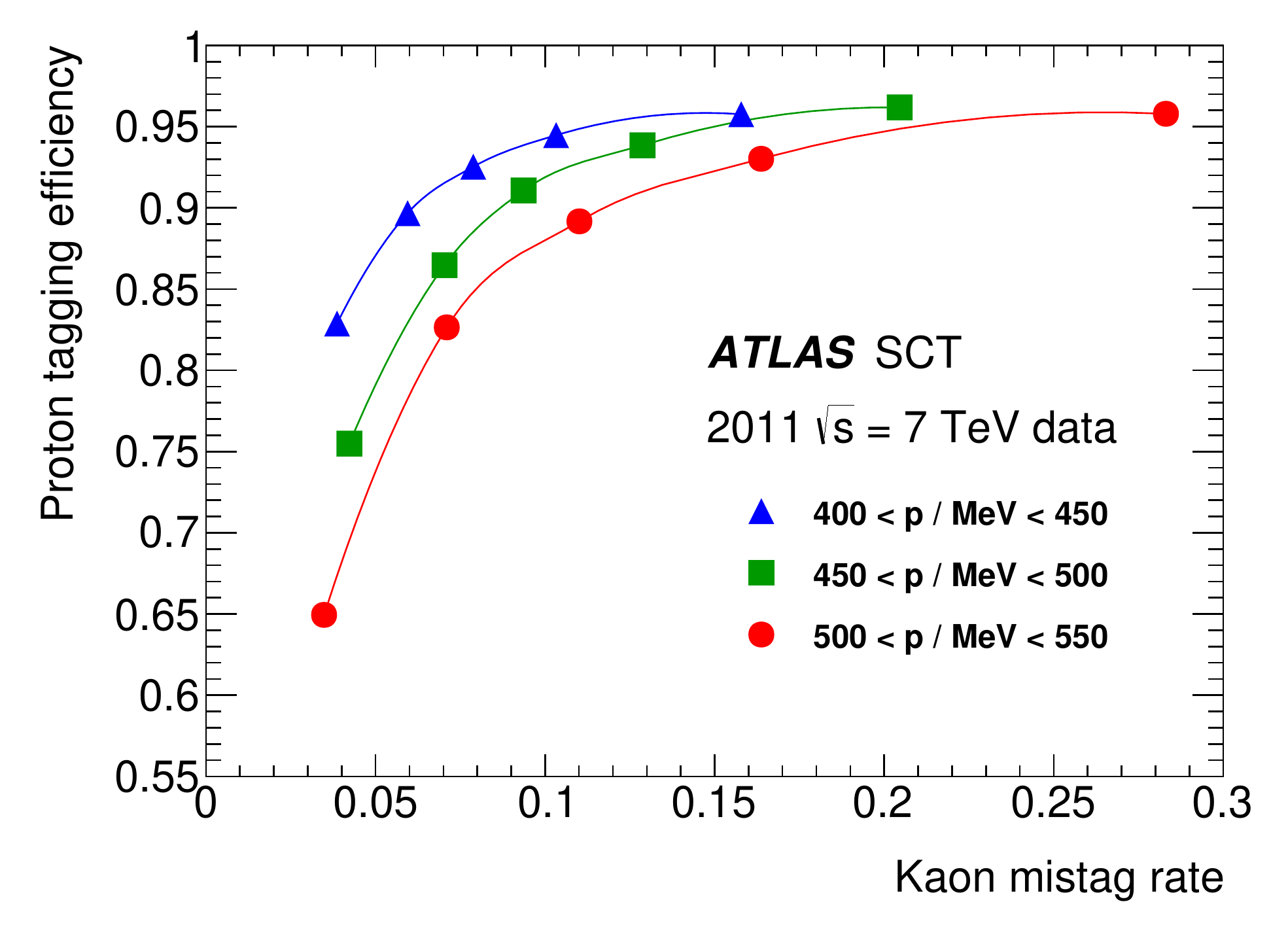} \label{fig:dedx_tag_eff_K}}
   \caption{(a) Energy loss measured in the SCT (see text for definition) as a function of momentum multiplied by charge for a
           sample of tracks enhanced in products of secondary interactions (see text). Tracks are 
           taken from 
           minimum bias data collected during 2010, and required to be in the barrel region of the
           detector.
           (b) Proton tagging efficiency as a function of the kaon mistag rate measured in 2011
            minimum bias data.}
\end{figure}

Particle identification is performed using a likelihood method. The probability density functions
for different particles are determined, in ranges of momentum, by fitting Gaussians to distributions 
of ${\rm d}E/{\rm d}x_{\rm SCT}$ for particles identified as protons, kaons or pions on the basis of 
${\rm d}E/{\rm d}x$ measured in the pixel detector~\cite{bib:Pixel_dEdx}. 
Figure~\ref{fig:dedx_tag_eff_K}
shows the efficiency for tagging protons, defined as the fraction of particles identified as protons
using the pixel detector which are tagged as protons using the SCT, as a function of the mistag rate for
kaons; the mistag rate for kaons is defined as the fraction of particles
identified as kaons by the pixel detector which are tagged as protons by the SCT. In the
momentum range 400--550~\MeV, a tagging efficiency for protons of more than 90\% can be
achieved, with a mistag rate of less than 30\% for kaons. The corresponding mistag rate for
pions is less than 4\%.

Although the SCT has been shown to have some discriminating power, its use for particle identification
is limited to momenta below about 600~\MeV. However, the peak position of the 
${\rm d}E/{\rm d}x_{\rm SCT}$ distribution for protons may become a useful tool for monitoring radiation 
damage to the sensors in future. The position of this peak was stable at the 5--10\% level during 
2010--2012 when radiation damage was negligible. Variations arising from drifts in detector timing 
relative to the ATLAS trigger time were observed. Future use of this peak for monitoring depends on the 
stability of the detector timing and threshold, and will require data recorded in level mode (see 
section~\ref{sec:DAQ}) with time-bin information.

\subsection{Measurement of $\delta$-ray production}
\label{sec:dray} 
Energy deposition by charged particles in silicon leads to the production of 
secondary electrons, called $\delta$-rays, that can travel distances of 
several hundred microns and produce secondary ionisation. These $\delta$-rays may 
give hits in neighbouring strips that were
not traversed by the primary particle, and thus broaden clusters and bias the
position measurement. It is thus important to check that the rate and spectrum
of $\delta$-rays are simulated correctly.
  
The production of $\delta$-rays in the barrel sensors has 
been measured and compared to simulation by counting the number of clusters that 
are broader than expected from the angle of the track traversing the sensor.
The measurement uses good-quality tracks with at least eight SCT hits from
about 257 million minimum-bias events, together with about 300 million
simulated events. Tracks sharing one or more clusters with another track
are removed from the analysis; the remaining contamination of clusters arising
from charge deposited by two or more primary particles is negligible. 

The idea is illustrated in figure~\ref{fig:dray_sketch}. Ignoring the effects of
charge diffusion, the expected width, $w_{\rm e}$, of a cluster from a track with 
incident angle $\phi_{\rm local}$ is given by
\begin{eqnarray}
w_{\rm e} = t (\tan\phi_{\rm local} - \tan\theta_{\rm L})
\end{eqnarray}
where $t$ is the sensor thickness and $\theta_{\rm L}$ is the Lorentz angle.
The emission of a $\delta$-ray may increase the observed width, $w_{\rm o}$, as
illustrated in figure~\ref{fig:dray_sketch}. To identify $\delta$-rays, 
clusters for which $w_{\rm e} < 80~\mu{\rm m}$ are selected. The primary charge
from such a particle can extend over at most two strips, and observed clusters
larger than this arise primarily from $\delta$-ray production. 
The production of
a single $\delta$-ray adds strips to one side of the cluster, leading to a
shift in the cluster centroid, and a shift in the track residual\footnote{
The cluster under consideration is removed from the track fit to remove any bias
in the residual.}
 of approximately $(w_{\rm o} - w_{\rm e})/2$.
An example residual distribution for clusters of width six strips is shown in 
figure~\ref{fig:dray_fit}(a). The peak at $\sim$0.2~mm corresponding to single
$\delta$-ray production is superimposed on a background centred around zero
which arises from effects such as multiple $\delta$-ray production.    
The number of single $\delta$-rays is estimated by fitting such residual
distributions to the sum of two Gaussian functions: one for the signal peak 
with a mean away from zero and one for the background with a mean near zero.
The choice of function is empirical, and is found to model the distributions
well, giving $\chi^2 / d.o.f$ values between 1.5 and 1.8.

\begin{figure}[htb]
\begin{center}
\includegraphics[width=0.5\columnwidth]{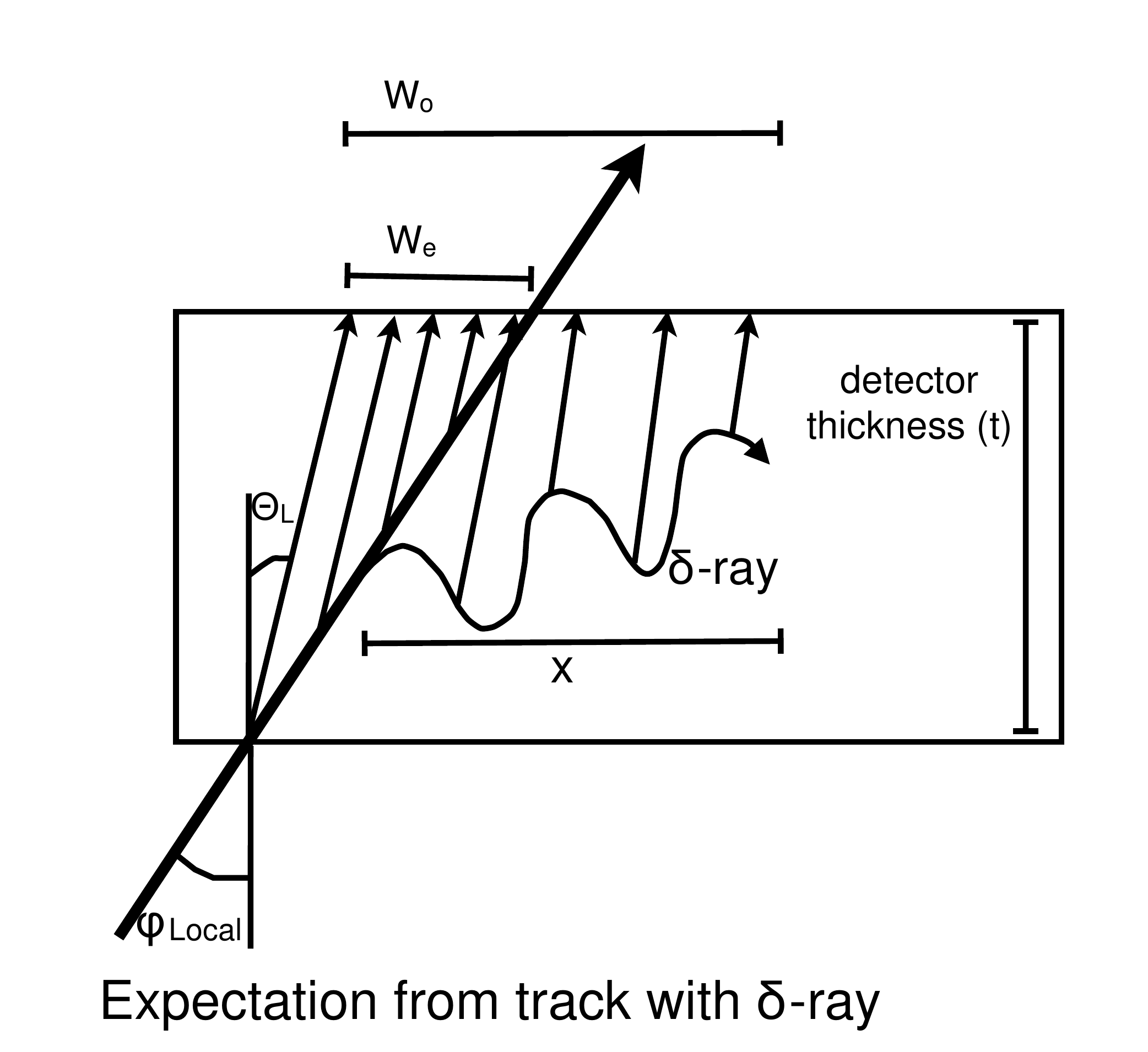}
\caption{Sketch of the geometric meaning of expected cluster width $w_{\rm e}$ and 
         observed width $w_{\rm o}$. 
         Only the two-dimensional projection of the
         three-dimensional path length $L$ is shown. Here, $x$ is the
         distance traversed by a particular $\delta$-ray as described in
         equation~(\protect\ref{eqn:dray1}).}
\label{fig:dray_sketch}
\end{center}
\end{figure}

In order to quantify $\delta$-ray production and study the dependence on
various parameters, the differential probability for a particle to emit a
$\delta$-ray that travels a distance $x$ in the $r$--$\phi$ plane (i.e. 
perpendicular to the strip direction) is modelled as:
\begin{eqnarray}
\label{eqn:dray1}
{\rm d}P_{\delta} = \frac{AL}{\rho}e^{-x/\rho}{\rm d}x
\end{eqnarray}
where $\rho$ is the range (in the $r$--$\phi$ plane) of the $\delta$-rays, to be 
determined. The integrated probability from $x = 0$ to $x = \infty$ is the
normalisation $AL$, where the dependence on path length in silicon, $L$, is
explicitly shown since $\delta$-ray production must scale with $L$.
The range $\rho$ is obtained by performing Gaussian fits to the residual
distributions to determine the
number of single $\delta$-rays in bins of observed width, and fitting the
resulting numbers to an exponential distribution. An example, for tracks
with momentum $p >$ 1.5~\GeV\ and path length in silicon $320 < L < 380~\mu{\rm m}$,
is shown in figure~\ref{fig:dray_fit}(b).
 
\begin{figure}
 \centering
 \subfigure[]{\includegraphics[width=0.49\columnwidth]{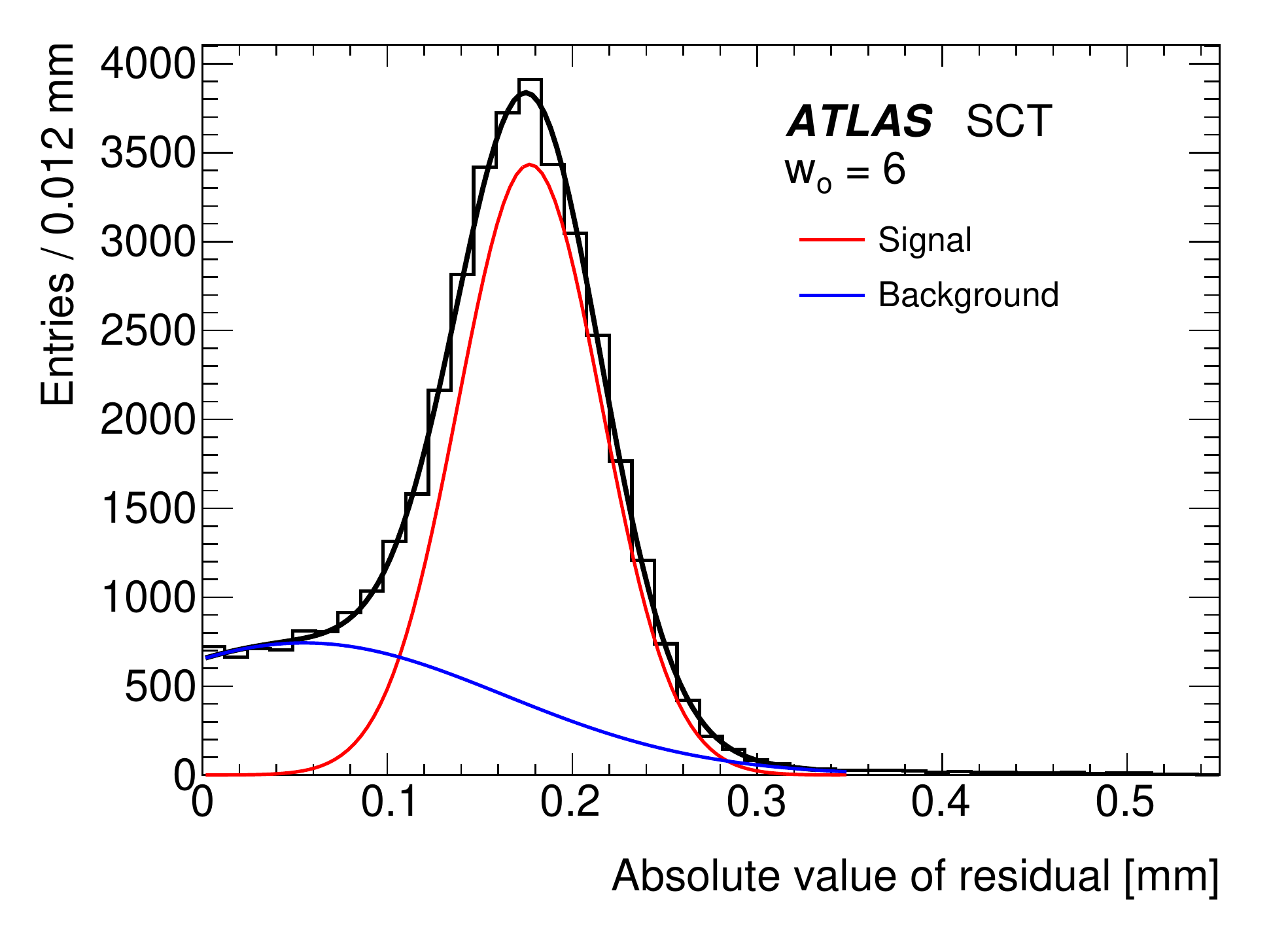}}
 \subfigure[]{\includegraphics[width=0.49\columnwidth]{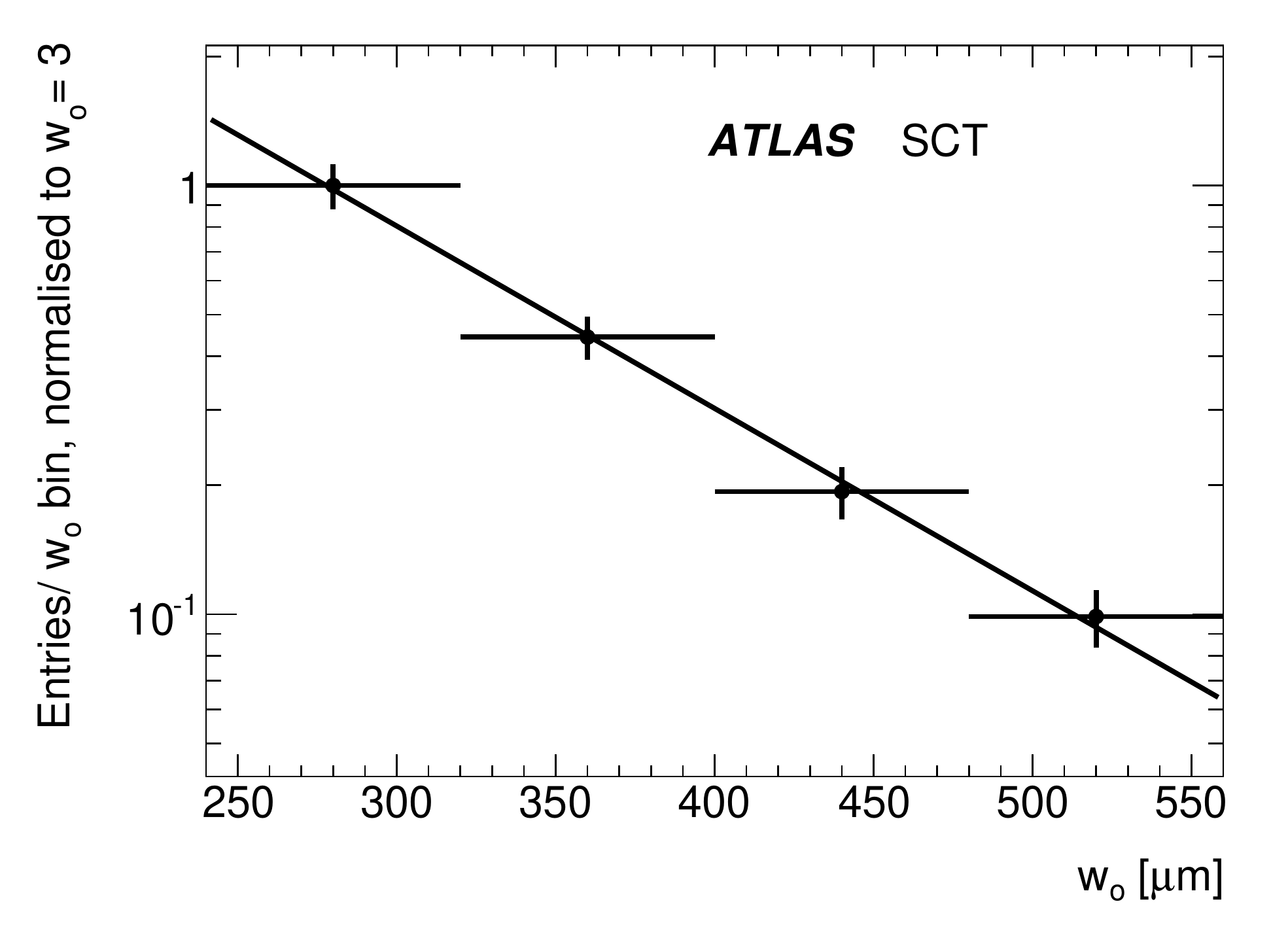}}
 \caption{(a) Distribution of residual magnitude in data for clusters 
         with expected width $w_{\rm e} < 80~\mu{\rm m}$, observed
         width $w_{\rm o}$ = 6 strips, track path length in silicon between 275~$\mu$m 
         and 450~$\mu$m and track momentum $p > 1$~\GeV. The curves show the fit to the 
         sum of two Gaussian distributions.
         (b) Distribution, normalised to the first bin, of the $\delta$-ray
         signal (from fits to residual peaks) in bins of $w_{\rm o}$
         for tracks with $p >$ 1.5~\GeV\ and path length in silicon between 
         320~$\mu$m and 380~$\mu{\rm m}$.
         The fit to an exponential function is superimposed.}
 \label{fig:dray_fit}
\end{figure}

The rate of $\delta$-ray production, $A$, is obtained by integrating 
equation~(\ref{eqn:dray1}) to obtain the expected number of single $\delta$-rays
in three consecutive $w_{\rm o}$ bins, $w_{\rm o} =$ 4, 5 and 6 strips. Clusters
with $w_{\rm 0} = 3$ strips are not used because the signal peak is not
well resolved leading to large uncertainties, while for larger values of
$w_{\rm o}$ the statistics are limited. 

The range and rate parameters, $\rho$ and $A$, are determined separately for
each barrel, in bins of $L$ and track momentum. 
Values of both range and rate measured in the separate barrels and in 
different ranges of $L$ are consistent, and are combined. 

The principal
systematic uncertainty on these measurements arises from modelling of the
background with a single Gaussian distribution, estimated to be 5\% on both
range and rate. An additional 3.5\% arises from comparison of the measured
values of $\rho$ for different barrels, $L$ and momentum ranges. The 
estimation of $A$ from the integral of equation~(\ref{eqn:dray1}) makes use
of the fraction of particles crossing one rather than two strips, which is
not directly observed, but is approximated as the ratio of one-strip to
two-strip clusters. This approximation, which is valid if the $\delta$-ray
production rate is low, adds an uncertainty of 2\% to the rate. The overall
systematic uncertainties on the range and rate measurements are 6.1\% 
and 6.4\% respectively.

Figure~\ref{fig:dray_results} shows the measured values of the rate of $\delta$-ray production
and their range as a function
of track momentum. Data and simulation are seen to be in good agreement,
validating the description of $\delta$-ray production in the simulation. 
The rate is expected to decrease with increasing momentum by about 12\% over the range
studied. While the measurements are consistent with this expectation, the
errors are too large to resolve this variation. By integrating equation~(\ref{eqn:dray1}),
the overall probability of a particle emitting a $\delta$-ray with a range of more than
40~$\mu$m (half the strip pitch) when traversing a typical silicon thickness of 300~$\mu$m
is found to be 1\%.
    
\begin{figure}
 \centering
 \subfigure[]{\includegraphics[width=0.49\columnwidth]{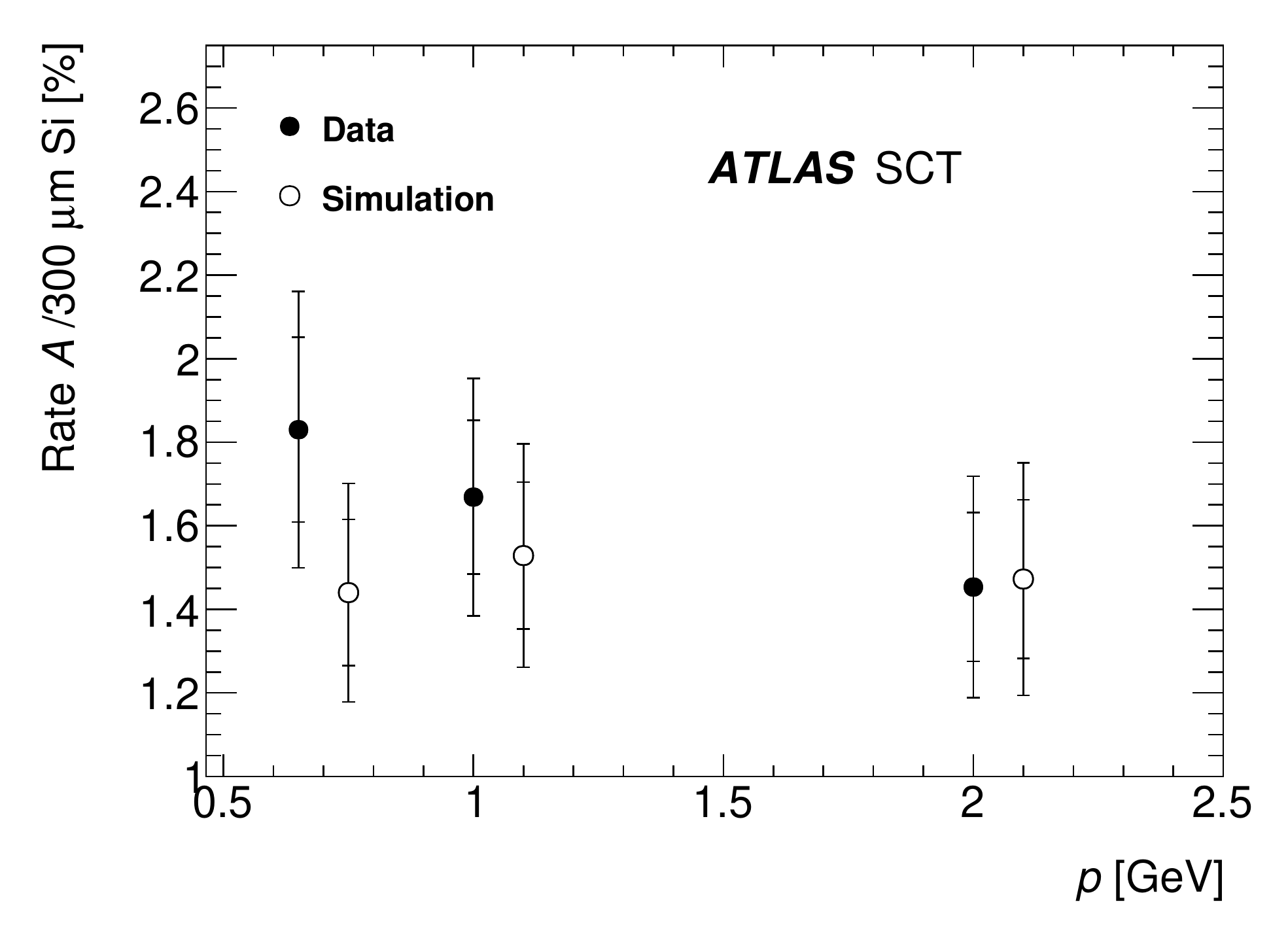}}
 \subfigure[]{\includegraphics[width=0.49\columnwidth]{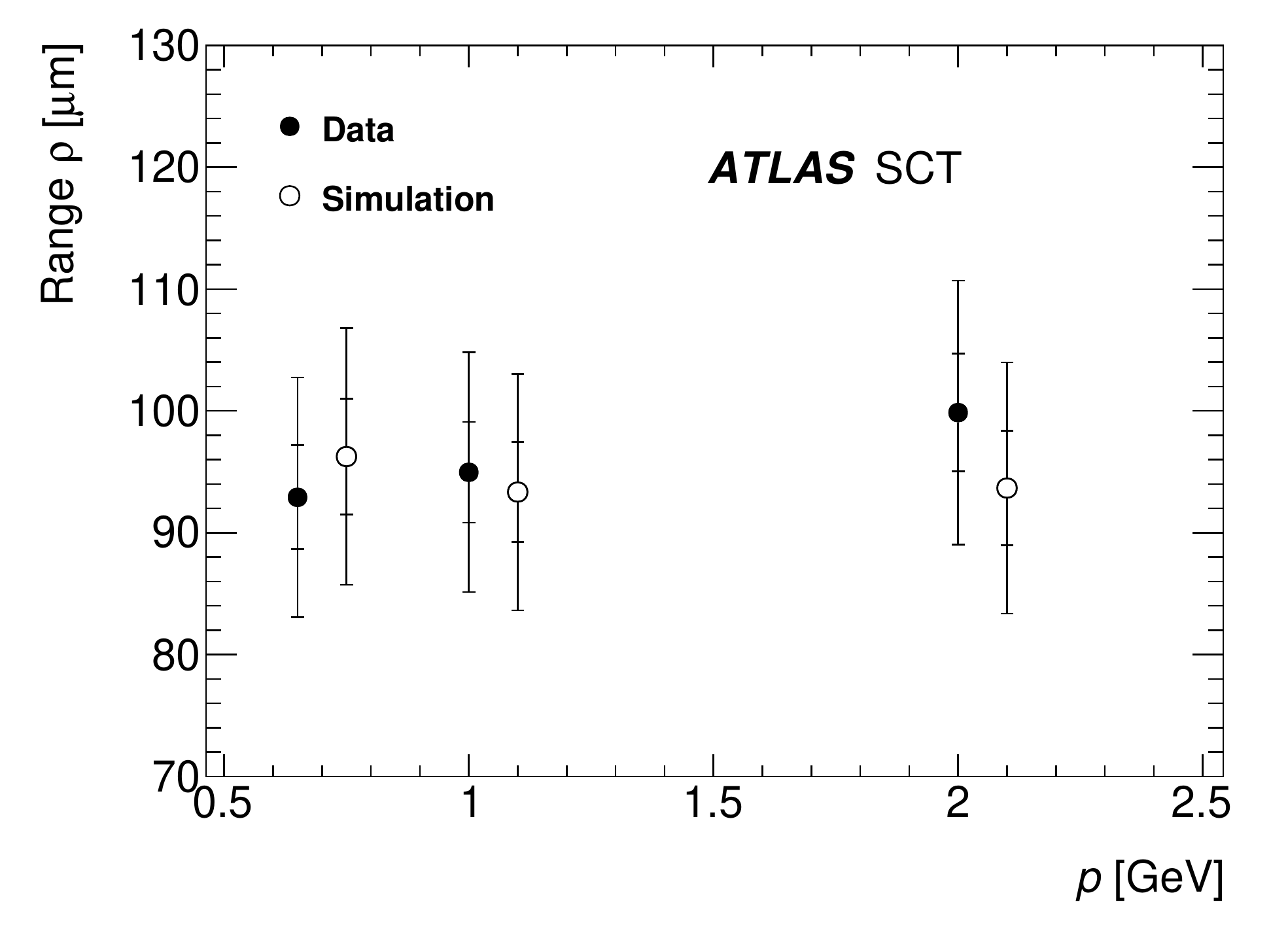}}
 \caption{Measured values of (a) rate per 300~$\mu$m path length and (b) range
         for $\delta$-ray production in silicon as a function of track
         momentum, $p$. The simulation points are artificially shifted by 0.1~\GeV\
         for clarity.}
 \label{fig:dray_results}
\end{figure}

\section{Radiation effects}
\label{sec:radiation}
The radiation environment in the ATLAS inner detector is complex and comprises a full spectrum of particles (pions, protons, neutrons, photons, etc.), with energies ranging from \TeV\ down to thermal in the case of neutrons. Close to the interaction point the environment is dominated by particles coming directly from the proton--proton collisions, but at larger radii albedo particles from high-energy hadron and electromagnetic cascades in the calorimeters dominate the radiation backgrounds in the inner detector. Advanced Monte Carlo event generator and particle transport codes are required to simulate these environments. A review of the radiation backgrounds expected in and around the ATLAS detector can be found in ref.~\cite{bib:ATLASDetectorPaper}. Results from the most recent simulations for the inner detector are given in section\,\ref{Rad:sec-radiation}.

The deleterious effects of radiation on the SCT include: increasing leakage currents; charge accumulation in silicon oxide layers; single-event upsets; decreasing signal-to-noise; changing depletion voltages; and radiation-induced activation of components. 
Typically both sensors and readout electronics are affected, but in the case of single-event upsets only the readout 
system is impacted.
The SCT system was designed to operate for fluences and doses corresponding to an integrated luminosity of 700\,fb$^{-1}$ at a centre-of-mass energy of 14~\TeV. 
Up to December 2012 an integrated luminosity of $\sim$\,29\,fb$^{-1}$ had been delivered to the ATLAS experiment, so the effects of radiation on the detectors are expected to be small. However, the large number of SCT modules 
has allowed robust statistical analyses to be performed. This is particularly true for the detector leakage-current measurements, described in section\,\ref{Rad:sec-leakage}, which in turn were used to verify the Monte Carlo fluence predictions. Some degradation effects, such as the change in detector depletion voltage 
due to radiation-induced impurity states in the silicon, are being observed but are small and still to be studied in detail. 
Fluence and dose measurements in the inner detector volume were also obtained using a dedicated
online radiation monitoring system, described in section\,\ref{Rad:sec-radmon}. 
The particular importance of the radiation monitoring system is to provide ionising-dose information, which is not available from SCT measurements and is important for predicting degradation of the read-out chips.

Single-event upsets have impacted SCT operations, and studies and measurements to understand these effects are described in section\,\ref{sec:SEU}. Some radiation effects were unexpected, such as the anomalous leakage-current increase observed in some CiS sensors leading to additional noise and DAQ issues. The impact of radiation effects on SCT operation and their mitigation is discussed in section\,\ref{Rad:sec-impact}.

\subsection{Simulations and predictions}
\label{Rad:sec-radiation}
Radiation background predictions in the inner detector have been performed with the FLUKA particle transport 
code~\cite{bib:FLUKA1,bib:FLUKA2}. A detailed description of the geometry and material of the ATLAS detector,
shielding and beam-line is constructed within the FLUKA framework. Simulations are performed using the same 
criteria as described in refs.~\cite{bib:ATLASDetectorPaper,bib:RadiationTaskForce}, which also give 
definitions and explanations of the radiation quantities used.
The PYTHIA8 event generator~\cite{bib:PYTHIA8} was used to simulate the inelastic proton--proton collisions, for both $\sqrt{s}$ = 7~\TeV\ and 8~\TeV\ corresponding to 2011 and 2012 running respectively. In past studies, the PHOJET~\cite{bib:PHOJET} event generator was used, but PYTHIA8 is now the preferred choice to take advantage of the significant effort by its authors and the experiments to model and tune PYTHIA8 to describe the $\sqrt{s}$ = 7--8~\TeV\ LHC collision data. The predictions for the fluences in the inner detector are typically $\sim$\,5\% higher with PYTHIA8 than with PHOJET. A proton--proton inelastic cross section of 71.4\,mb at $\sqrt{s}$ = 7~\TeV\ is predicted by PYTHIA8, consistent with measurements~\cite{bib:rad-xsAtlas,bib:CMS_sigtot,bib:TOTEM_sigtot,bib:ALICE_sigtot}.

The 1~\MeV\ neutron-equivalent fluences~\cite{bib:RadiationTaskForce}  for $\sqrt s$\,=\,7~\TeV, 
normalised to one fb$^{-1}$, are shown in 
figure\,\ref{Rad:fig-ID-FLUKA}. The rise in the contours towards the endcaps is due to the increasing 
fluence of albedo neutrons from secondary interactions in the endcap calorimeter material.  
Average values of the fluences by region are given in table\,\ref{Rad:tab-flukasct}. The barrel values 
are averaged over the full length in $z$ of the layers, while the endcap values are averaged over each 
ring of a disk (see figure~\ref{detector:Quadrant}). 
Statistical uncertainties on the predicted fluences are typically a few percent.
Comparisons were also made with simulations from the ATLAS GEANT4 framework. Although higher fluence 
values were predicted, differences were typically less than 30\%\,\cite{bib:rad-Vertex2012}.
Table\,\ref{Rad:tab-fluka-radmon} shows the corresponding FLUKA predictions for 1~\MeV\ 
neutron-equivalent fluence, ionising dose and thermal neutron fluence at the positions of the four 
radiation monitors, the locations of which can be seen in figure\,\ref{Rad:fig-ID-FLUKA}. 

\begin{figure}[htb]
  \centering
   \includegraphics[width=0.9\textwidth]{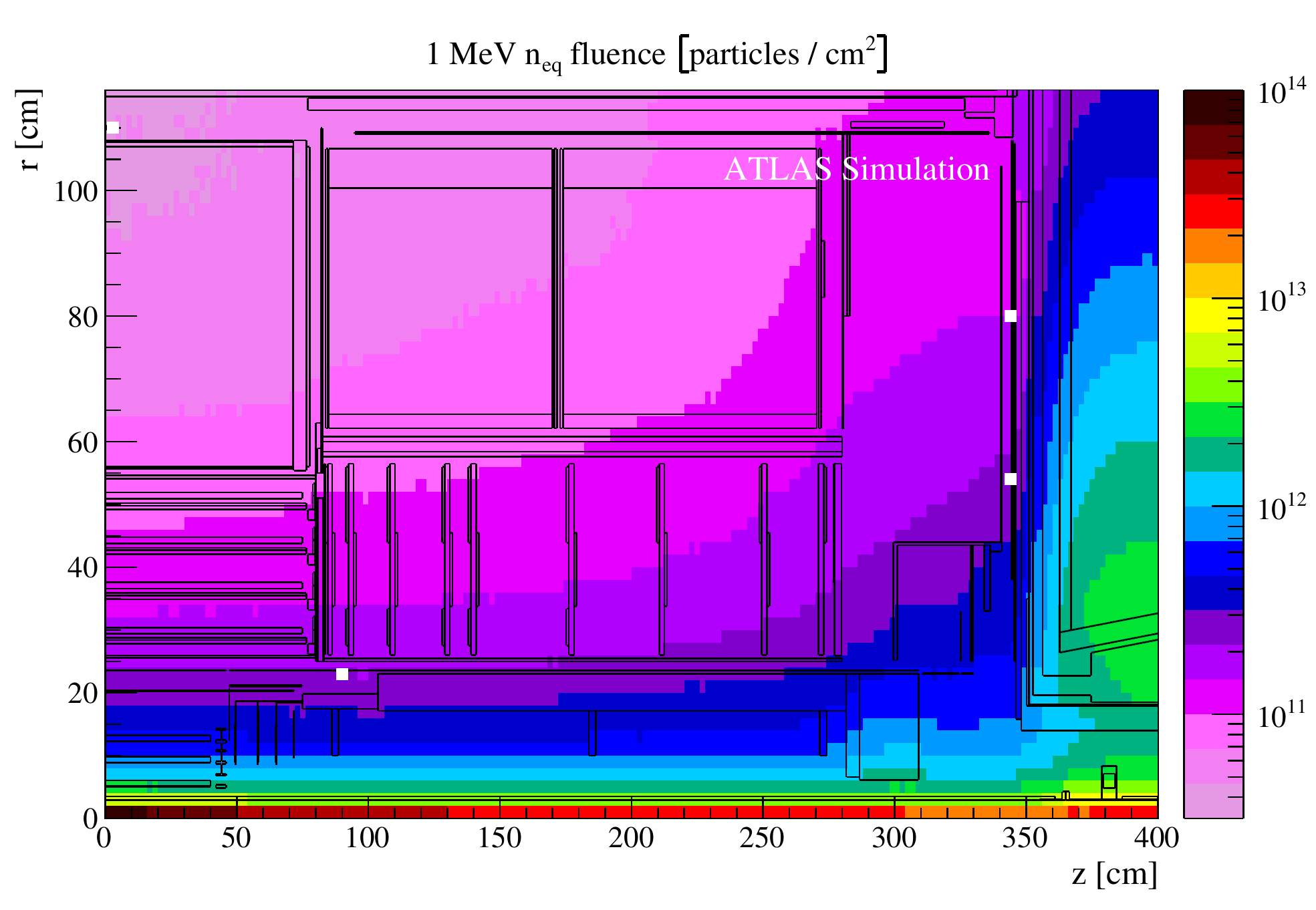}
  \caption{Quarter slice through the inner detector region of the geometry constructed in the FLUKA simulations. The colour 
                 contours represent the 1~\MeV\ neutron-equivalent fluences, corresponding to an integrated luminosity of 1\,fb$^{-1}$. 
                The four white squares show the positions of the radiation monitors.
}
  \label{Rad:fig-ID-FLUKA}
\end{figure}

\begin{table}[htb]
  \caption{FLUKA predictions ($\sqrt s$\,=\,7~\TeV) for 1~\MeV\ neutron-equivalent fluences in the SCT. 
           The radiation backgrounds originating from proton--proton collisions are symmetric in $\pm$\,$z$ 
           in the inner detector. Statistical uncertainties are typically 3--5\%. Fluences and doses at 
           $\sqrt s$\,=\,8~\TeV\ are predicted to be $\sim$\,5\% higher than for 
           $\sqrt s$\,=\,7~\TeV\,\cite{bib:rad-Vertex2012}.}
  \begin{center}
    \begin{tabular}{ccccccc}
      \hline \hline
      \multicolumn{7}{c}{1~\MeV\ fluences per fb$^{-1}$ ($\times$\,10$^{11}$)  [cm$^{-2}$]} \\
      &&\phantom{00}&\multicolumn{4}{c}{Endcap} \\
      Barrel & & & Disk  & Outer & Middle & Inner \\
      \hline 
        3 & 1.66 && 1 &  1.04 & 1.29  &   --      \\
        4 & 1.30 && 2 &  1.03 & 1.26  & 1.66 	\\
        5 & 1.07 && 3 &  1.02 & 1.27  & 1.66 	\\
        6 & 0.92 && 4 &  1.05 & 1.30  & 1.68	\\
	  &      &    & 5 &  1.08 & 1.32  & 1.73 	\\
	  &      &    & 6 &  1.15 & 1.41  & 1.81 	\\
	  &      &    & 7 &  1.26 & 1.54  &   --	\\
	  &      &    & 8 &  1.45 & 1.79  &   --	\\
	  &      &    & 9 &  1.66 &   --     &   --	\\
     \hline \hline
    \end{tabular}
  \label{Rad:tab-flukasct}
  \end{center}
\vspace{1cm}
  \caption{FLUKA predictions ($\sqrt s$\,=\,7~\TeV) for 1~\MeV\ neutron-equivalent fluences, ionising dose 
           and thermal neutron fluence at the four different radiation-monitor locations; the positions of
           these are given in the second and third columns. Fluences and doses at $\sqrt s$\,=\,8~\TeV\ are 
           predicted to be $\sim$\,5\% higher than for $\sqrt s$\,=\,7~\TeV.}
 \begin{center}
    \begin{tabular}{lccccc}
      \hline \hline
     &&&  \multicolumn{3}{c}{Fluence and ionising-dose predictions per fb$^{-1}$} \\
    Location &$r$ [cm] &$z$ [cm] & \multicolumn{1}{c}{$\phi_{\rm 1MeV}$\,[cm$^{-2}$]} & \multicolumn{1}{c}{Dose [Gy]} & \multicolumn{1}{c}{$n_{\rm thermal}$\,[cm$^{-2}$]} \\
      \hline 
	Pixel support tube &23  &90  & $2.3\times$\,10$^{11}$ &  109  &  $0.63\times$\,10$^{11}$  \\
	ID end plate       &54  &344 & $2.4\times$\,10$^{11}$ &   48  &  $1.02\times$\,10$^{11}$  \\
	ID end plate       &80  &344 & $1.5\times$\,10$^{11}$ &   23  &  $0.87\times$\,10$^{11}$  \\
	Cryostat wall      &110 &1   & $0.46\times$\,10$^{11}$ &  5.1 &  $0.40\times$\,10$^{11}$  \\
     \hline \hline
    \end{tabular}
  \label{Rad:tab-fluka-radmon}
  \end{center}
\end{table}

\subsection{Detector leakage currents}
\label{Rad:sec-leakage} 
The increase in detector leakage current is proportional to the 1~\MeV\
neutron-equivalent fluence and is sensitive to temperature and annealing
effects. Leakage-current predictions are obtained from two models, the
Hamburg/Dortmund model~\cite{Rad:ref-mollPhd, Rad:ref-kraselPhd} and the 
Harper model~\cite{Rad:ref-harperPhd}, each using different assumptions for the 
temperature-dependent annealing behaviour (see appendix~\ref{LCmodels} for details).
Both models include self-annealing effects with various time constants. 
As a consequence, the history of sensor temperature must be carefully tracked 
including warm-up periods, even if they are short, and this is described in 
section\,\ref{Rad:sec-temp}. The sensor temperature during the periods without cooling 
was assumed to be the same as the enviromental gas temperature, which was 17.5$^{\circ}$C 
during the 2011--2012 winter shutdown.

The largest uncertainty in the leakage-current predictions comes from the 1~\MeV\ neutron-equivalent fluence 
simulation, so the comparison of measurement with prediction is an important check of the fluence simulations. 

\subsubsection{Temperature and leakage-current measurement}
\label{Rad:sec-temp} 
The measured leakage-current values $I_{\rm HV}$ are normalised to those at a temperature of $0^{\circ}$C, 
$I_{\rm norm}$, using the temperature scaling formula for the silicon bulk generation current:
\begin{equation}
\left(\frac{I_{\rm norm}}{I_{\rm HV}}\right) = \left(\frac{T_{\rm norm}}{T_{\rm meas}}\right)^2\exp\left[-\frac{E_{\rm gen}}{2k_{\rm B}}\left(\frac{1}{T_{\rm norm}}-\frac{1}{T_{\rm meas}} \right)\right]
\label{tempcorrection}
\end{equation}

\noindent
where $T_{\rm meas}\,(T_{\rm norm})$ is the sensor (normalisation) temperature, 
$E_{\rm gen}$ the effective 
generation energy of 1.21~\eV~\cite{Rad:ref-Chilingarov} 
and $k_{\rm B}$ is Boltzmann's constant. 
The sensor temperature is deduced from 
the temperature measured by thermistors mounted 
on the hybrid circuits. 
For the barrel modules, FEM thermal simulations 
predict the average sensor temperature to be 
3.7$\pm$1.0$^{\circ}$C below the thermistor temperature,
and to vary by only $\sim$0.2$^{\circ}$C across
the sensor. Predicted temperature differences for
the endcap sensors have large uncertainties. 
Furthermore a systematic difference in the raw leakage current 
between the two sides A and C exists, the reason for which is 
currently under investigation. Therefore
results are presented below for the barrel only.  

Figure~\ref{barrel_temperature} shows module-by-module 
distributions of high voltage, hybrid temperature and 
raw leakage current per module as well as the 
normalised leakage current per unit volume 
at 0$^{\circ}$C for the innermost barrel of
the SCT. Periodic increases in raw leakage current,
which arise from higher temperatures in one cooling loop,
disappear after temperature normalisation, resulting in
a distribution which is quite flat along $z$, a
reflection of the flat pseudorapidity distribution of secondary 
particles in minimum-bias events. Similar flat distributions
are seen in all other barrel layers.

\begin{figure}[htb]
\begin{center}
\includegraphics[width=\textwidth]{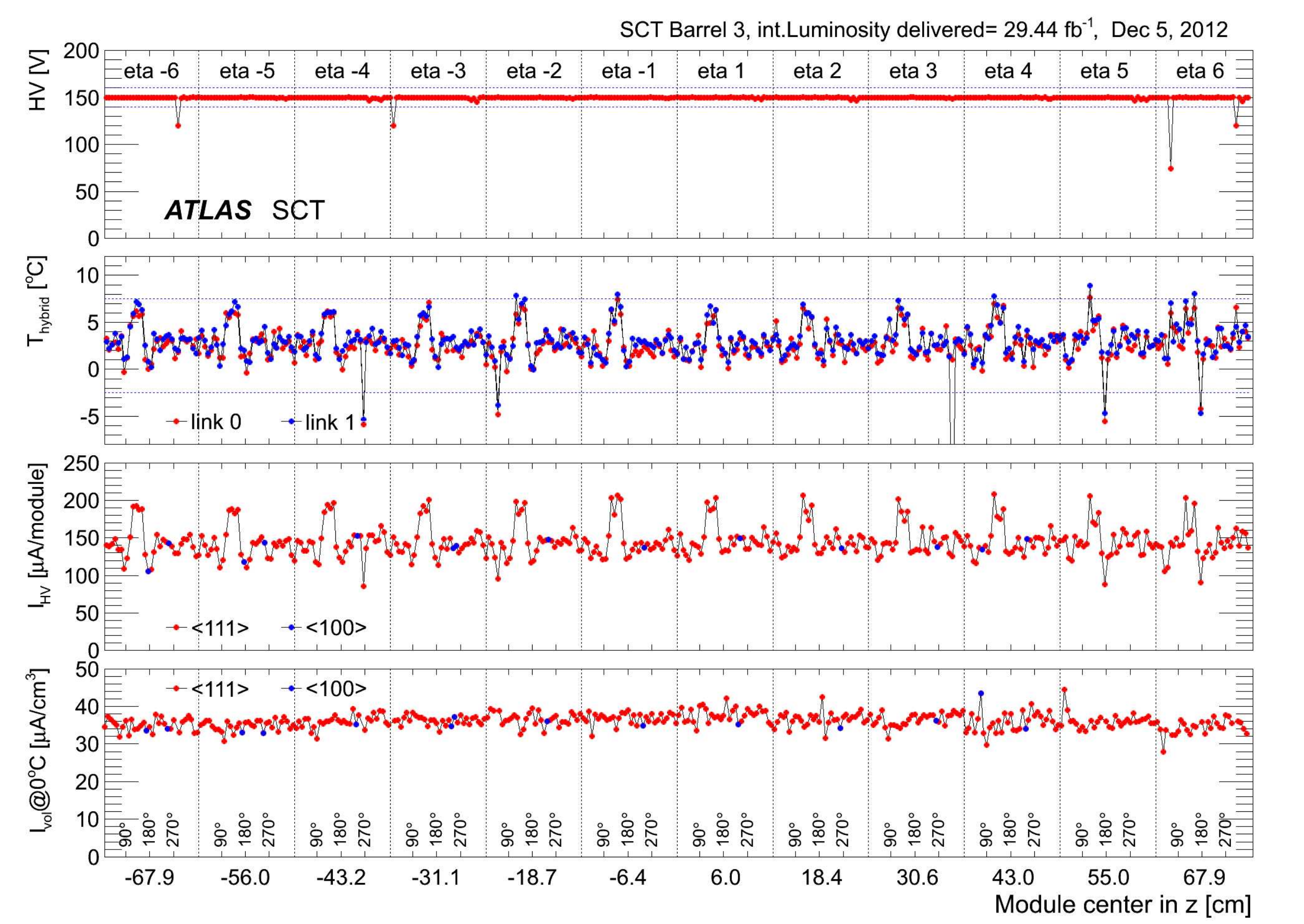}
\end{center}
\caption{From top to bottom: high voltage (HV), hybrid temperature ($T_{\rm hybrid}$), 
         measured leakage current ($I_{\rm HV}$) and normalised leakage current per unit
         volume at 0$^{\circ}$C ($I_{\rm vol}$) for each module of barrel 3 as of December 2012.
         The temperature of each module is measured by two thermistors; values from those mounted 
         on the inner (outer) sides of modules are shown in blue (red).}
\label{barrel_temperature}
\end{figure}

\begin{figure}
\begin{center}
\includegraphics[width=\textwidth]{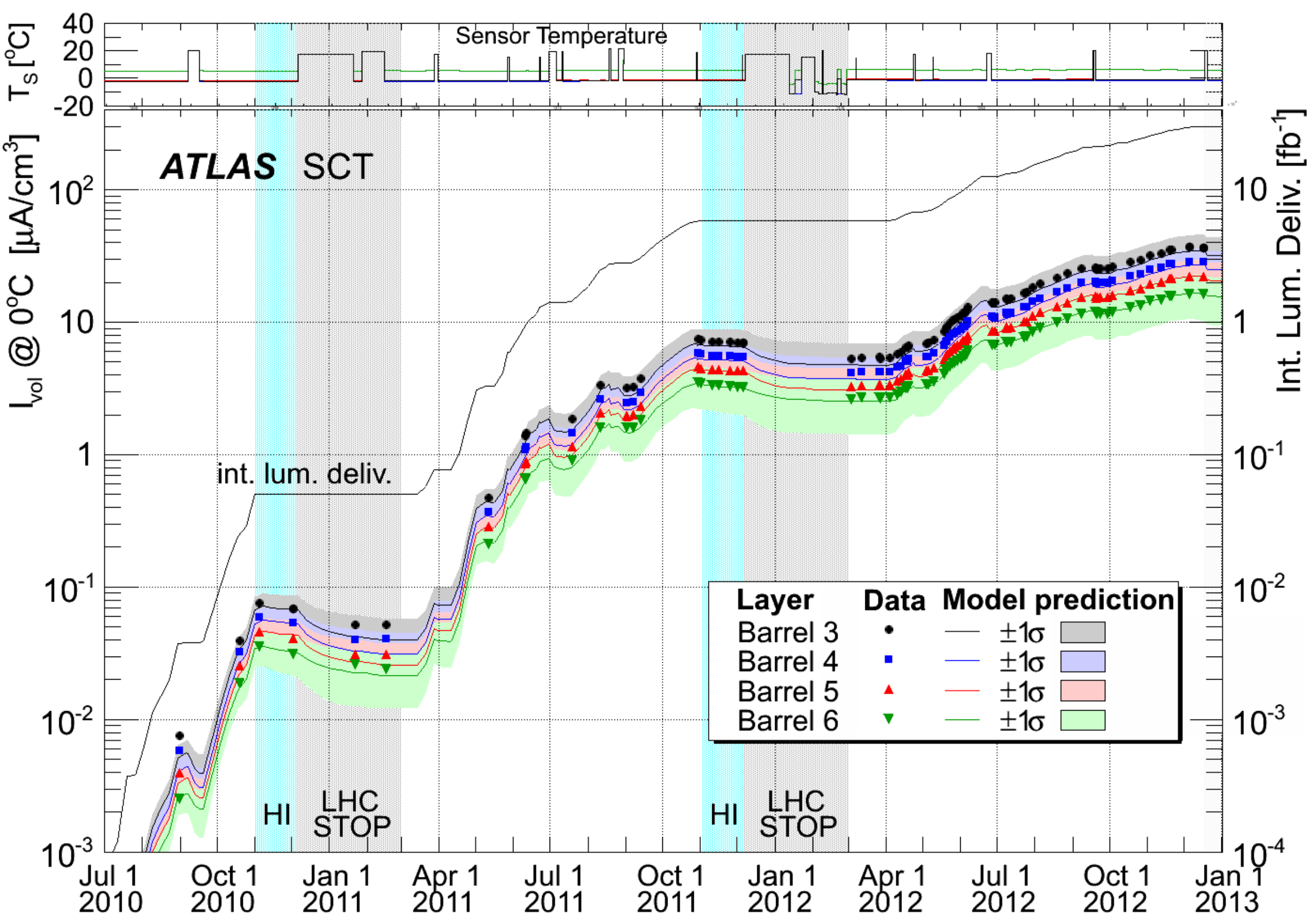}
\end{center}
\caption{Comparison between data (points) and Hamburg/Dortmund model 
predictions (lines with uncertainties shown by the coloured bands)
of the leakage current per unit volume at 0$^{\circ}$C ($I_{\rm vol}$)
of the four barrel layers. The integrated luminosity and the average sensor 
temperatures are also shown. The blue shading and label `HI' indicate
periods of heavy-ion running, while extended periods with no beam in 
the LHC during which the SCT was off are shaded grey.}
\label{Rad:fig-logplot}
\end{figure}

\subsubsection{Leakage-current evolution and comparison with predictions}
In 2010 and 2011, when the leakage currents were small, electrical noise from the FSI alignment system 
interfered with the measurements,  and the study of the leakage currents was limited to periods when the FSI system was off.
No such restrictions were necessary in 2012. 

Figure~\ref{Rad:fig-logplot} shows the average 
leakage current of each barrel layer compared with  
the predictions of the Hamburg/Dortmund model~\cite{Rad:ref-mollPhd, Rad:ref-kraselPhd}.
The coloured bands indicate $1\sigma$ 
uncertainties on the model predictions. They were calculated by 
varying each parameter of the model by $\pm1\sigma$ and 
then adding the deviations in quadrature. 
Uncertainties on temperature measurements ($\pm1^{\circ}$C) 
and delivered luminosities ($\pm$3.7\%) are also taken into account;
the uncertainty on the FLUKA fluence simulations is not included. 
The Harper model~\cite{Rad:ref-harperPhd} predicts quite similar profiles but with about 15\% 
larger values systematically.
The total uncertainties on the model predictions are at the level of $\pm$20\%. 
The major uncertainties come from two parameters ($k_{\rm 0I}$ and $E_{\rm I}$)
in the Hamburg/Dortmund model (see appendix~\ref{LCmodels} for details), but from one current-damage
constant ($A_{1}$) with an annealing time of 833 days at $-7^{\circ}$C
in the Harper model.
For all barrel layers, the data are within the uncertainties 
of the model predictions
over a period of 2.5 years and four orders of magnitude in leakage current. 
This strongly suggests that the radiation fluences are well described by
the FLUKA predictions. 

\subsection{Online radiation monitor measurements}
\label{Rad:sec-radmon}
In addition to the SCT leakage-current measurements, the online radiation monitoring system also offers important information on fluences and doses in the inner detector volume. 
Details can be found in refs.~\cite{bib:radmon_ref1,bib:radmon_ref2} and only a brief description is given here. 
The radiation-monitor packages provide measurements of non-ionising energy loss (NIEL), total ionising dose (TID) and thermal neutron fluence. 
NIEL is monitored using {\it p-i-n} diodes and two methods are employed. The first measures the increase in the forward voltage for a given forward current and is used for 
determining NIEL in the fluence range $10^{9}$ to $5\cdot10^{12}$~n/cm$^{2}$. This method exploits the increase in silicon resistivity 
from radiation due to degradation in minority-carrier lifetime. 
The second method measures the increase in leakage current in a way similar to the SCT silicon sensors. 
TID is measured with radiation-sensitive p-MOSFET transistors (RadFETs). Ionising radiation creates electron--hole pairs in the silicon oxide; positive charge is trapped in the gate 
oxide, and an increasing negative gate voltage must be applied for a given drain current. The increase in gate voltage provides a measure of the total ionising dose. The thickness
of the oxide layer defines the sensitivity and range of doses that can be measured and different RadFETs are used to cover the required dynamic range. 
Measurements of TID are of particular interest for predicting degradation of the SCT front-end chips, and the radiation monitors provide 
the only online information about TID near to the SCT.
Thermal neutron fluence is estimated from the decrease in common-emitter-current 
gain of {\it n-p-n} bipolar transistors, which are the same as
those used in the SCT read-out chips. The gain in these transistors is
degraded due to bulk damage caused by fast hadrons and by thermal
neutrons. Since the 1~\MeV\ equivalent fluence of fast hadrons is known from
measurements with diodes, the contribution of thermal neutrons can be
estimated from the gain degradation~\cite{bib:radmon_ref1}.

The measured voltages and leakage currents are converted to fluences and doses using parameterisations based on calibrated 
irradiation data~\cite{bib:radmon_ref1}. 
Figure \,\ref{Rad:fig-comp} shows measurements of TID and NIEL at the four sets of locations 
shown in figure\,\ref{Rad:fig-ID-FLUKA} and listed in table \ref{Rad:tab-fluka-radmon}. 
The NIEL measurements for the pixel support tube and inner detector end-plate locations are obtained from the reverse current increase, whereas 
the values for the cryostat wall are obtained using the forward bias method.
The measurements are average values from sensors at the same $r$ and $|z|$ placed at different azimuth angles $\phi$. The shaded bands 
represent the r.m.s.\ of the measurements with a 20\% systematic error added in quadrature. 
The systematic error is dominated by the sensor calibration precision from the irradiations and the measured sensor temperature.
The predicted fluence and dose are shown as dotted lines and are obtained by multiplying the values from table~\ref{Rad:tab-fluka-radmon} by the delivered integrated luminosity at a certain date. 
Very good agreement between measurement and simulation can be seen, especially for TID. 

\begin{figure}[htp]
 \centering
 \subfigure[]{\includegraphics*[width=0.49\textwidth]{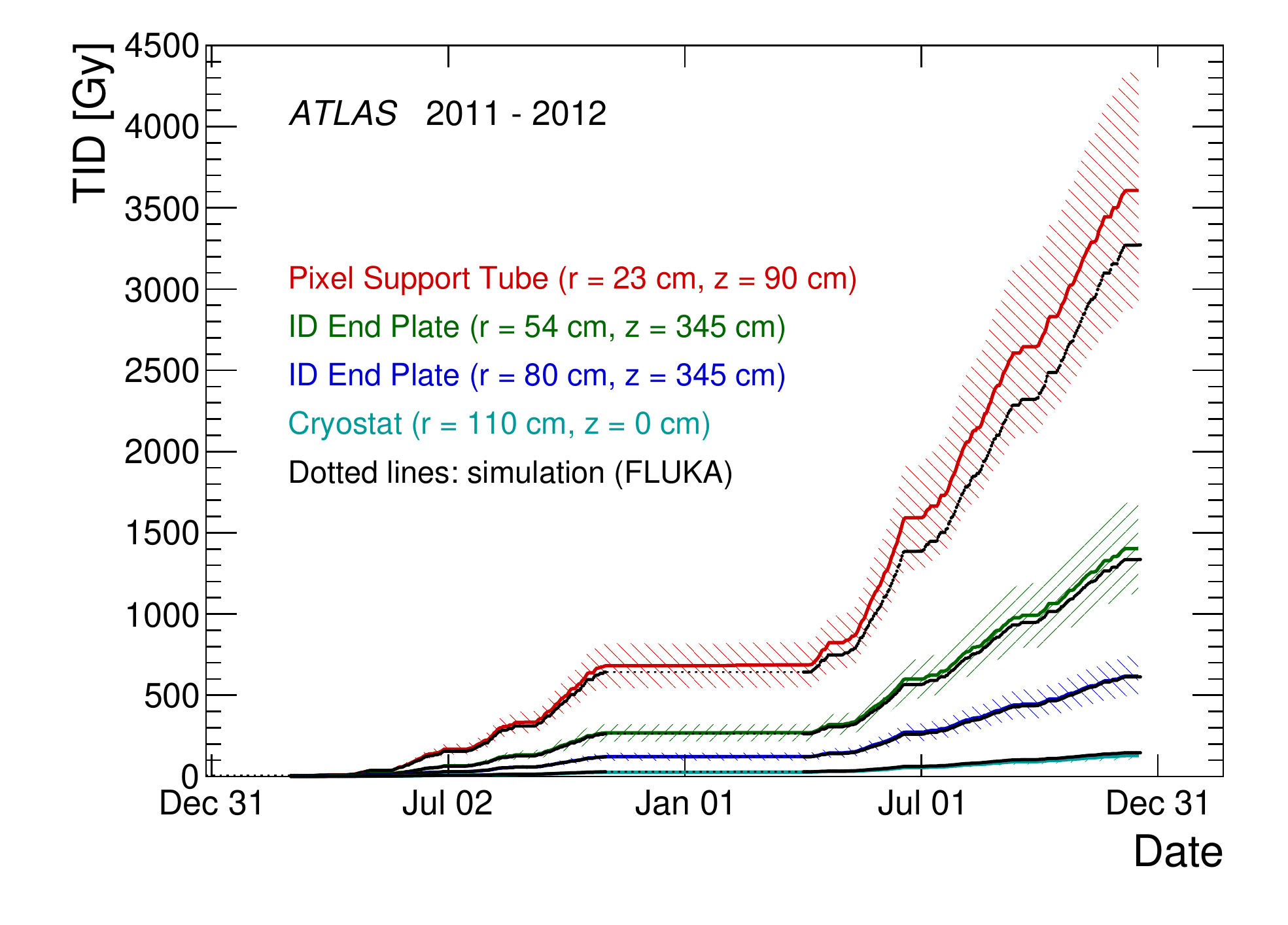}}
 \subfigure[]{\includegraphics*[width=0.49\textwidth]{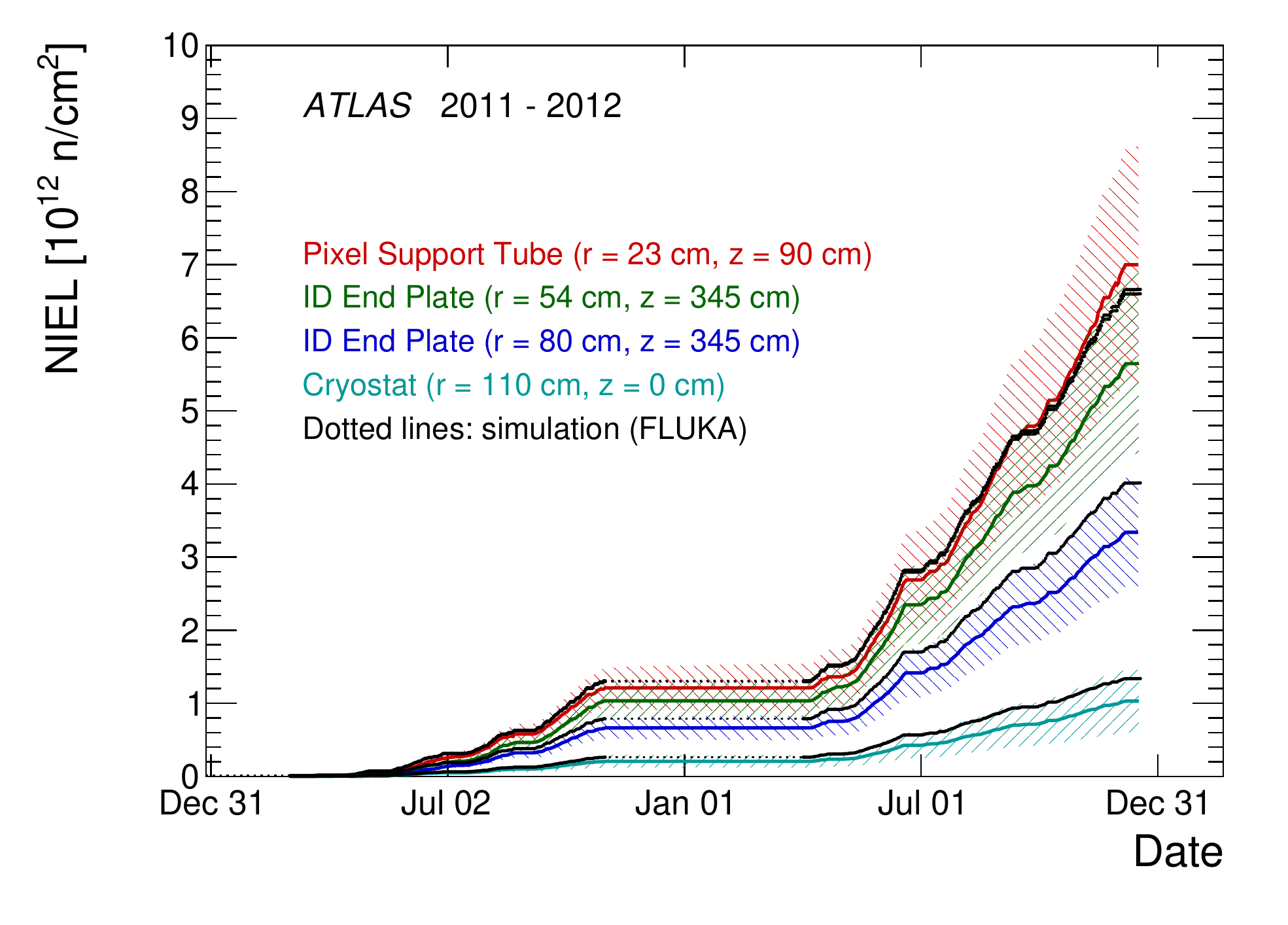}}
 \caption{(a) Total ionising dose (TID) and (b) non-ionising energy loss (NIEL) in 1~\MeV\ 
          neutron-equivalent fluence measured at four different sets of locations in the inner detector. 
          Each location is drawn with a different colour. Coloured bands represent measurement 
          uncertainties. The simulated values at each location are shown with black dotted lines. 
          The simulated values of NIEL for the pixel support tube and ID end plate ($r$ = 54 cm) are 
          very similar.} 
\label{Rad:fig-comp}
\end{figure}

The base current in transistors increases with increasing integrated luminosity. At the end of 2012
it has increased by about 10 nA from an initial value of 70 nA (at 10 $\mu$A collector current) in the most exposed transistors on the pixel support tube and inner detector end-plates.
This increase is consistent with a thermal neutron fluence of the order of $10^{12}$~n/cm$^{2}$, which is in agreement with
the FLUKA calculations in table~\ref{Rad:tab-fluka-radmon}.

\subsection{Single-event upsets}
\label{sec:SEU}
As well as causing damage to the sensors and on-detector readout electronics, radiation can also cause `soft errors', or single-event upsets (SEUs), in the SCT on-detector electronics. SEUs can induce bit flips in the static registers of the ABCD chips, and can also affect the on-detector {\it{p-i-n}} diodes. In both cases clear SEU signals have been identified in data from the barrel modules, as summarised below; more details can be found in ref.~\cite{bib:ATLAS_SEU}.
 
The registers in the ABCD chips are not protected against SEU, and test-beam studies have measured a small but non-zero SEU cross-section~\cite{bib:Eklund_SEU}. In the SCT, a direct read-back of register values is not possible, so the analysis of SEU rates uses the sudden increase in occupancy from individual chips which can arise when a bit is flipped in the threshold registers. In particular, the threshold registers use an 8-bit DAC, with the fifth bit usually set to one. When this bit is flipped, the chip outputs the maximum number of strips (128) for each trigger. A few problematic chips which gave spurious SEU-like noise bursts were removed from the analysis.
A verification of the SEU hypothesis is provided by studying the correlation between SEU rate and module occupancy. Figure~\ref{fig:SEU_ABCD} shows the SEU rate as a function of cluster occupancy in a module; occupancy is proportional to flux. The data show the expected linear behaviour. The apparent saturation observed at high occupancy partially arises from the requirement 
in the analysis that there be at most one SEU per chip per run.

\begin{figure}[h]
 \centering
 \subfigure[]{\includegraphics[width=0.49\textwidth]{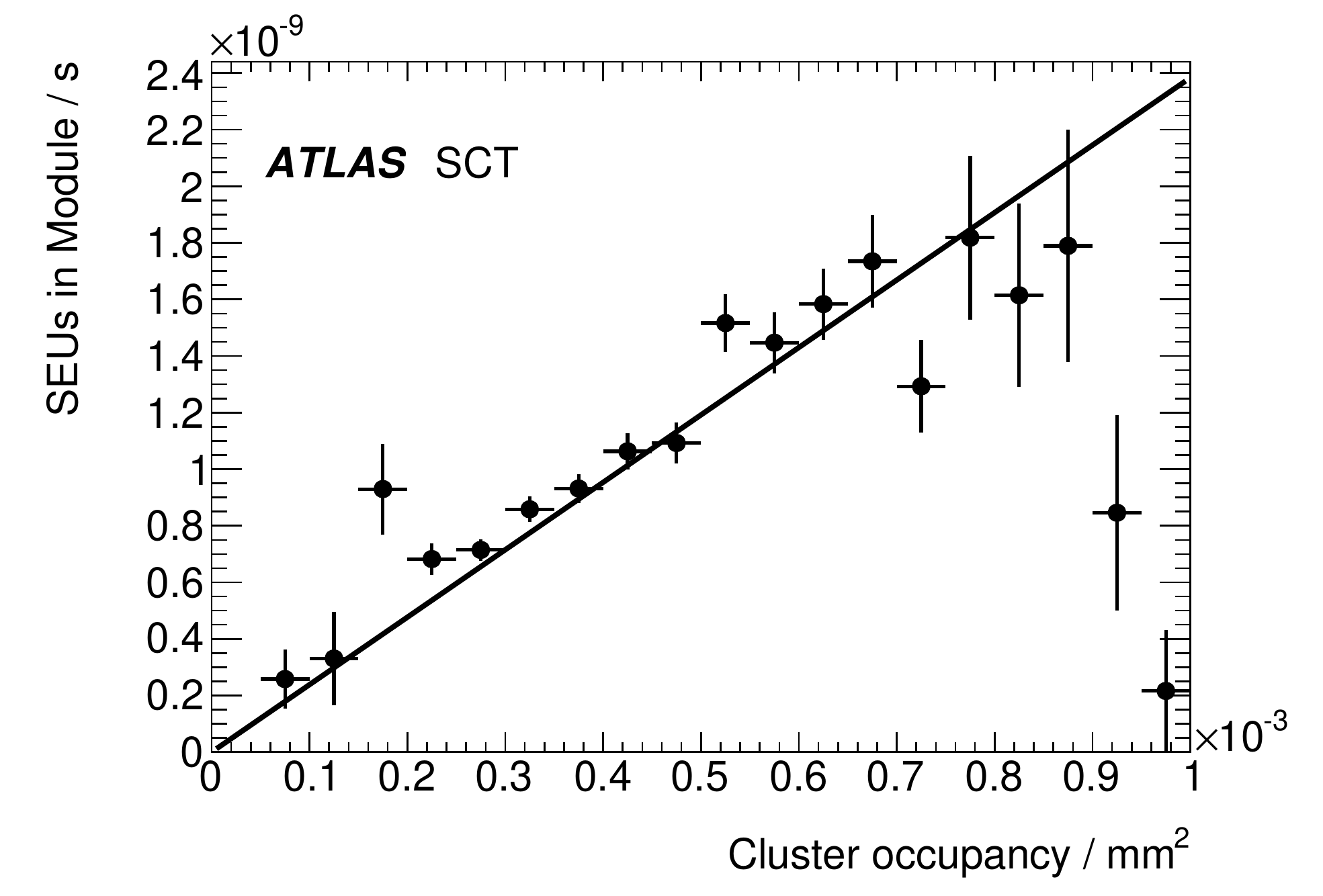} \label{fig:SEU_ABCD}}
 \subfigure[]{\includegraphics[width=0.49\textwidth]{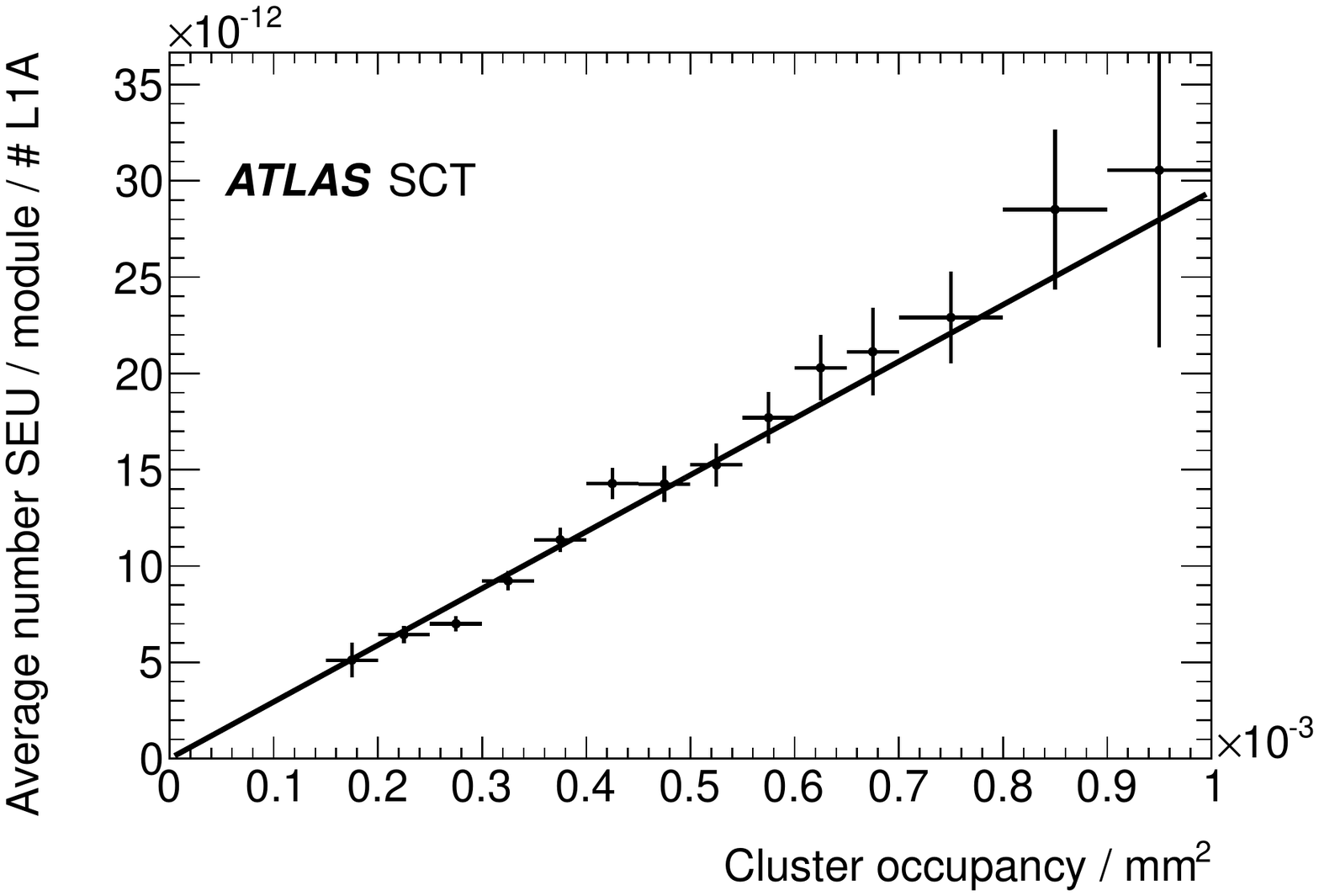} \label{fig:avgNrSEU_vs_occu}}
 \caption{(a) SEU rate in the ABCD DAC register versus module cluster occupancy in a run. 
          (b) Mean number of SEUs per module per level-1 trigger (L1A) in a run versus cluster occupancy in that module and run. 
         Note that in each case same module contributes to several bins depending on the cluster occupancy recorded in different runs. The lines show the results of linear fits through the origin.}
\end{figure}

Another measurable effect of SEUs is the desynchronisation of individual modules from the rest of the ATLAS readout. Triggers are transmitted to the modules via optical links, and
test-beam studies have demonstrated that an SEU arising from a relatively small energy deposition in the receiving {\it p-i-n} diode can cause a bit flip in a discriminator~\cite{bib:Dowell_SEU} and may lead to a missing trigger. Desynchronisation is flagged when the trigger count in the ROD differs from the trigger count received from the module, and the mismatch persists until the module is reset. Figure~\ref{fig:avgNrSEU_vs_occu} shows the average number of SEUs per module per level-1 trigger in a run versus cluster occupancy for that module in the run. This shows the expected linear trend, confirming that these errors are genuine SEUs.
Further confirmation of the SEUs hypothesis comes from the observation that the number of SEUs in a module correlates with decreasing {\it p-i-n} diode current, as expected and measured in the test-beam studies~\cite{bib:Dowell_SEU}.

\subsection{Impact of radiation on detector operation and its mitigation}
\label{Rad:sec-impact}
The SCT power-supply system incorporates adjustable trip limits for various parameters which are set to ensure that faults are quickly identified and addressed. Most parameters are stable, but as described in section~\ref{Rad:sec-leakage}, significant increases in 
leakage currents were observed in the SCT throughout 2011 and 2012. 
Throughout 2010, module trip limits were maintained at 5\,$\mu$A, which was comfortably above leakage 
currents for unirradiated modules, but low enough to flag power-supply issues.
The limits were increased to 50\,$\mu$A and then 100\,$\mu$A in 2011 and 2012 respectively, to track the evolution of
the current. 

During running early in 2012 a significant fraction (27\%) of 
the modules with CiS-manufactured sensors started to exhibit 
anomalously high and varying leakage currents 
after a few hours of proton--proton collisions. 
Figure~\ref{cis_current} shows the leakage-current 
profile with time for one such module, illustrating
the leakage current with the module at standby (50\,V)
and fully on (150\,V).
In this example the current increases significantly during the course of 
each run; at the same time there was an abnormally high increase in noise 
sufficient to provoke a ROD busy signal and prevent data taking.
The behaviour was mitigated by reducing the HV by typically 30--40\,V specifically for the modules associated with the high noise; for those modules the HV still remained above the highest depletion voltage of the sensors within the module so that hit efficiency was not impacted.

Empirically it was found that setting all CiS modules to 5\,V during standby (instead of 50\,V) prevented the more severe current increases associated with the noise problems, though the current profile during the run still varied widely.
The modules which had previously provoked ROD busy signals continued to be operated with the reduced HV.
Figure~\ref{cis_current_Hamamatsu-CiS} shows typical
beam-time behaviours of HV and current in late 2012. 
Modules with Hamamatsu sensors (left) show the 
expected flat current profiles 
while those with CiS sensors (right) exhibit
varying current behaviours during beam time.
The last run in these plots corresponds to a day-long
calibration run with no beam and all currents stayed constant.
The cause of the current anomalies has not been 
understood, though it is clearly correlated with 
the presence of radiation. The behaviour may evolve as the sensors experience further radiation damage in future.

\begin{figure}[htb]
\begin{center}
\subfigure[]{\includegraphics[width=0.49\textwidth]{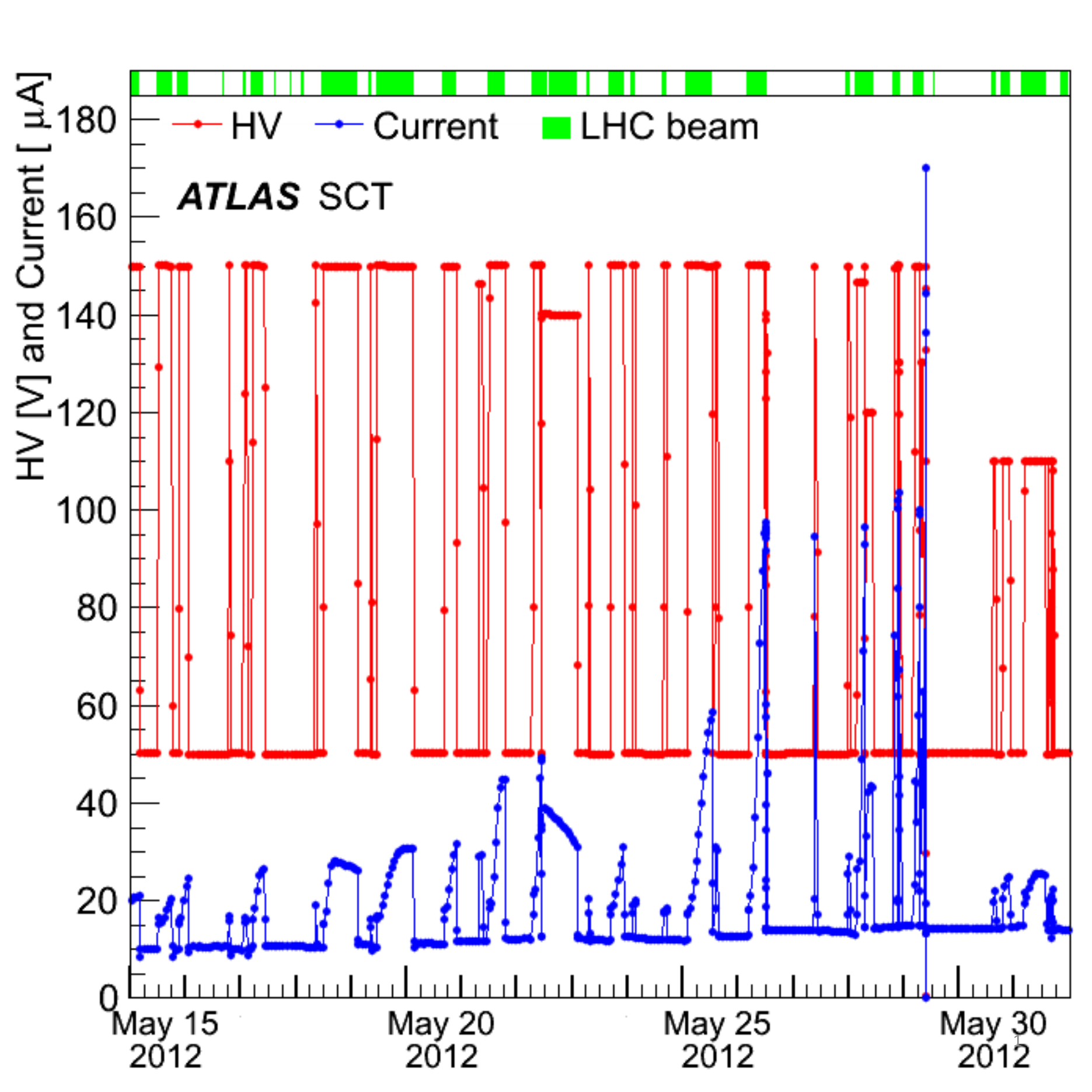} \label{cis_current}}
\subfigure[]{\includegraphics[width=0.49\textwidth]{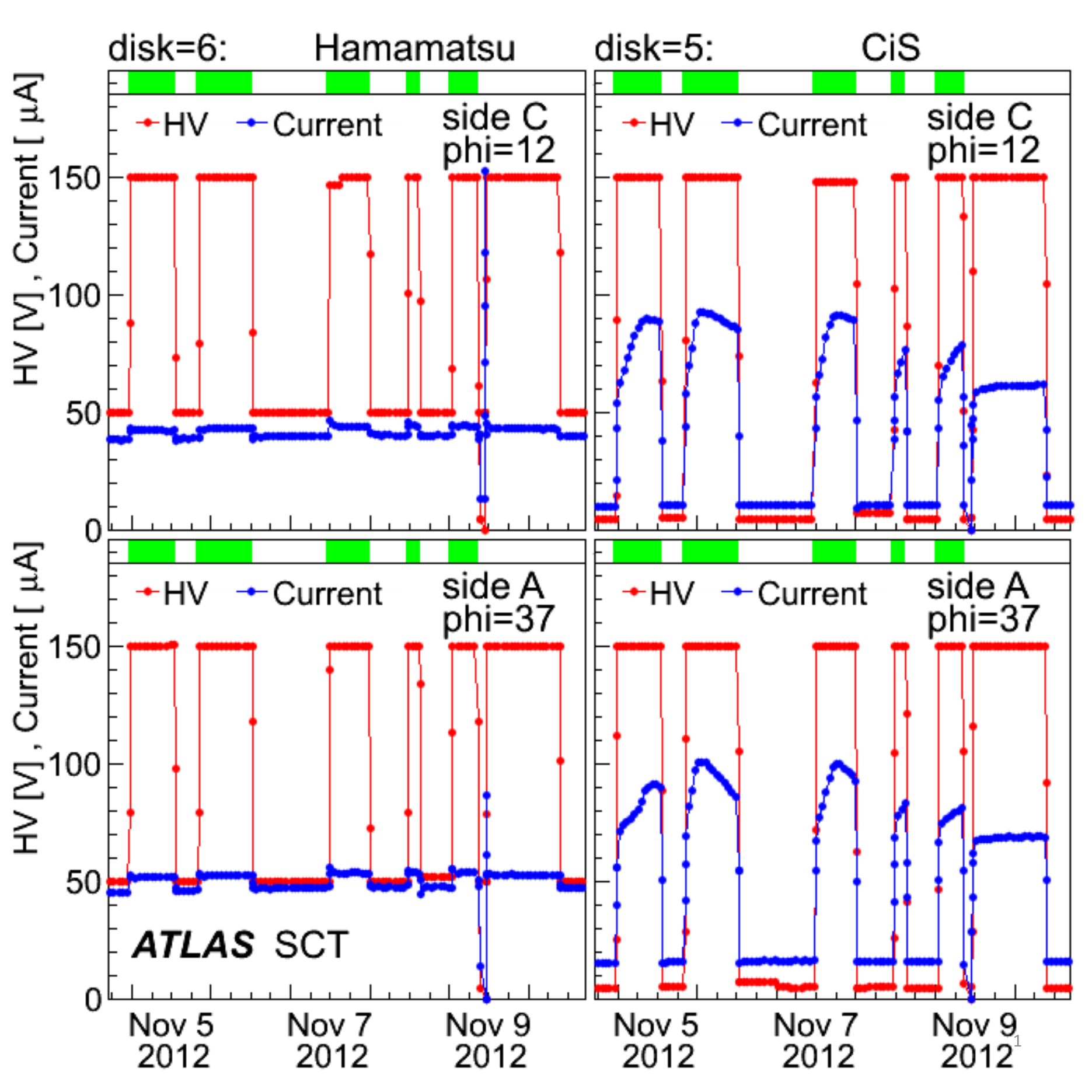} \label{cis_current_Hamamatsu-CiS}}
\caption{(a) Leakage current (blue) observed in May 2012 together with high voltage (red) for one endcap 
         module showing anomalous behaviour. 
         (b) Leakage current (blue) and high voltage (red), observed in November 2012, of typical endcap 
         modules with sensors manufactured by Hamamatsu (left) and CiS (right).
         Periods with LHC colliding beams are indicated by the green bands at the 
         top of the figure in each case.} 
\end{center}
\end{figure}

The online monitoring flags very noisy modules caused by SEUs in the ABCD registers, allowing shifters to do prompt resets. Even if a shifter fails to respond, the regular module resets limit the data loss to a tolerable level (typically the loss of one chip for up to 30~minutes).
The DAQ detects trigger synchronisation errors and resets the counters and pipelines for the module with errors. This procedure takes about 30~s and therefore reduces the data loss resulting from SEUs in the {\it p-i-n} diodes to a negligible level, as illustrated in figure~\ref{fig:error_rate}.

\section{Conclusions}
The operation and performance of the ATLAS semiconductor tracker during 2009--2013
are described in this paper. During this period, more than 99\% of detector modules were
operational, and more than 99\% of data collected by the ATLAS experiment had good SCT data quality.
The frequency-scannning interferometry system showed the position of the detector
to be stable at the micron level over long periods of time. Measurements of the increase 
in leakage currents with time are consistent with the radiation-damage predictions.
The differences between data and simulation are typically less than 30\%. This level
of agreement exceeds expectations, and provides confidence in the fluence predictions.
The verification of the simulations will be repeated at higher beam energies in future. 
Single event upsets have been identified and measured in the barrel module data, and a 
strategy for their mitigation implemented.

The detector occupancy was found to vary linearly with the number of interactions
per bunch crossing, up to the maximum of 70 interactions per crossing, where it is
less than 2\% in the innermost barrel layer (which has highest occupancy). 
The intrinsic hit efficiency of the detector was measured to be (99.74$\pm$0.04)\%,
and the noise occupancies of almost all chips remained below the design requirement of 
$5 \times 10^{-4}$.

Measured values of the Lorentz angle are compatible with model predictions within at most 
twice the estimated uncertainties on those predictions. 
The measured values for sensors with $<$100$>$ crystal orientation are approximately $1^{\circ}$ 
lower than for those with $<$111$>$ crystal orientation, contrary to the expectation that a higher 
expected charge-carrier mobility in the sensors with $<$100$>$ crystal orientation should result in 
a higher value of the Lorentz angle.

Despite the binary readout, some particle identification from energy-loss measurements
is possible: the discriminating power arises from the number of time bins above threshold
and cluster widths. 
The position of the proton energy-loss peak was found to be stable at the 5--10\% level during
2010--2012. The position of this peak may become a useful tool for monitoring radiation
damage in future.   
The production of $\delta$-rays in the silicon sensors was measured, and is found to be
in good agreement with expectations.
 
\acknowledgments




We thank CERN for the very successful operation of the LHC, as well as the
support staff from our institutions without whom ATLAS could not be
operated efficiently.

We acknowledge the support of ANPCyT, Argentina; YerPhI, Armenia; ARC,
Australia; BMWF and FWF, Austria; ANAS, Azerbaijan; SSTC, Belarus; CNPq and FAPESP,
Brazil; NSERC, NRC and CFI, Canada; CERN; CONICYT, Chile; CAS, MOST and NSFC,
China; COLCIENCIAS, Colombia; MSMT CR, MPO CR and VSC CR, Czech Republic;
DNRF, DNSRC and Lundbeck Foundation, Denmark; EPLANET, ERC and NSRF, European Union;
IN2P3-CNRS, CEA-DSM/IRFU, France; GNSF, Georgia; BMBF, DFG, HGF, MPG and AvH
Foundation, Germany; GSRT and NSRF, Greece; ISF, MINERVA, GIF, I-CORE and Benoziyo Center,
Israel; INFN, Italy; MEXT and JSPS, Japan; CNRST, Morocco; FOM and NWO,
Netherlands; BRF and RCN, Norway; MNiSW and NCN, Poland; GRICES and FCT, Portugal; MNE/IFA, Romania; MES of Russia and ROSATOM, Russian Federation; JINR; MSTD,
Serbia; MSSR, Slovakia; ARRS and MIZ\v{S}, Slovenia; DST/NRF, South Africa;
MINECO, Spain; SRC and Wallenberg Foundation, Sweden; SER, SNSF and Cantons of
Bern and Geneva, Switzerland; NSC, Taiwan; TAEK, Turkey; STFC, the Royal
Society and Leverhulme Trust, United Kingdom; DOE and NSF, United States of
America.

The crucial computing support from all WLCG partners is acknowledged
gratefully, in particular from CERN and the ATLAS Tier-1 facilities at
TRIUMF (Canada), NDGF (Denmark, Norway, Sweden), CC-IN2P3 (France),
KIT/GridKA (Germany), INFN-CNAF (Italy), NL-T1 (Netherlands), PIC (Spain),
ASGC (Taiwan), RAL (UK) and BNL (USA) and in the Tier-2 facilities
worldwide.

\appendix
\section{Estimation of the bulk leakage current using models}\label{LCmodels}

It is assumed that the whole period from the beginning of the experiment to the present time is divided into $n$ time segments.  During each time segment with a
time interval of $\delta t_i[{\rm s}]$, a sensor at a bulk temperature of $T_{i}$ is supposed to receive a 1~MeV neutron-equivalent fluence of $\delta 
\Phi^{\rm eq}_i [{\rm cm^{-2}}]$. The prediction of the leakage current $I_n(T_{\rm ref}) [{\rm A}/{\rm cm^3}]$ at the end of time segment $n$ can be 
calculated using the annealing formulae given below, where $T_{\rm ref}$ is the reference temperature used in the model.  For the comparison with data, 
the temperature scaling in equation~\ref{tempcorrection} is used to convert the predicted values to those at $0^{\circ}{\rm C}$.   

\begin{enumerate}

\item Hamburg/Dortmund model:

The Hamburg/Dortmund model for the bulk leakage current is based on the work by M.~Moll~\cite{Rad:ref-mollPhd} and O.~Krasel~\cite{Rad:ref-kraselPhd}. 

\begin{equation}
I_n(T_{\rm ref})\equiv I_n^{\rm exp}+I_n^{\rm log} =  \sum_{i=1}^{n}\delta \Phi^{\rm eq}_i\cdot \alpha_{\rm I} \cdot 
\exp\left(-\frac{t_{n,i}^{\rm I}}{\tau_{\rm I}(T_{\rm ref})}\right)
+ \sum_{i=1}^{n}\delta \Phi^{\rm eq}_i(\alpha_0^*-\beta\log(t_{n,i}^{\rm log}/t_0))
\label{iequation4}
\end{equation}

\begin{equation}
t_{n,i}^{\rm I}=\sum_{j=i}^n\delta t_j\frac{\tau_{\rm I}(T_{\rm ref})}{\tau_{\rm I}(T_j)},\,\,\,\,\,
t_{n,i}^{\rm log}= \sum_{j=i}^n\delta t_j\Theta_{\rm A}(T_j)
\label{iequation5}
\end{equation}

\begin{equation}
\frac{1}{\tau_{\rm I}(T)}=k_{\rm 0I}\cdot \exp(-\frac{E_{\rm I}}{k_{\rm B}T}),\,\,\,\,
\Theta_{\rm A}(T) = \exp\left(-\frac{E_{\rm I}^{*}}{k_{\rm B}} \left[ \frac{1}{T} -\frac{1}{T_{\rm ref}}\right] \right)
\label{iequation6}
\end{equation}

\noindent
where $k_{\rm B}$ is Boltzmann's constant, $\alpha_{\rm I}=(1.23\pm0.06)\times10^{-17}$~A/cm,
$k_{\rm 0I}=1.2_{-1.0}^{+5.3}\times10^{13}$~s$^{-1}$,
$E_{\rm I}=(1.11\pm0.03)$~eV, $\alpha_0^*=7.07\times10^{-17}$~A/cm,
$\beta=3.29\times10^{-18}$~A/cm,
$E_{\rm I}^{*}=(1.30\pm0.14)$~eV,
$T_{\rm ref}=21^{\circ}$C and $t_0=1$~min. The uncertainties on $\alpha_0^*$ and $\beta$ cannot be found in the references and are set to $\pm$10\%.

\item Harper model:

The Harper model for the bulk leakage current is based on the work by R.~Harper~\cite{Rad:ref-harperPhd}. 

\begin{equation}
I_n(T_{\rm ref})\equiv \sum_{i=1}^{n}g_{n,i}\cdot \alpha(T_{\rm ref}) \cdot \delta \Phi^{\rm eq}_i
\label{iequation1}
\end{equation}

\begin{equation}
g_{n,i} = \sum^{5}_{k=1} \left\{A_{k}
  \frac{\tau_{k}}{\Theta_{\rm A}(T_i)\delta t_i}\left[ 1-\exp\left(
      \frac{-\Theta_{\rm A}(T_i)\delta t_i}{\tau_{k}}\right)\right] 
      \exp\left(-\frac{1}{\tau_k}\sum^n_{j=i+1} \Theta_{\rm A}(T_j)\delta t_j \right)\right\}
\label{iequation2}
\end{equation}

\noindent
where $\alpha(T_{\rm ref}=20^{\circ}{\rm C})=(4.81\pm0.13)\times10^{-17}$~A/cm is the current-related damage constant. The annealing effects are 
divided into five different components ($k=$ 1 to 5) with annealing time 
constants of $\tau_k$ and amplitudes $A_k$ as given in table~\ref
{Rad:parametersHarper}. $\Theta_{\rm A}(T)$ is the Arrhenius function used to scale the annealing 
time. It gives the annealing rate at a temperature $T$ relative to the rate at the reference 
temperature of $T_{\rm ref}$:

\begin{equation}
\Theta_{\rm A}(T) = \exp\left( \frac{E_{\rm I}}{k_{\rm B}} \left[ \frac{1}{T_{\rm ref}} -\frac{1}{T}
\right] \right),\,\,\,\,\,\,E_{\rm I}=1.09\pm0.14\,{\rm eV}.
\label{iequation3}
\end{equation}

\begin{table}[htb]
\caption{Annealing time constants and amplitudes at $T_{\rm ref}=20^{\circ}$C of the Harper model.}
\label{Rad:parametersHarper}
\begin{center}
\begin{tabular}{lcc}
\hline\hline
$k$  &  $\tau_{k}$ [min]   & $A_{k}$ \\
\hline
1  &    (1.2 $\pm$ 0.2) $\times 10^6$  &  0.42 $\pm$ 0.11\\
2  &    (4.1 $\pm$ 0.6) $\times 10^4$  &  0.10 $\pm$ 0.01\\
3  &    (3.7 $\pm$ 0.3) $\times 10^3$  &  0.23 $\pm$ 0.02\\
4  &    124 $\pm$ 2.5     &  0.21 $\pm$ 0.02\\
5  &    8 $\pm$ 5         &  0.04 $\pm$ 0.03\\
\hline\hline
\end{tabular}\\[2pt]
\end{center}
\end{table}
\end{enumerate}

\clearpage
\bibliographystyle{JHEP}
\bibliography{sct2011}

\onecolumn
\clearpage
\begin{flushleft}
{\Large The ATLAS Collaboration}

\bigskip

G.~Aad$^{\rm 84}$,
B.~Abbott$^{\rm 112}$,
J.~Abdallah$^{\rm 152}$,
S.~Abdel~Khalek$^{\rm 116}$,
O.~Abdinov$^{\rm 11}$,
R.~Aben$^{\rm 106}$,
B.~Abi$^{\rm 113}$,
M.~Abolins$^{\rm 89}$,
O.S.~AbouZeid$^{\rm 159}$,
H.~Abramowicz$^{\rm 154}$
% ...
\bigskip
\\
% ...
$^{11}$ Institute of Physics, Azerbaijan Academy of Sciences, Baku, Azerbaijan\\
% ...
$^{84}$ CPPM, Aix-Marseille Universit{\'e} and CNRS/IN2P3, Marseille, France\\
% ...
$^{89}$ Department of Physics and Astronomy, Michigan State University, East Lansing MI, United States of America\\
% ...
$^{106}$ Nikhef National Institute for Subatomic Physics and University of Amsterdam, Amsterdam, Netherlands\\
% ...
$^{112}$ Homer L. Dodge Department of Physics and Astronomy, University of Oklahoma, Norman OK, United States of America\\
$^{113}$ Department of Physics, Oklahoma State University, Stillwater OK, United States of America\\
% ...
$^{116}$ LAL, Universit{\'e} Paris-Sud and CNRS/IN2P3, Orsay, France\\
% ...
$^{152}$ Institute of Physics, Academia Sinica, Taipei, Taiwan\\
% ...
$^{154}$ Raymond and Beverly Sackler School of Physics and Astronomy, Tel Aviv University, Tel Aviv, Israel\\
% ...
$^{159}$ Department of Physics, University of Toronto, Toronto ON, Canada\\
% ...
\end{flushleft}


\end{document}